\documentclass[usenatbib,fleqn,letterpaper]{mn2e} % letterpaper needed for arxiv

\pdfoutput=1

\newcommand{\lyat}{Ly$\alpha$}
\newcommand{\lya}{Ly$\alpha~$} 
\newcommand{\msun}{\ifmmode M_{\odot} \else M$_{\odot}$\fi}
\newcommand{\msunyr}{\ifmmode M_{\odot} {\rm yr}^{-1} \else
M$_{\odot}$ yr$^{-1}$\fi}
\newcommand{\kms}{\ifmmode {\rm km s}^{-1} \else km s$^{-1}$\fi}
\newcommand{\cmtwo}{\ifmmode {\rm cm}^{2} \else cm$^{2}$\fi}
\newcommand{\ergs}{\ifmmode {\rm erg s}^{-1} \else erg s$^{-1}$\fi}
\newcommand{\ergscm}{\ifmmode {\rm erg s}^{-1}{\rm cm}^{-2} \else erg s$^{-1}$ cm$^{-2}$\fi}
\newcommand{\sqarcm}{\ifmmode {\rm arcmin}^{2} \else arcmin$^{2} \:$\fi}
\newcommand{\hi}{\ifmmode {\textrm{H\textsc{i}}} \else H\textsc{i} \fi}
\newcommand{\hii}{\ifmmode {\textrm{H\textsc{ii}}} \else H\textsc{ii} \fi}

\newcommand{\xhi}{\ifmmode {x_{\scaleto{\rm HI}{3.5pt}}} \else $x_{\scaleto{\rm HI}{3.5pt}}$ \fi}
\newcommand{\xhit}{\ifmmode {x_{\scaleto{\rm HI}{3.5pt}}} \else $x_{\scaleto{\rm HI}{3.5pt}}\!\!$ \fi}

\newcommand{\affil}[1]{$^{\rm #1}$}

\newcommand{\xlya}{\ifmmode {\mathpzc{X}_{\scaleto{\rm \: Ly\alpha}{5.5pt}}} \else $\mathpzc{X}_{\scaleto{\rm \: Ly\alpha}{5.5pt}}$ \fi}
\newcommand{\xlyat}{\ifmmode {\mathpzc{X}_{\scaleto{\rm \: Ly\alpha}{5.5pt}}} \else $\mathpzc{X}_{\scaleto{\rm \: Ly\alpha}{5.5pt}}\!\!$ \fi}
\newcommand{\fesc}{$f_{\scaleto{\rm esc}{3.3pt}}\:$}
\newcommand{\tigm}{$T_{\scaleto{\rm IGM}{4pt}}\:$}
\newcommand{\tigmt}{$T_{\scaleto{\rm IGM}{4pt}}$}

\newcommand{\sphinxd}{{\scaleto{\textsc{sphinx}}{5pt}}\:}
\newcommand{\sphinxdt}{{\scaleto{\textsc{sphinx}}{5pt}}}

\newcommand{\sphinx}{{\scaleto{\textsc{sphinx}}{5pt}}\:}
\newcommand{\nhii}{\ifmmode n_{\scaleto{\rm HII}{4pt}} \else $n_{\scaleto{\rm HII}{4pt}}\:$ \fi}
\newcommand{\nhi}{\ifmmode n_{\scaleto{\rm HI}{4pt}} \else $n_{\scaleto{\rm HI}{4pt}}\:$ \fi}
\newcommand{\rascas}{{\scaleto{\textsc{rascas}}{5pt}}\:}
\newcommand{\rascast}{{\scaleto{\textsc{rascas}}{5pt}}}

\usepackage[pdftex]{graphicx}
\usepackage{epsfig}
\usepackage{amssymb}
\usepackage{natbib}
\usepackage[latin1]{inputenc}
\graphicspath{plots/}
\usepackage{amsmath}
\usepackage{fixltx2e}
\usepackage{array}
\usepackage{wrapfig}
\usepackage{relsize}
\usepackage{color}
\usepackage{lipsum}
\usepackage{subfig}
\usepackage[section] {placeins}
\usepackage{caption}
\usepackage{hhline}
\usepackage{multirow}
\usepackage{colortbl}
\usepackage[table]{xcolor}
\usepackage{hyperref}
\usepackage{afterpage}
\usepackage{scalerel}
\usepackage{newtxtext,newtxmath}
\usepackage{stmaryrd}
\DeclareFontFamily{OT1}{pzc}{}
\DeclareFontShape{OT1}{pzc}{m}{it}{<-> s * [1.100] pzcmi7t}{}
\DeclareMathAlphabet{\mathpzc}{OT1}{pzc}{m}{it}

\title[\lya as a tracer of cosmic reionisation in SPHINX]{Lyman-$\alpha$ as a tracer of cosmic reionisation in the SPHINX radiation-hydrodynamics cosmological simulation}
\author[Garel et al.]{\parbox{0.85\linewidth}{Thibault Garel\affil{1,2}\thanks{Email: \href{mailto:thibault.garel@unige.ch}{thibault.garel@unige.ch}}, Jérémy Blaizot\affil{2}, Joakim Rosdahl\affil{2}, Léo Michel-Dansac\affil{2}, Martin G. Haehnelt\affil{3}, Harley Katz\affil{4}\thanks{Visitor}, Taysun Kimm\affil{5} and Anne Verhamme\affil{1}}\\\\
\vspace{-0.1cm}
{\affil{1}\, Observatoire de Genève, Université de Genève, 51 Ch. des Maillettes, 1290 Versoix, Switzerland}\\
\vspace{-0.1cm}
{\affil{2}\,Univ Lyon, Univ Lyon1, Ens de Lyon, CNRS, Centre de Recherche Astrophysique de Lyon UMR5574, F-69230, Saint-Genis-Laval, France}\\
\vspace{-0.1cm}
{\affil{3}\,Kavli Institute for Cosmology and Institute of Astronomy, Madingley Road, Cambridge CB3 0HA, UK}\\
\vspace{-0.1cm}
{\affil{4}\,Sub-department of Astrophysics, University of Oxford, Keble Road, Oxford OX1 3RH, UK}\\
\vspace{-0.1cm}
{\affil{5}\,Department of Astronomy, Yonsei University, 50 Yonsei-ro, Seodaemun-gu, Seoul 03722, Republic of Korea}}
\vspace{-0.4cm}

\begin{document}

\pdfminorversion=5
\pdfobjcompresslevel=2

\pagerange{\pageref{firstpage}--\pageref{lastpage}} \pubyear{2021}

\maketitle
\label{firstpage}

\begin{abstract}
The \lya emission line is one of the most promising probes of cosmic reionisation but isolating the signature of a change in the ionisation state of the IGM is challenging because of intrinsic evolution and internal radiation transfer effects. We present the first study of the evolution of \lya emitters (LAE) during the epoch of reionisation based on a full radiation-hydrodynamics cosmological simulation that is able to capture both the large-scale process of reionisation and the small-scale properties of galaxies. We predict the \lya emission of galaxies in the $10^3$ cMpc$^3$ \sphinx simulation at $6\leq z\leq9$ by computing the full \lya radiation transfer from ISM to IGM scales. \sphinx is able to reproduce many observational constraints such as the UV/\lya luminosity functions and stellar mass functions at z $\gtrsim$ 6 for the dynamical range probed by our simulation ($M_{\rm 1500}\gtrsim-18$, $L_{\rm Ly\alpha}\lesssim10^{42}$ \ergs, $M_{\star}\lesssim10^9$\msun). As intrinsic \lya emission and internal \lya escape fractions barely evolve from $z=6$ to 9, the observed suppression of \lya luminosities with increasing redshift is fully attributed to IGM absorption. For most observable galaxies ($M_{\rm 1500}\lesssim-16$), the \lya line profiles are slightly shifted to the red due to internal radiative transfer effects which mitigates the effect of IGM absorption. Overall, the enhanced \lya suppression during reionisation traces the IGM neutral fraction \xhi well but the predicted amplitude of this reduction is a strong function of the \lya peak shift, which is set at ISM/CGM scales. We find that a large number of LAEs could be detectable in very deep surveys during reionisation when \xhi is still $\approx 50\%$.
\end{abstract}


\section{Introduction}
\label{sec:intro}

Cosmic reionisation is one of the most fundamental stages in the history of the Universe, marking the end of the Dark ages and the formation of the first luminous sources. A patchy scenario in which \hii regions expand around ionising sources until filling up the entire Universe is currently favoured but a thorough understanding of this process remains challenging. In spite of intense research over the last decades, there is still no consensus regarding the nature of the objects which reionised the intergalactic medium and the timeline over which it occurred.

While active galactic nuclei certainly contributed to the global ionising photon budget, there is growing evidence that stellar emission within galaxies is the dominant source \citep{Kulkarni_2019b,Parsa_2017,Finkelstein_2019}. Nevertheless, the relative contribution of low-mass versus massive galaxies still needs to be assessed due to uncertainties in the abundance of faint dwarfs during the epoch of reionisation \citep[EoR;][]{Livermore_2017,Bouwens_2015,Atek_2018,Bhatawdekar_2019} and in the ability of ionising photons to escape \citep{Robertson_2015,Wise_2014,Rosdahl_2018,Ma_2016,Kimm_2014,Paardekooper_2015}. The direct measurement of the ionising Lyman continuum (LyC) escape fraction is impossible at high redshift because of the high opacity of the intergalactic medium (IGM) but observations of low-redshift analogs suggest that the typical fraction of photons able to escape galaxies is low \citep[$\lesssim$ 10 \%; ][]{Steidel_2018,Izotov_2016,Grazian_2015}, even though a handful of strong leakers have been reported \citep[$\gtrsim 50 \%$; e.g.][]{Vanzella_2016b,Izotov_2018}. In parallel, quasar absorption spectra suggest that the Universe was almost fully ionised at z $\approx$ 5-6 \citep{fan2006a,Mesinger_2010,Kulkarni_2019} and still partially neutral at z $\gtrsim$ 7 \citep[e.g.][]{Ba_ados_2017,Davies_2018,durovcikova_2020}. 

In addition to future 21 cm observations, one of the most promising routes to probe the EoR resides in Lyman-$\alpha$ (hereafter \lyat) surveys. It is well known that the strong \lya line produced in galaxies can be used as an indirect measurement of the neutral IGM component since \lya photons can be scattered off the line of sight by intervening \hi atoms. As the Universe becomes more neutral towards higher redshifts, the visibility of \lya emitters (hereafter LAEs) will drop and the imprint of reionisation should translate into a shift of the \lya luminosity function \citep[LF;][]{haiman05,dijk07b}. Hints for such behaviour have been indeed reported in various narrow-band surveys at z $\gtrsim$ 6 \citep{ouchi2010a,Konno_2014,Zheng_2017}. A similar signature of reionisation is also seen in UV-selected samples where the fraction of objects with strong \lya emission, \xlyat, is first found to increase from z $\approx$ 3 to 6 and then to decline at higher redshift \citep{stark10,Pentericci_2018,Hoag_2019}. While this trend is often interpreted as a rapid increase of the volumetric IGM neutral fraction (\xhit) at z $\gtrsim$ 6, it is noticeable that the significance and the redshift of the drop often differ from one study to another \citep[see e.g.][]{Kusakabe_2020,Stark_2016,Fuller_2020}. This may be a consequence of the patchiness of the reionisation process, or simply due to the different depths, selections and limited statistics of the samples used to compute \xlyat. 

Altogether, these diagnostics can be used to assess the variation of the visibility of LAEs and therefore probe the evolution of the ionisation state of the IGM. However it is not necessarily straightforward to disentangle the impact of IGM attenuation from the intrinsic evolution of the \lya emission and galactic radiative transfer (RT) effects \citep{dayal2012,laursen2011a,jensen2013a,garel2015a,Hassan_2021}. Intrinsic \lya luminosities are usually assumed to scale linearly with star formation rate. Still, this relation may evolve at high redshift if recombination of photoionised gas is no longer the dominant production channel of \lya photons, or if very low metallicities are involved \citep[][]{Laursen_2019,Smith_2018,Raiter_2010}. In addition, the \lya line is very sensitive to resonant scattering in the interstellar medium (ISM) and circumgalactic medium (CGM), i.e. the material inside and close to galaxies. First, the enhanced distance travelled by \lya photons due to local scatterings increases dust absorption which can significantly suppress the flux emerging from galaxies. Second, RT in the optically thick regime can strongly affect the line profile and shift it away from resonance, especially in non-static media. In the presence of outflows, this effect can tremendously reduce the relative impact of the IGM on the visibility of LAEs \citep{santos04, dijkstra2010a, garel2012a,Mason_2018}. 
 
The modelling of the LAE population during the EoR is therefore a multi-scale problem which ideally requires to self-consistently describe the production and transfer of \lya photons at small scales in galaxies as well as their propagation in the intergalactic medium. Such simulations are computationally expensive because of (i) the wide dynamical range involved, (ii) the need for radiation-hydrodynamics (RHD) to account for the interplay between ionising radiation and the gas, and (iii) the full post-processing with \lya RT. Several studies have focused on individual objects but neglecting the IGM component \citep{verhamme2012,yajima2014a,Smith_2018}. Alternatively, the transmission of the \lya line through the IGM has been investigated in representative simulation volumes \citep{dayal2011a, Hutter_2014,Gronke_2020, Jensen_2014,Inoue_2018}. This is however at the expense of the physical and mass resolution which is needed to model the \lya emission and transfer within the ISM and the CGM. To overcome some of these issues, \citet{Laursen_2019} have recently built a new hybrid framework to model hundreds of \lya sources at z $\approx$ 9. Their approach combines a semi-analytical scheme to predict the halo mass function with high-resolution hydrodynamic zoom simulations in which both ionising and \lya RT are performed as a post-processing step.

In this paper, we present a new study of the evolution of LAEs during the EoR based on the \sphinx simulation project \citep{Rosdahl_2018}. \sphinx is a set of full RHD cosmological simulations of the formation and evolution of galaxies at $z > 6$. In the current study, we use exclusively the $10^3$ cMpc$^3$ version of \sphinx which includes the effect of binary stars with BPASS v2.0 to fully reionise the simulated volume before redshift six \footnote{Although there are now several \sphinx simulations, we will for simplicity refer to this $10$ cMpc simulation with BPASS v2.0 throughout this paper as \sphinxdt.}. Taking advantage of the adaptative mesh refinement code {\scaleto{\textsc{ramses-rt}}{5.1pt}} \citep{teyssier02,Rosdahl_2013}, we are able to capture a wide range of scales with \sphinxdt. Here, we intend to assess the relative impact of intrinsic evolution, absorption at galaxy scales and IGM transmission to predict to which extent the visibility of LAEs is tracing the IGM neutral fraction during the EoR. The \sphinxd simulation is therefore well-suited since it allows us to investigate the transport of \lya photons from the ISM to the IGM for a large sample of objects.

The outline of the article is as follows. Section 2 describes the \sphinx simulation project and our modelling of \lya emission and transfer. In Section 3, we compare our results with statistical observational constraints (stellar mass, UV/\lya LFs, LAE fraction) and assess the relative evolution of the \lya IGM transmission compared to \lya intrinsic emission and escape fraction during the EoR. Then we attempt to characterise the imprint of the IGM on the \lya LF, equivalent width (EW) distribution, LAE fraction, and spectra as a function of \xhit. We discuss our results in Section 4 and we give a summary in Section 5.  

\section{Simulation and method}
In this section, we describe the \sphinx simulation suite and the \rascas radiation transfer code that we use to post-process the \sphinxd outputs.

\subsection{The SPHINX simulation}
\label{subsec:sphinx}

\sphinx is a set of cosmological radiation-hydrodynamics simulations of galaxy formation during the epoch of reionisation. It has been run with the 3D adaptive mesh refinement code RAMSES-RT \citep{Rosdahl_2013} to describe the evolution of dark matter, baryons, and ionising radiation via gravity, hydrodynamics, RT, and non-equilibrium radiative cooling/heating. The \sphinx simulation suite has been presented in \citet{Rosdahl_2018} and \citet{Katz_2020} and here we recall the main features that are relevant to our study.

\subsubsection{Numerical setup}

\citet{Rosdahl_2018} have explored several simulations with various sizes, mass resolutions and SED models. Here, we make use of the fiducial simulation of the \sphinx project which describes a $V_{\rm box} = 10^3$ cMpc$^3$ volume, and includes the effects of binary stars, a maximum physical resolution of $10.9$ pc (at z$=$6), and $512^3$ dark matter particles of mass $m_{\rm DM} = 2.5 \times 10^5$ \msun. 

In \sphinxdt, the hydrodynamics are solved using the HLLC Riemann solver \citep{Toro_1994} and a MinMod slope limiter. An adiabatic index of $5/3$ is assumed to close the relation between gas pressure and internal energy. Gravitational interactions for DM and stellar particles are computed with a particle-mesh solver and cloud-in-cell interpolation following \citet{Guillet_2011}. The radiation is advected between cells using the M1 closure method \citep{Levermore_1984} and the Global-Lax-Friedrich intercell flux function.

\begin{figure*}
\includegraphics[width=0.9\textwidth]{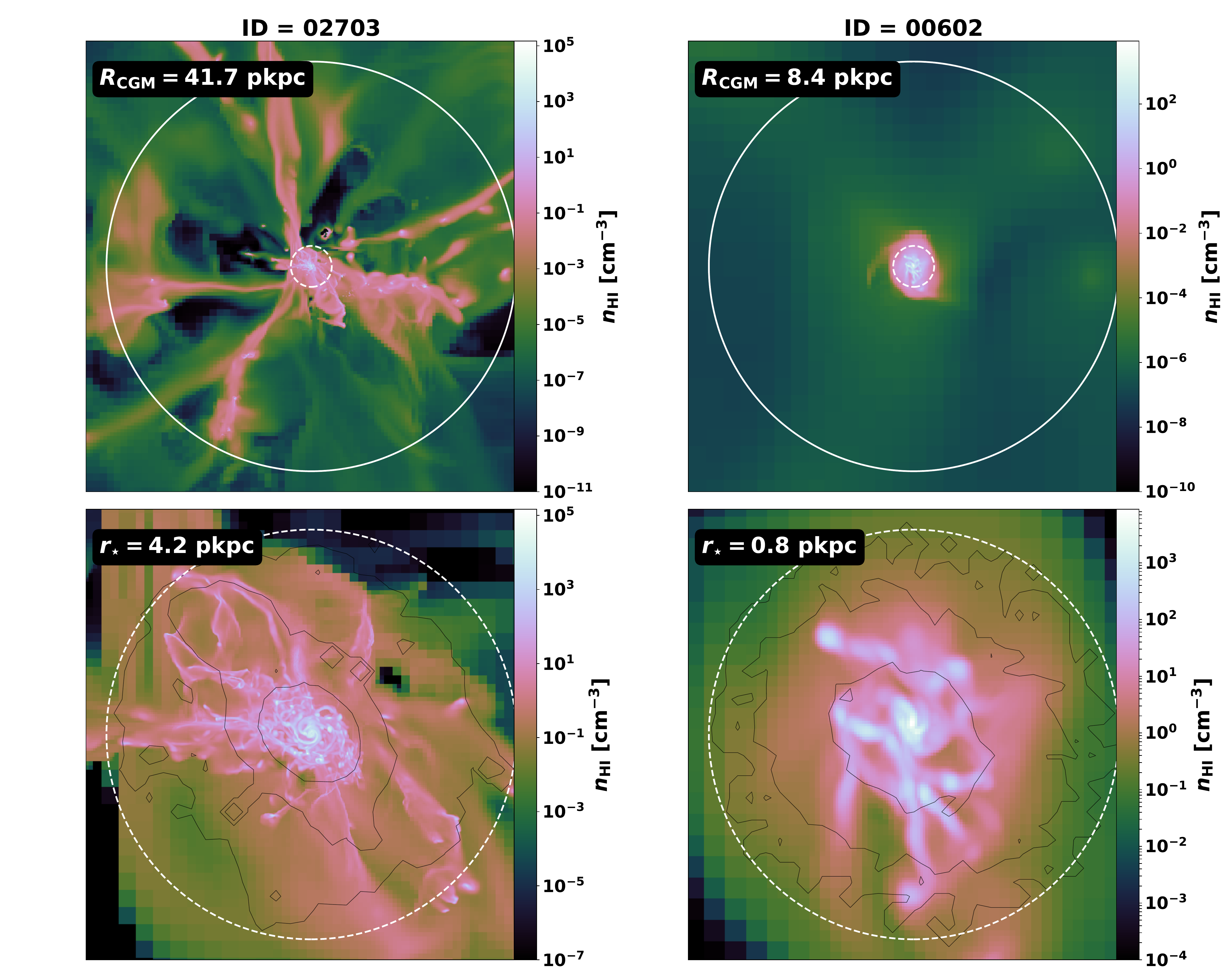}
\vskip-2ex
\caption{Projections of two galaxies from \sphinxd at $z=6$ showing the \hi density maps at the CGM scale (top) and at the ISM scale (bottom). The left panels correspond to one of the most massive galaxies in our simulation (ID2703) while the right panels present a relatively low-mass object (ID602). The black contours in the lower panels indicate the distribution of stars within the ISM, with stellar surface density levels of $10^5$, $10^6$ and $10^7$ \msun{} kpc$^{-2}$. The white solid and dashed circles depict the CGM radius ($R_{\scaleto{\rm CGM}{3.5pt}}$) and the stellar radius ($r_\star$) respectively (the values are given in the legend of each panel; see Section \ref{sec:internal_rt} for the definition of $R_{\scaleto{\rm CGM}{3.5pt}}$). The stellar mass and star formation rate (computed over the last 10 Myr) for ID2703 (resp. ID602) are $1.5 \times 10^9$ \msun{} (resp. $10^7$ \msun) and $3$ \msunyr{} (resp. $0.005$ \msunyr).}
\label{fig:maps_gals}
\end{figure*}

The initial conditions (IC) were generated with MUSIC \citep{Hahn_2011} assuming cosmological parameters consistent with the Planck results \citep[][$h_{100} = 0.6711$, $\Omega_\Lambda = 0.6825$, $\Omega_{\rm m} = 0.3175$, $\Omega_{\rm b} = 0.049$, and $\sigma_8 = 0.83$]{Planck_2014} and they were chosen from a set of DM-only simulations so as to obtain a representative sample of sources that minimises the effect of cosmic variance on the ionising radiation budget. This was done using a large number of simulations with different IC realisations and the chosen set of ICs corresponds to the one yielding a sample of objects which generates an average ionizing luminosity budget \citep[Section 2.2.1 in][]{Rosdahl_2018}.
Regarding the primordial abundance of chemical elements, we have adopted a mixture of hydrogen ($X=0.76$), helium ($Y=0.24$) and metals ($Z=6.4 \times 10^{-6}$) where the initial metallicity value was chosen to account for the lack of molecular cooling at early stages, such that first stars can start forming by z $\approx15-20$.

\subsubsection{Baryonic physics and stellar library}

The gas cooling implementation includes the contribution of both primordial species and metal lines, following the prescription presented in \citet{Rosdahl_2013}.  
Star formation (SF) is modelled using a recipe adapted from \citet{Federrath_2012} where turbulent gas motions act as an additional pressure support against gravitational collapse. As described in \citet{Rosdahl_2018}, stars can form in a grid cell when the local density corresponds to a maximum and is greater than 200 times the cosmological mean, the gas motion is locally convergent, and the turbulent Jeans length is less than one cell width. Gas cells meeting these criteria can produce stars according to a Schmidt law with a varying SF efficiency that depends on the local thermoturbulent properties of the gas \citep[see][for details]{Kimm_2017}. In each simulation cell, the gas is stochastically converted into stellar particles by sampling the Poisson probability distribution for gas to star conversion over the timestep \citep[see][for details]{Rasera_2005}, such that on average, the conversion rate follows the Schmidt law \citep[Eq. 3 in][]{Rosdahl_2018}. Initially, stellar particles, each representing a stellar population, are allocated masses equal to integer multiples of $10^3$ \msun. An upper limit is set such that no more than $90\%$ of the cell gas can turn into stars. As shown in \citet{Trebitsch_2017}, this recipe leads to a much more bursty SF than typical models based on a constant SF efficiency.

Stellar evolution and feedback is modelled following \citet{Kimm_2015} by injecting mass, metal and momentum to surrounding gas cells. In practice, Type II SN explosions are stochastically sampled from the delay-time distribution for the \citet{Kroupa_2001} IMF over the first 50 Myr of the lifetime of each star particle. We assume that each star particle hosts 4 SN events per 100 \msun, which is four times larger than the typical SN frequency computed for the Kroupa IMF (1 per 100 \msun), in order to avoid overcooling and reproduce observational constraints at $z\approx6$ \citep{Rosdahl_2018}. 

Spectral energy distributions are computed using the BPASS library \citep[][]{Eldridge_2008} which includes the effect of interacting binary stars (assuming $100\%$ of stars are in binary systems) with metallicities and ages in the range $0.001-0.4$ and 1 Myr $-$10 Gyr respectively. As shown in \citet{Rosdahl_2018}, this choice of stellar library produces a much earlier reionisation history than an identical simulation with single stars only as it can fully reionise the box by z $\approx$ 7 whereas the IGM is still $\approx 50\%$ neutral at z$=$6 with the single star model. The discrepancy is mainly due to two factors. First, the binary model produces more ionising photons for a given stellar population, especially at low metallicities. Second, the ionising emission is prolonged for interacting binaries with respect to single stars (e.g. 25 Myr after a starburst, the ionising luminosity is $\gtrsim 10$ times larger with binaries) which leaves more time for SN feedback to clear the gas away from dense regions, allowing photons to escape more easily into the IGM. 

The ionising radiation is injected directly into the cells hosting stellar particles in each simulation step and propagated through the volume using the so-called M1 moment method. The radiation is split into three monochromatic groups bracketed by the \hi, He\textsc{i}, and He\textsc{ii} ionisation energies. The simulation tracks the local non-equilibrium ionisation fractions of hydrogen and helium and radiation interacts with the gas via photoionisation, heating, and momentum transfer.

As shown in \citet{Rosdahl_2018}, \sphinxd starts ionising the Universe subsequently to the formation of the first stars and reionisation proceeds through the growth of \hii bubbles until filling the whole volume with ionised hydrogen by redshift $\approx 7$. The reionisation history in \sphinxd seems to occur over a similar timescale as estimated from observations but completes slightly too early (i.e. by $\Delta z \approx 0.5$) with respect to these observational constraints \citep[see Figure 9 in][]{Rosdahl_2018}.

\subsubsection{Galaxy catalog}
 
The present study aims at following the evolution of LAEs during the EoR and thus we decide to focus on the four snapshots of the simulation corresponding to z $=$ 6, 7, 8 and 9. At each snapshot, we identify individual galaxies with ADAPTAHOP \citep{Aubert_2004,tweed09} and select groups with at least 100 star particles ($M_{\star,min} = 10^5$ \msun) and a local density threshold $\rho_{\rm th}=1000$ following the notation of \citet{Aubert_2004}. These values have been chosen so as to avoid spurious identifications and to maximise the association of star particles with galaxies. The galaxy size is returned by the galaxy finder and corresponds to the distance from the furthest star particle to the mass center. This value, defined as the stellar radius $r_\star$, ensures that it encompasses the bulk of the photon budget produced within the interstellar medium in order to compute the \lya and UV intrinsic emissivities of each galaxy (see Section \ref{subsubsec:emission}). This methodology allows to construct a statistical sample of simulated galaxies at each redshift of interest, yielding 2911, 2357, 1867 and 1353 sources at z $=$ 6, 7, 8, and 9 respectively.

In Figure \ref{fig:maps_gals}, we present examples of a bright/massive galaxy (left) and a relatively faint and less massive galaxy (right) from \sphinxd at $z=6$. The images show their \hi density maps at the CGM scale (top) and at the ISM scale (bottom). The black contours in the lower panels represent the distribution of stars within the ISM, with increasing levels of stellar surface density from $10^5$ to $10^7$ \msun{} kpc$^{-2}$. Figure \ref{fig:maps_gals} highlights the high level of details that can be resolved in the internal structure of our galaxies and their surrounding medium, as well as the large dynamical range that can be probed with gas densities spanning many orders of magnitudes. Note that galaxies in our simulation display a wide diversity of morphologies so these two objects, which have been chosen arbitrarily, are not necessarily representative of the global population. 

\subsection{\lya and UV post-processing}
\label{subsec:lya_uv}

\subsubsection{The \rascas code}
\label{subsubsec:rascas}

The emission and transport of \lya and (non-ionising) UV photons is performed in post-processing using the 3D Monte-Carlo RT code \rascas \citep{Michel_Dansac_2020}. \rascas has been specifically designed to ingest large simulations like \sphinxd using full MPI parallelization, domain decomposition and adaptive load balancing in order to predict intrinsic emissivities from the gas/stars and the transfer of resonant lines (as well as non-resonant lines or continuum) in the presence of dust. 

\rascas generates the intrinsic emission for each source (i.e. a gas cell or a star particle) of interest with a given number of photon packets according to its luminosity, each photon packet being assigned a constant weight. Photon packets are cast isotropically from the source with a probability $P$ which is given by $P = \dot{N}^{\rm intr}_{\lambda} / \dot{N}^{\rm intr}_{\lambda, \rm tot}$ where $\dot{N}^{\rm intr}_{\lambda}$ is the true number of emitted photons per unit time by the source and $\dot{N}^{\rm intr}_{\lambda, \rm tot}$ is the sum over all sources, such that $\dot{N}^{\rm intr}_{\lambda, \rm tot} = $ \( \sum\limits_{i}\) $\dot{N}^{\rm intr}_{\lambda, i}$.

The subsequent propagation of photon packets through the mesh is performed based on a Monte-Carlo procedure which includes the core-skipping algorithm of \citet{Smith_2015}. The interaction with matter is set by the optical depth of a mixture of hydrogen and dust \citep[see Section 3 in][]{Michel_Dansac_2020}. While \lya photons can interact with \hi atoms and dust, UV continuum photons only interact with the latter. When a photon interacts with a dust grain, it can either be absorbed or scattered with a probability set by the albedo, $A$. Following \citet{Li_2001}, we assume $A=0.32$ at \lya and $A=0.38$ at $1500$ \AA.

The formation of dust grains is not modelled in \sphinxd so we use the default dust model implemented in \rascast, and based on the formulation of \citet{laursen09}, to compute the effective dust content of each cell. With this prescription, the dust absorption coefficient is given by $(\nhi + f_{\rm ion} \nhii)\sigma_{\rm dust}(\lambda)Z/Z_0$ in each cell, where $Z$ is the gas metallicity and $f_{\rm ion}=0.01$ is a free parameter representing the relative dust abundance in ionised gas. The effective dust cross-section per H atom $\sigma_{\rm dust}(\lambda)$ and the $Z_0=0.005$ parameter are normalised to the Small Magellanic Cloud (SMC) extinction curve, as in e.g. \citet{laursen09} and \citet{Smith_2018}. As noted in \citet{laursen09}, the SMC is hosting younger stellar populations than the Milky-Way (MW) or Large Magellanic Cloud (LMC) so the SMC normalisation is presumably more appropriate when applied to low-mass galaxies at high redshift like in \sphinxd \citep[see also][]{Reddy_2012}. These authors also show that the \lya escape fraction from galaxies varies only by a few percent when switching from the SMC to the LMC normalisation\footnote{Shallower extinction curves \citep[based on e.g. the MW or SN-like dust formation scenarios;][]{Gallerani_2010} would further increase escape fractions compared to the SMC or LMC cases by a few percent, so the impact on our results would be almost negligible.}. \\

In the present study, we run \rascas on all galaxies identified in the catalog at z $=$ 6, 7, 8, and 9, both for \lya and the UV continuum 1500 \AA{} band in order to compare \sphinxd results with existing observational data. The main goal of the current study being the analysis of the co-evolution of the \lya intrinsic properties, internal attenuation by dust and IGM transmission, we explicitly describe these three steps separately in the following subsections. 

\subsubsection{\lya and UV emission}
\label{subsubsec:emission}

\begin{figure*}
\includegraphics[width=0.9\textwidth]{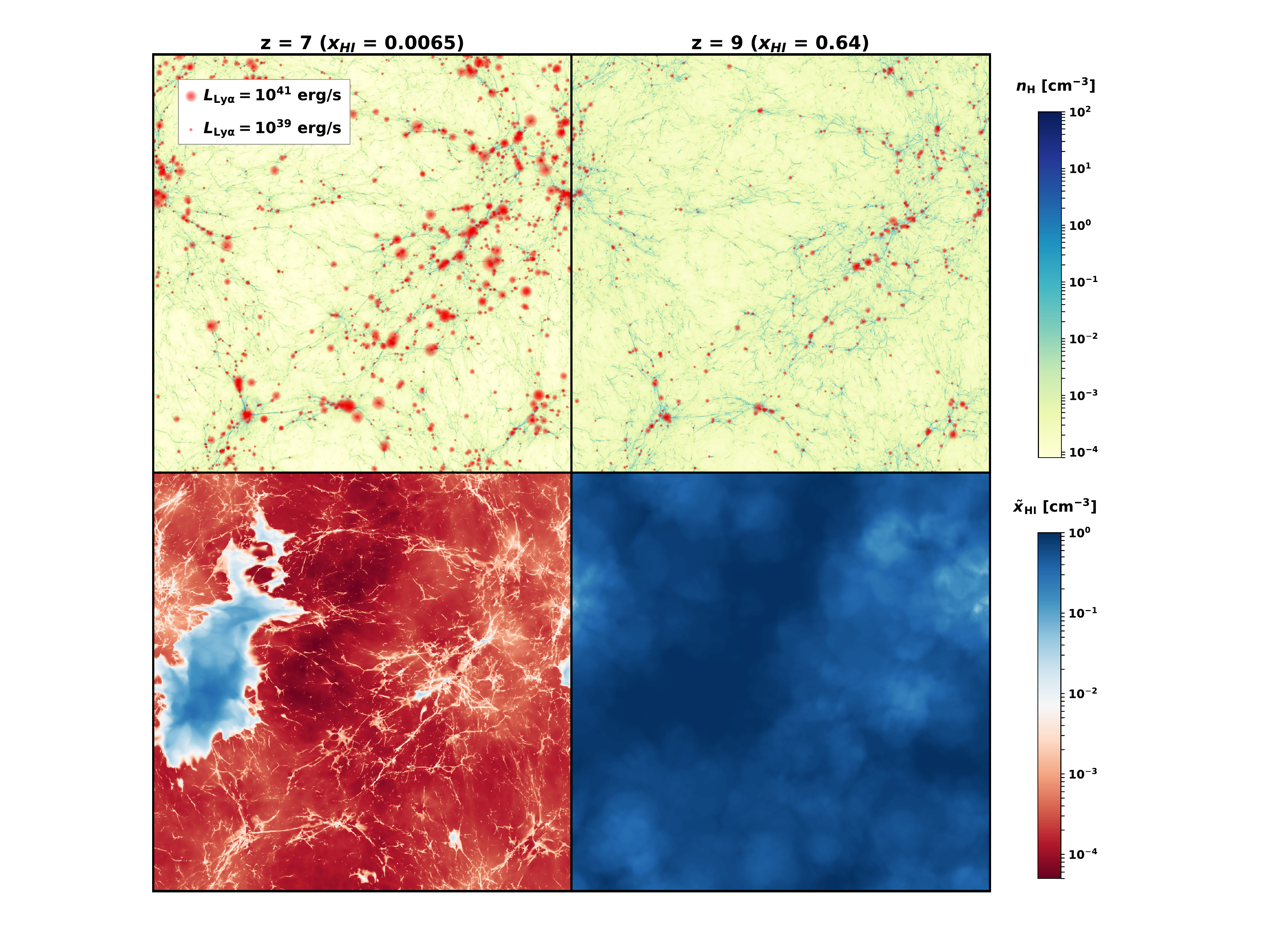}
\vskip-2ex
\caption{Projected maps of the \sphinxd volume at $z=7$ (left) and $z=9$ (right). The top panels represent the hydrogen mass-weighted density distribution ($n_{\rm H}$). LAEs are painted on top of the density map as red dots. The size of the dots scales with the \lya luminosity after internal and IGM transfer, ranging from $10^{38}$ to $10^{42}$ \ergs. The bottom panels show the local volume-weighted hydrogen neutral fraction, $\tilde{x}_{\scaleto{\rm HI}{3.5pt}}$.}
\label{fig:maps}
\end{figure*}

The intrinsic emission of \lya and UV photons from each galaxy is computed from the gas and stars within $r_\star$ respectively. The \lya production occurs through two different channels, namely recombinations and collisions, arising from the gas cells. The total number of isotropically-emitted \lya photons per unit time in a gas cell is given by $\dot{N}^{\rm intr}_{\rm Ly\alpha} = \dot{N}^{\rm intr}_{\rm Ly\alpha, \rm rec} + \dot{N}^{\rm intr}_{\rm Ly\alpha, \rm coll}$ where:
\begin{flalign}
\dot{N}^{\rm intr}_{\rm Ly\alpha, \rm rec} &= n_e \: \nhii \: \epsilon^{\rm B}_{\rm Ly\alpha}(T) \: \alpha_{\rm B}(T) \: \mathrm{d}V && \\ \nonumber
\dot{N}^{\rm intr}_{\rm Ly\alpha, \rm coll} &= n_e \: \nhi \: C_{\rm Ly\alpha}(T) \: \mathrm{d}V
\end{flalign}
For the recombination term $\dot{N}^{\rm intr}_{\rm Ly\alpha, \rm rec}$, $n_e$ and $\nhii$ are respectively the electron and proton number densities,
directly predicted by the simulation. $\alpha_{\rm B}(T)$ is the
case B recombination coefficient \citep{Hui_1997}, $\epsilon^{\rm B}_{\rm Ly\alpha}(T)$
is the fraction of recombinations leading to a \lya emission \citep{Cantalupo_2008}, and
$\mathrm{d}V$ is the volume of the cell. For the collision term $\dot{N}^{\rm intr}_{\rm Ly\alpha, \rm coll}$, $\nhi$ is the number density of neutral H atoms, and $C_{\rm Ly\alpha}(T)$
is the rate of collisional excitations from $1s$ to $2p$. 
In practice, we cast 20,000 photon packets per galaxy and sample the frequencies in the rest-frame of the cells according to a Gaussian distribution with a width set by the local thermal velocity of the gas and centred on the \lya resonance wavelength $\lambda_\alpha = 1215.67$ \AA. 

For the UV continuum, the intrinsic stellar emission (directly given by the BPASS library) is distributed over $10^6$ photon packets per galaxy emitted in the rest-frames of the star particles. A detailed description of the spatial and spectral sampling procedures is given in Section 2 of \citet{Michel_Dansac_2020}.

\subsubsection{Internal \lya and UV radiative transfer}
\label{sec:internal_rt}

\begin{figure*}
\hspace*{-0.7cm}  
\includegraphics[width=0.98\textwidth]{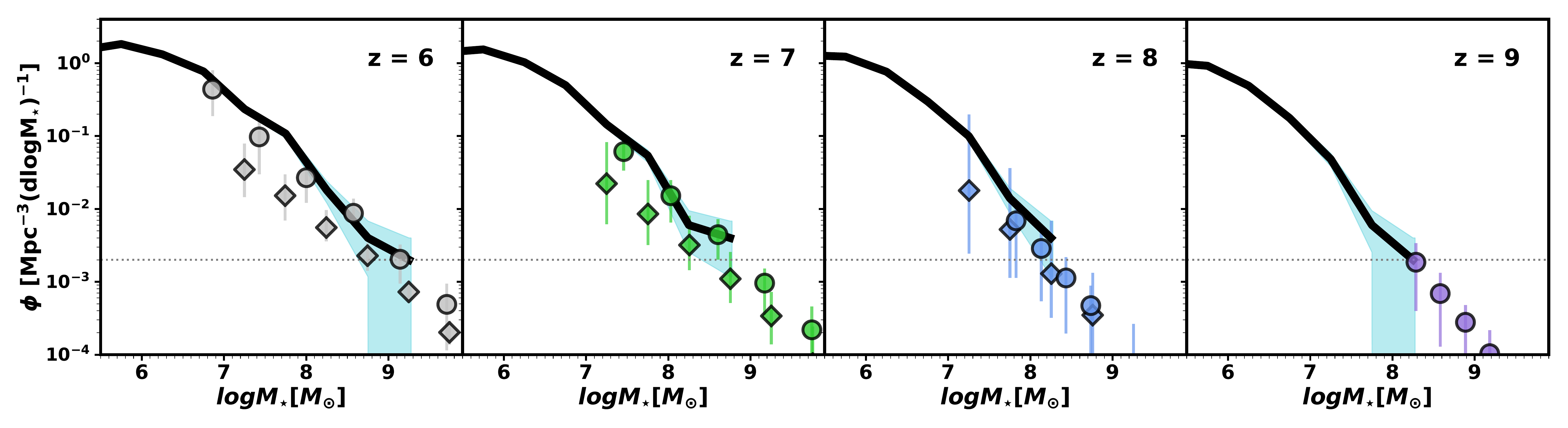}
\vskip-2ex
\caption{Stellar mass functions at z $=$ 6, 7, 8 and 9. We compare our results (black curves) with observational data from \citet{Song_2016} (diamonds) and \citet{Bhatawdekar_2019} (circles). The cyan shaded area in each panel represents the statistical error ($\propto \sqrt{N}$) on the number counts in each bin of log$M_{\star}$. The grey horizontal dotted line indicates our volume limit of one object per bin.}
\label{fig:smf}
\end{figure*}

To follow the radiation transport through the ISM and CGM, we define a characteristic radius, $R_{\scaleto{\rm CGM}{3.5pt}}= 10r_\star$, at which we evaluate the escape fraction for \lya and the UV continuum before photons enter the IGM. The choice of this particular value for the CGM radius is twofold. First, this radius needs to be large enough such that the escape fraction is converged and that dust attenuation is no longer effective beyond $R_{\scaleto{\rm CGM}{3.5pt}}$. The second reason is inherent to \lya RT numerical experiments in the IGM in which we assume that a \lya photon is removed from the line of sight (i.e. \textit{not transmitted} to the observer) if it scatters once along its path in the IGM (see the next section). It is well-known that \lya photons keep being resonantly scattered in and out of the line of sight in the densest parts of the CGM, leading to the observed extended emission around star forming galaxies \citep{steidel2011a,Wisotzki_2016}. Thus, a common assumption is to choose $R_{\scaleto{\rm CGM}{3.5pt}}$ such that the number of photons that still scatter at this scale becomes small \citep[see e.g.][]{laursen2011a,Gronke_2020}. We perform a series of convergence tests that are presented in Appendix \ref{appendix:rcgm} from which we choose to set $R_{\scaleto{\rm CGM}{3.5pt}}$ to $10r_\star$ to separate the internal RT (i.e. ISM and CGM) and the IGM RT regions. Note that the shape of the \lya profiles and the \lya escape fractions are only weakly dependent on our choice of the $R_{\scaleto{\rm CGM}{3.5pt}}$ values (see Section \ref{subsec:spectra}). 

In practice, the angle-averaged escape fractions after internal transfer in the ISM/CGM are computed as follows : $f_{\scaleto{\rm esc}{3.3pt}} = \int \dot{N}^{\scaleto{\rm CGM}{3.5pt}}_{\lambda} (hc/ \lambda) \mathrm{d}\lambda / \int \dot{N}^{\rm intr}_{\lambda} (hc/ \lambda) \mathrm{d}\lambda = L^{\scaleto{\rm CGM}{3.5pt}}_\lambda / L^{\scaleto{\rm intr}{4pt}}_\lambda$ where $\dot{N}^{\rm intr}_\lambda$ and $\dot{N}^{\scaleto{\rm CGM}{3.5pt}}_\lambda$ are the total number of emitted photons per unit time and the total number of emitted photons per unit time that escape at $R_{\scaleto{\rm CGM}{3.5pt}}$. $L^{\scaleto{\rm intr}{4pt}}_\lambda$ and $L^{\scaleto{\rm CGM}{3.5pt}}_\lambda$ refer to the intrinsic luminosity and the dust-attenuated luminosity respectively. We do not choose any particular direction to estimate the escape fractions, so $f_{\scaleto{\rm esc}{3.3pt}}$ is an angle-average quantity computed by summing photons over all directions. Therefore, $L^{\scaleto{\rm CGM}{3.5pt}}_\lambda$ corresponds to the mean escaping luminosity.

\subsubsection{IGM \lya radiative transfer}
\label{sec:igm_rt}

Once they reach $R_{\scaleto{\rm CGM}{3.5pt}}$, \lya photons continue their propagation through the IGM in the whole simulation volume using the periodic boundary conditions.
We modify the \rascas code to account for the Hubble flow by adding a velocity component, $V_{\rm hub}$, to the cell gas velocity. The \lya scattering probability is a sharp function centred at the \lya resonance (i.e. a Voigt profile $\Phi(x)$) that varies sensitively with $x$, the frequency shift expressed in Doppler units. It is therefore important to compute it at the correct $x$, especially when the Hubble flow within a given cell becomes non negligible compared to the thermal gas velocity $V_{\rm th}$ \citep[see e.g.][]{jensen2013a,Behrens_2019}.
Hence, we introduce an adaptive scheme to propagate photons within cells in order to accurately evaluate the probability of interaction between a \lya photon and an H atom. In practice, photons are walked over substeps in velocity space that remain small compared to the gas thermal motion in the cell and the local variation of the Voigt profile. 

If a photon ever happens to scatter, we consider that it is removed from the line of sight and will not transmit to the observer. Alternatively, a photon is transmitted if it can travel a sufficiently large distance without being absorbed. To assess this stopping criterion, we follow \citet{Loeb_1999} and compute the proper distance, $d_{\rm tr}(z)$, at which an expanding, homogeneous, and neutral IGM becomes transparent to \lya photons (i.e. where the opacity equals one). Interestingly, this distance is nearly independent of $z$ and corresponds to $\approx 1$ pMpc at z $= 6-9$, i.e. roughly the physical size of the \sphinxd box. We have performed a series of tests to make sure that our \lya transmissions are not affected by the exact $d_{\rm tr}(z)$ value. We find that the results are well converged if we use $10d_{\rm tr}$ (i.e. a Hubble-velocity shift of $V_{\rm hub} \gtrsim 6000$ \kms) and we thus opt for this value. For photons emerging blueward of \lya from the CGM, we require an additional travelled distance, $d_{\mathrm{blue} \shortrightarrow \mathrm{red}}$, corresponding to the time needed to red-shift past the resonance such that blue photons need to travel $(10d_{\rm tr} + d_{\mathrm{blue} \shortrightarrow \mathrm{red}})$ to be transmitted.

As mentioned in the previous section, the escape fractions from the CGM are computed by averaging over all directions. Similarly, we define the IGM transmission as the ratio of the total transmitted luminosity to the total of escaped luminosity : $T_{\scaleto{\rm IGM}{4pt}} = \int \dot{N}^{\scaleto{\rm IGM}{3.5pt}}_{\rm Ly\alpha} (hc/ \lambda) \mathrm{d}\lambda / 
\int \dot{N}^{\scaleto{\rm CGM}{3.5pt}}_{\rm Ly\alpha} (hc/ \lambda) \mathrm{d}\lambda$ where $\dot{N}^{\scaleto{\rm IGM}{3.5pt}}_{\rm Ly\alpha}$ is the total number of IGM transmitted \lya photons per unit time. The transmitted \lya luminosity is therefore given by $L^{\scaleto{\rm IGM}{3.5pt}}_{\rm Ly\alpha} = T_{\scaleto{\rm IGM}{4pt}} f_{\scaleto{\rm esc}{3.3pt}} L^{\scaleto{\rm intr}{4pt}}_{\rm Ly\alpha}$.

\section{Results}

To begin with, we present visualisations of the \sphinxd simulation at $z=7$ and $z=9$ (Figure \ref{fig:maps}). The top panels illustrate the filamentary structure of the hydrogen gas density distribution over the 10 cMpc scale spanned by our simulation. The red dots represent individual LAEs with the dot sizes reflecting the observed \lya intensity of each object (i.e. after internal and IGM transfer) which vary from $10^{38}$ to $10^{42}$ \ergs{} here. Along with Figure \ref{fig:maps_gals}, these images emphasise the broad range of physical scales probed by \sphinxdt. The two bottom panels of Figure \ref{fig:maps} illustrate the patchy reionization process captured by \sphinxd by showing maps of the volume-weighted hydrogen neutral fraction. While the IGM is still highly neutral at $z=9$ ($\xhi \approx 0.64$), it ionises rapidly over about 200 Myr to reach $\xhi \approx 0.007$ by $z=7$ .  

\begin{figure*}
\hspace*{-0.4cm}  
\includegraphics[width=0.98\textwidth]{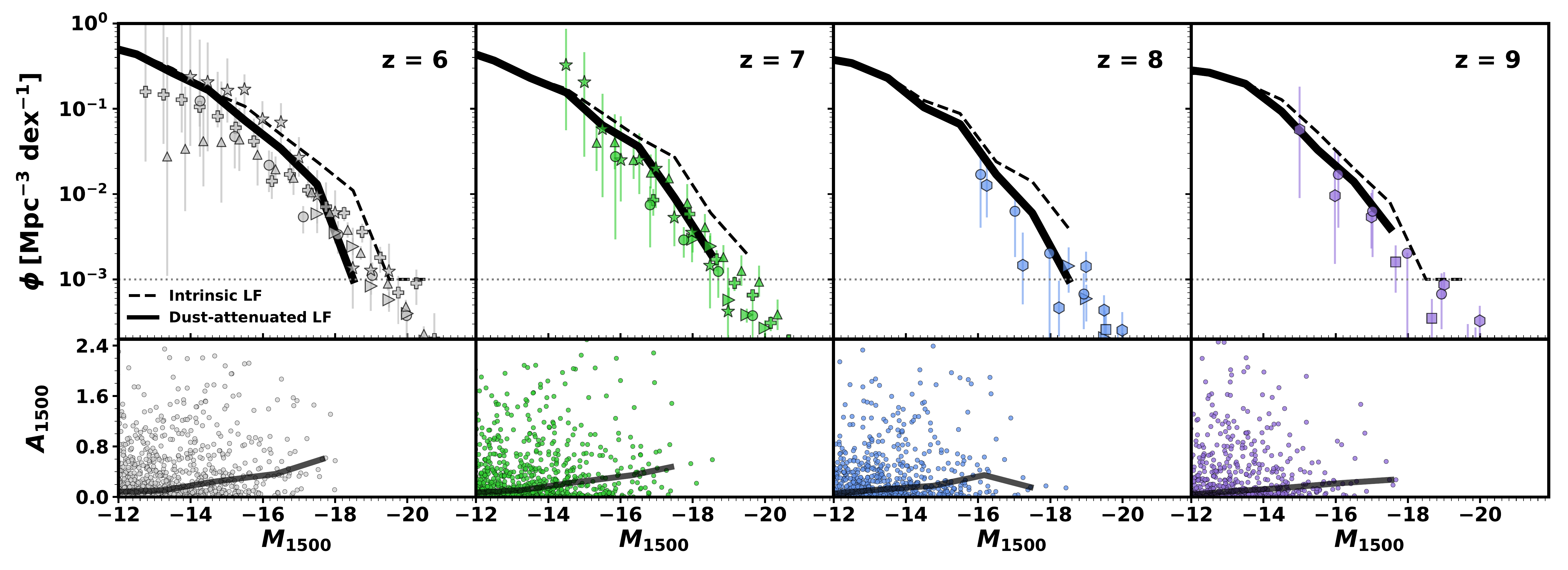}
\vskip-2ex
\caption{UV luminosity functions (LF; $1500$ \AA{} rest-frame) at z $=$ 6, 7, 8 and 9 (top). The dashed and solid black curves correspond to the intrinsic and dust-obscured LFs respectively. We compare with observations represented by the symbols: \citet[plus signs][]{Bouwens_2015,Bouwens_2017}, \citet[circles][]{Bhatawdekar_2019}, \citet[right triangles][]{Finkelstein_2015}, \citet[][stars]{Livermore_2017}, \citet[][upward triangles]{Atek_2015,Atek_2018}, \citet[hexagons][]{Ishigaki_2018}, and \citet[squares][]{Oesch_2013}. The data points of Atek et al. are shifted by $-0.1$ mag for clarity. The grey horizontal dotted line indicates our volume limit of one object per bin. We also show the dust attenuation, $A_{1500}$, as a function of UV magnitude (bottom). Dots correspond to individual galaxies and the line shows the median $A_{1500}$ per $M_{\rm 1500}$ bin.}
\label{fig:uvlf}
\end{figure*}

In the following, we present the main results of the \lya post-processing of \sphinxd at $z=6-9$, starting with an analysis of the galaxy properties and a comparison with statistical observational constraints. Then we focus on the redshift evolution of relevant \lya quantities (luminosity function, EWs, LAE fraction and spectra) and assess the relative effects of IGM transmission and dust attenuation on the visibility of LAEs in the context of cosmic reionisation in \sphinxdt.

\subsection{Stellar mass and UV luminosity functions}

As explained in \citet{Rosdahl_2018}, \sphinx is calibrated on the stellar mass-to-halo mass relation at $z=6$ by boosting the number of SN explosions compared to the fiducial value for a Kroupa IMF. Here, we extend the comparison to observational constraints by presenting the stellar mass function (SMF) and the dust-attenuated UV luminosity functions (LF) at $z=6, 7, 8$ and 9.

From Figure \ref{fig:smf}, we see that \sphinxd can well reproduce the abundance of galaxies in the stellar mass range probed by \sphinxdt. Indeed, because of the limited box size of our simulation, rare bright/massive galaxies are missed which means that we do not predict the massive end of the SMF for $M_{\star} \gtrsim 2\times10^9$ \msun{} ($M_{\star} \gtrsim 2\times10^8$ \msun) at $z=6$ ($z=9$). The lack of massive objects is highlighted by the shaded regions in Figure \ref{fig:smf} which represent the statistical error in each bin of log$M_{\star}$. 

In Figure \ref{fig:uvlf}, we show the UV luminosity functions before dust attenuation (dashed lines) and after dust attenuation (solid lines) and compare with existing constraints (top panels). At all redshifts, the dust-attenuated LF is in good agreement with the observational data at magnitudes $M_{\rm 1500} \gtrsim -18$. Due to the same finite-volume effect already mentioned above, the brightest intrinsic magnitudes found in \sphinxd are $\approx -20$. Nevertheless, recent deep surveys have pushed the observational limit down to extremely faint magnitudes ($M_{\rm 1500} \gtrsim -13$) which allows us to compare our results over a wide dynamical range \citep[$\approx 6$ mag;][]{Livermore_2017,Bouwens_2015}. We find that the abundance of galaxies increases steeply towards faint magnitudes which is in good agreement with observations, although error bars remain large at $M_{\rm 1500} \gtrsim -15$. Here it is worth pointing out that the apparent flattening of the simulated UV LF (and SMF) at the faint (low-mass) end does not necessarily represent a physical turnover, and may in part be due to mass resolution effects. At the faint-end, the LFs are incomplete because of our selection on stellar mass (we only analyse galaxies with more than $10^5$ \msun{} in stars). Concerning the low-mass end of the SMF, as discussed in \citet{Rosdahl_2018} and \citet{Katz_2020} the simulation only barely describes the formation of galaxies in halos at the atomic cooling limit, which are resolved with only $\approx 100$ DM particles, and we may thus miss some of the smallest objects.  

As highlighted in the bottom panels of Figure \ref{fig:uvlf}, the effect of dust is stronger for bright sources. This is a consequence of UV bright galaxies being on average more massive, more star-forming and therefore more metal- and gas-rich. The median dust attenuation $A_{\rm 1500}$, represented by the curves, is approximatively $0.5$ dex at the bright end while it becomes negligible at the faint end. This trend is similar to the observed one reported for bright Lyman-Break galaxies at high redshift where the magnitude attenuation evolves from $0.5$ dex at $M_{\rm 1500} \approx -19$ to $1.5$ dex at $M_{\rm 1500} \approx -22$ \citep{Bouwens_2016}. Interestingly, despite this correlation, the $A_{\rm 1500}$ values are widely spread around the median value at all magnitudes and redshifts, and galaxies as faint as $M_{\rm 1500} \approx -14$ can suffer an attenuation up to $\approx 2.5$. These outliers typically correspond to objects which experienced a very recent starburst ($t \lesssim 5$ Myr), indicating the presence of high gas densities and an ongoing production of metals in the SF sites, and thus increasing the attenuation according to our dust model (see Sec. \ref{sec:internal_rt}). \\  

\subsection{\lya luminosity functions}

\begin{figure*}
\hspace*{-0.5cm}  
\includegraphics[width=0.98\textwidth]{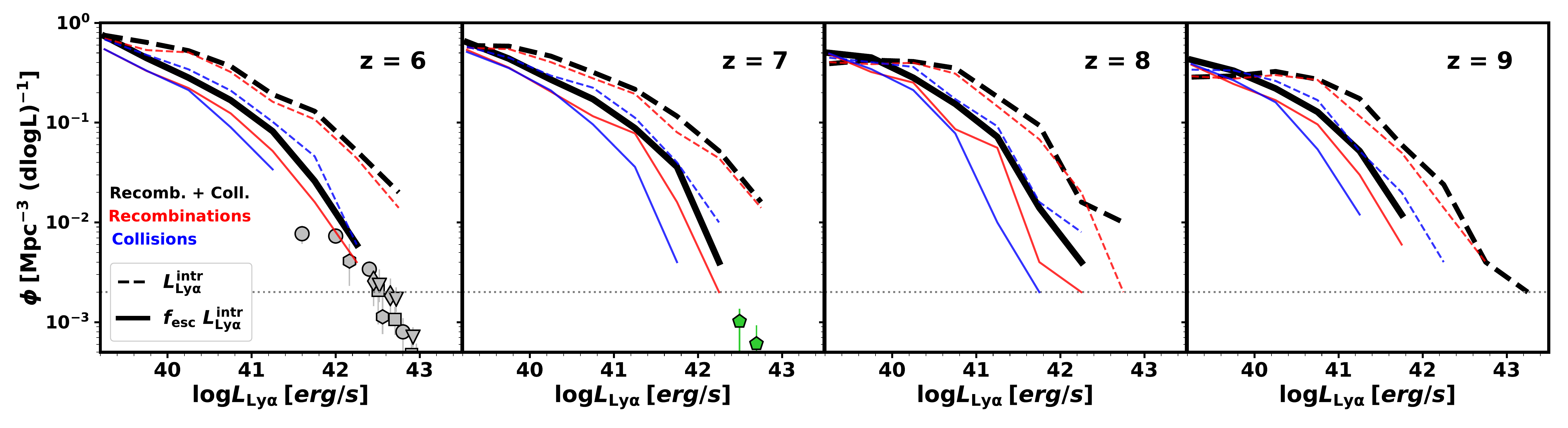}
\vskip-2ex
\caption{\lya luminosity functions at z $=$ 6, 7, 8 and 9 without IGM transmission. In each panel, the dashed and solid black lines account for the total \lya luminosity before and after dust attenuation respectively. The relative contribution of recombinations (in red) and collisions (in blue) are also highlighted. The grey data points represent $z\approx5.5-6.5$ observational data from \citet[][squares]{Konno_2017}, \citet[][diamonds]{Herenz_2019}, \citet[][hexagons]{cassata2011a}, and \citet[][downward triangles]{Santos_2016}. At this redshift, the only constraints on the faint-end of the \lya LF which are deep enough to be directly comparable to our predictions come from the MUSE-Deep survey \citep[][grey circles]{Drake_2017} ($L_{\rm Ly\alpha}^{\scaleto{\rm intr}{4pt}} \gtrsim 3 \times 10^{41}$ \ergs). The green pentagons show the LF measured by \citet[][]{Konno_2017} at $z\approx6.6$. The grey horizontal dotted line indicates our volume limit of one object per bin.}
\label{fig:lyalf}
\end{figure*}

The \lya luminosity function (\lya LF) is a fundamental quantity used to probe cosmic reionisation since the \lya line is expected to be increasingly suppressed by the neutral IGM towards higher redshifts. Statistical samples of LAEs at $z\gtrsim6$ have allowed us to put constraints on the bright-end of the \lya LF, i.e. $L_{\rm Ly\alpha} \gtrsim 10^{42}$ \ergs. While mild evolution is seen below $z\approx6$ \citep{ouchi2010a,cassata2011a}, the characteristic luminosity parameter $L^*$ appears to drop by a factor 1.4 at $z = 6.6$ and by a factor $2-3$ at $z \approx 7-7.5$ \citep{Zheng_2017,Itoh_2018} compared to $z=5.7$. Here, we present our predicted \lya LFs at $z=6-9$ before dust attenuation, after dust attenuation and after IGM transmission. We estimate the relative impact of the IGM on the redshift evolution of the \lya LF in our simulation to assess to which extent the observed suppression of the LF can be interpreted as an imprint of reionisation.

\subsubsection{Intrinsic \lya emission}
\label{subsubsec:intr_lya}

We begin with Figure \ref{fig:lyalf} that shows our predicted \lya LFs at $z=6, 7, 8$ and 9 ignoring the effect of IGM. In each panel, the black dashed curves represent the intrinsic LF. Although there are no very massive objects in our sample, we see that intrinsically bright LAEs can be produced, with \lya luminosities as high as $\approx 10^{43}$ \ergs. This is mostly caused by (i) the burstiness of star formation in our simulation \citep[see e.g.][]{Trebitsch_2017} which gives rise to brief but intense \lya emission episodes and (ii) the use of the BPASS stellar library which boosts the ionising photon budget for a given SF event compared to stellar evolution models without binary stellar systems, and hence the \lya production under case B recombination. For a constant SFR and a Kroupa IMF (with single stars only), the intrinsic \lya luminosity from recombination is often estimated to be $1.7 \times 10^{42} \times (SFR/[\msunyr])$ erg s$^{-1}$ \citep[e.g.][]{Dijkstra_2017}. For the two reasons mentioned above, we instead find an average relation of $L_{\rm Ly\alpha}^{\scaleto{\rm intr}{4pt}} \approx 3-4 \times 10^{42} \times (SFR/[\msunyr])$ erg s$^{-1}$ in \sphinxdt.  

As explained in Section \ref{subsubsec:emission}, \lya photons are emitted through two different channels in our study. In Figure \ref{fig:lyalf}, we also show the relative contribution of recombinations (in blue) and collisions (in red) to the \lya LFs. At all redshifts, recombinations strongly dominate the \lya intrinsic budget over collisions in brighter LAEs ($L_{\rm Ly\alpha} \gtrsim 10^{41} \ergs$) whereas both channels contribute equivalently in fainter objects. Note that the collisional excitation rate, $C_{\rm Ly\alpha}(T)$, is highly sensitive to the temperature so its exact contribution will depend on the subgrid physics that can affect the thermal properties of the gas, in particular the feedback model. In addition, the collisional excitation rate is poorly estimated in gas cells where the net cooling time is small compared to the simulation timestep. We therefore make the conservative approximation of setting \lya collisional emission to zero in cells which net cooling time is less than five times the timestep value. We tested that our results are not sensitive to this choice and that it has a minor impact on the total budget of \lya collisional emission (Blaizot et al., in prep). 

\subsubsection{\lya transfer in the ISM and CGM}

Due to the complex nature of the \lya resonant line, it is paramount to account for the radiative transfer of \lya photons in the ISM and CGM (which we refer to as internal RT for simplicity) to realistically model the evolution of LAEs during reionisation. Based on the procedure detailed in Section \ref{sec:internal_rt}, we construct the dust-attenuated \lya LF and show our results in Figure \ref{fig:lyalf} (black solid curves).

At all redshifts, the internal RT suppresses \lya emission by a factor $1.5-3$ on average. Most of \sphinxd LAEs are too faint to be compared with observations except at $z=6$, where our LF is in reasonable agreement with the deep MUSE constraints ($L_{\rm Ly\alpha} \lesssim 10^{42}$ \ergs), though slightly above (but we remind that we have ignored IGM transmission for now). Similarly to the UV LF, we predict that the LF keeps rising steeply at $L_{\rm Ly\alpha} \lesssim 10^{42}$ \ergs{} despite the mass resolution effect discussed earlier which implies that the number density is even under-estimated at the very faint end ($L_{\rm Ly\alpha} \lesssim 10^{40}$ \ergs). 

\lya photons produced by recombinations dominate the bright-end of our LF after RT in the ISM and the CGM, as was already the case for the intrinsic emission. Nevertheless, \lya radiation emitted through collisional excitation has a somewhat higher escape fraction. This is because recombinations mainly occur in dense, metal-rich, star-forming regions where dust extinction is generally strong while collisional emission can also be generated in the more diffuse and metal-poor parts of the ISM. 

\subsubsection{Impact of IGM transmission on the \lya LF}
\label{subsubsec:igm_lya}

As discussed in the introduction, only a fraction of the \lya flux escaping galaxies can reach the observer due to \hi absorption by the IGM. 
Yet, understanding how much of the observed \lya suppression is connected to the IGM neutral fraction at a given redshift remains elusive. In the current and following sections, we intend to quantify the impact of IGM transmission on the \lya LF. 

\begin{figure*}
\includegraphics[width=0.7\textwidth]{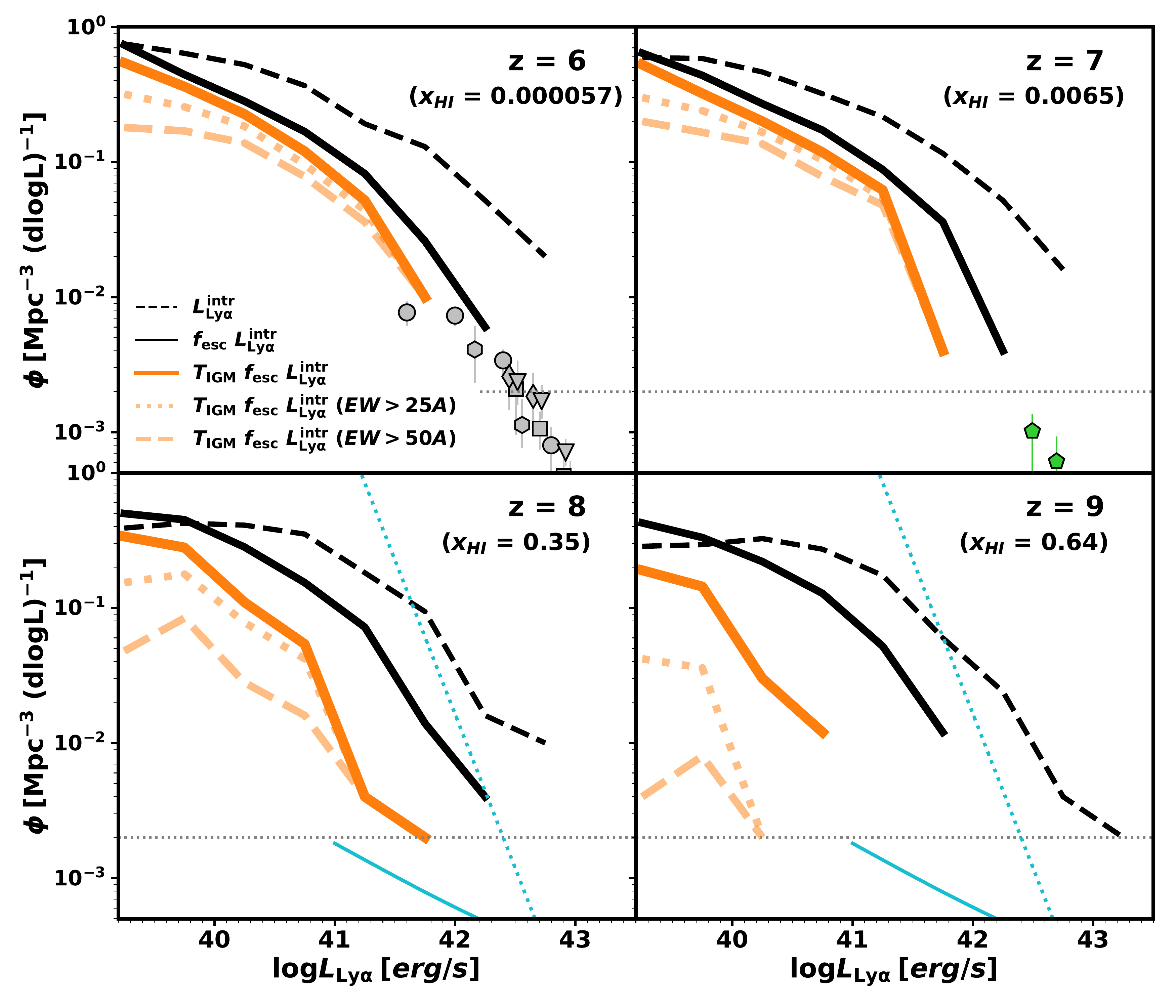}
\vskip-2ex
\caption{Effect of IGM transmission on the \lya luminosity functions at z $=$ 6, 7, 8 and 9 shown by the solid orange curves. For comparison, we plot again the \lya LF before and after dust attenuation (dashed and solid black curves respectively). To mimic roughly the narrow-band selection in wide-field surveys, we also construct the IGM-transmitted LFs with two different EW cuts: $EW_{\scaleto{\rm thresh}{4pt}} > 25$ \AA{} (dotted orange line) and $> 50$ \AA{} (dashed orange line). The observed data points at $z\approx6$ and 7 are identical to those shown in Figure \ref{fig:lyalf}. There are no constraints on the faint-end of the \lya LF at $z \geq 8$ but, as a guide, we add the power-law and Schechter function best-fits reported by \citet{Matthee_2014} extrapolated at low luminosities (dotted and solid cyan curves respectively). The grey horizontal dotted line indicates our volume limit of one object per bin.}
\label{fig:lyalf-igm}
\end{figure*}

From Figure \ref{fig:lyalf-igm}, we see that the IGM has quite a significant impact on the LF (solid orange curves), in particular towards higher redshifts. Of course, this is expected because \xhi increases from $10^{-4}$ at $z=6$ to $0.6$ at $z=9$ in the simulation. At $z=6$, the IGM transmission \tigm is about 50\% whereas it drops to $\approx$ 5-10\% at $z=9$, clearly reflecting the evolution of the ionisation state of the diffuse IGM (see Section \ref{subsec:tigm}). 

Although the comparison with observational data is obviously dubious at these high redshifts and low luminosities, we note that our $z=6$ IGM-attenuated LF falls near the MUSE-deep constraints at $L_{\rm Ly\alpha} \lesssim 10^{42}$ \ergs. Due to our limited box size, it is impossible to draw any conclusion regarding the bright-end but it is worth noting that a crude extrapolation "by eye" of our LFs at $z=6$ and 7 does not seem inconsistent with the data at $L_{\rm Ly\alpha} \approx 10^{42-43}$ \ergs. While there is no compelling observational constraints at $z=8-9$, we nevertheless plot as a guide the Schechter and power-law fits derived by \citet{Matthee_2014} and extrapolated to $L_{\rm Ly\alpha} \approx 10^{41}$ \ergs. Based on this (uncertain) comparison, our LF falls in the expected range of densities at such low luminosities. 

Our predicted LFs indicate that numerous LAEs should be detectable during the heart of reionisation era, assuming that detection limits are pushed further down by a couple orders of magnitude. The identification of LAEs in typical narrow-band surveys is however often based on colour selections, or equivalently, EW thresholds. We therefore also show in Figure \ref{fig:lyalf-igm} the effect of EW cuts ($EW_{\scaleto{\rm thresh}{4pt}} > 25$ \AA{} and $> 50$ \AA). As discussed in Appendix \ref{appendix:scal_rel}, the \lya EWs are larger for brighter LAEs in our simulation (see Figure \ref{fig:scal_rel_llya}), so that EW selections will predominantly remove galaxies at the faint end of the LF.  Even so, taking a conservative cut of $EW_{\scaleto{\rm thresh}{4pt}} > 50$ \AA, we predict a high number density of LAEs at $L_{\rm Ly\alpha} \approx 10^{40}$ \ergs{} at $z=8$ where the IGM neutral fraction is still $\approx 35 \%$. 

\subsubsection{\lya LF evolution with redshift}
\label{subsubsec:lyalf_evolution}

\begin{figure}
\hspace{-0.3cm}  
\includegraphics[width=0.45\textwidth]{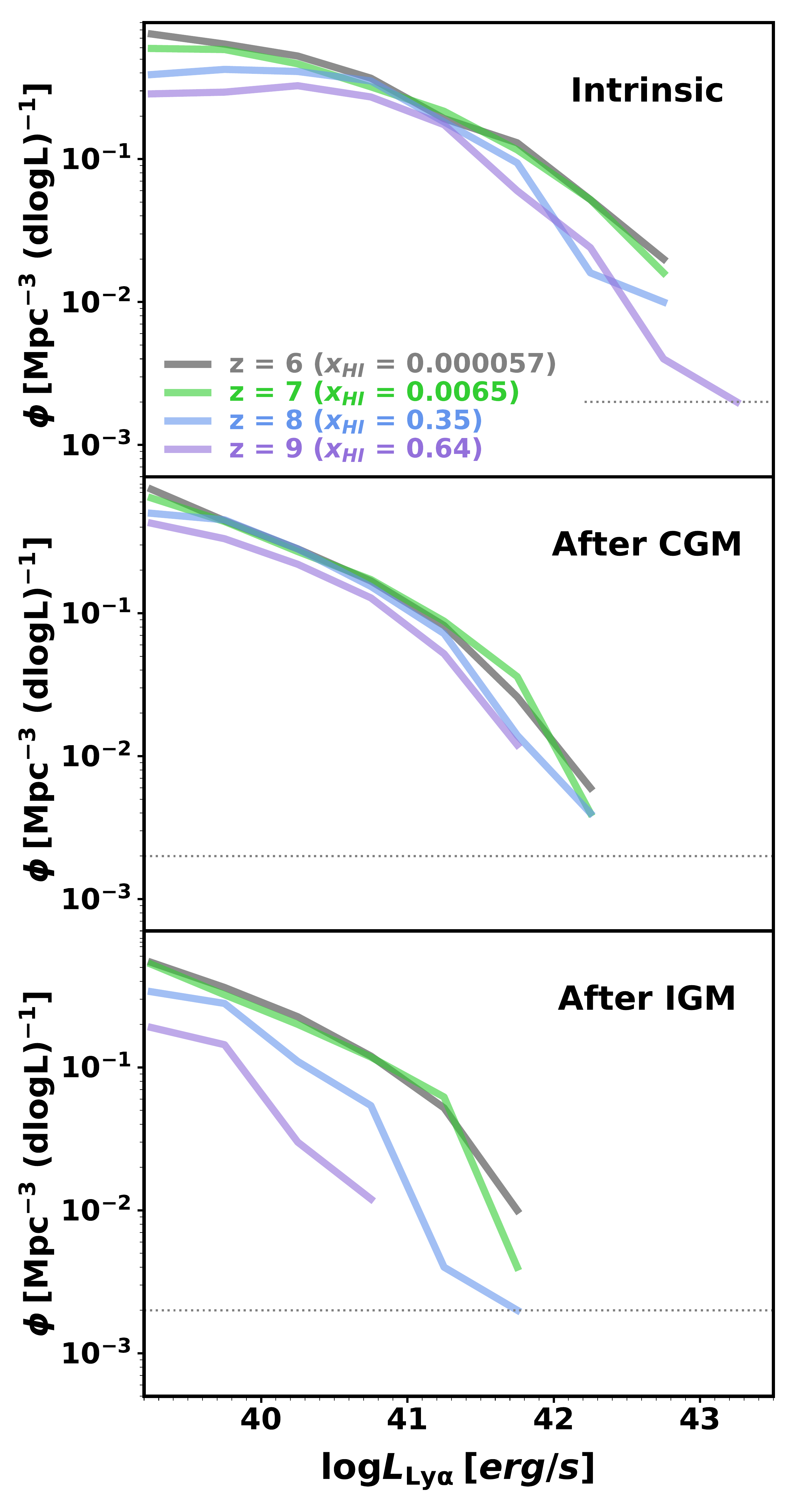}
\vskip-2ex
\caption{Redshift evolution of the \lya luminosity functions highlighting the contribution of intrinsic emission, internal absorption and IGM transmission. The grey horizontal dotted line indicates our volume limit of one object per bin.}
\label{fig:lyalf-zevol}
\end{figure}

In Figure \ref{fig:lyalf-zevol}, we highlight the redshift evolution of the \lya LF by plotting together the LFs based on intrinsic luminosities (top panel), dust-attenuated luminosities (middle panel), and IGM-transmitted luminosities (bottom panel). On the one hand, we clearly see that the intrinsic and dust-attenuated LFs remain nearly constant from $z=6$ to $z=9$, highlighting the very weak evolution of the internal properties driving \lya emission and escape from galaxies during the EoR. On the other hand, the substantial effect of IGM is completely dominating the variation of the visibility of LAEs during this period. While the \lya LF is unchanged at $z \lesssim 7$ as long as the IGM is highly ionised ($\xhi < 0.01$), the \lya transmission drops significantly from $z=7$ to $z=9$. This strong suppression is directly due to the IGM neutral fraction increasing rapidly at $z \gtrsim 7$ in \sphinxdt, i.e. $\xhi \approx 0.35$ at $z=8$ and $\xhi \approx 0.65$ at $z=9$ \citep[see Figure 9 of][]{Rosdahl_2018}.

\begin{figure}
\includegraphics[width=0.48\textwidth]{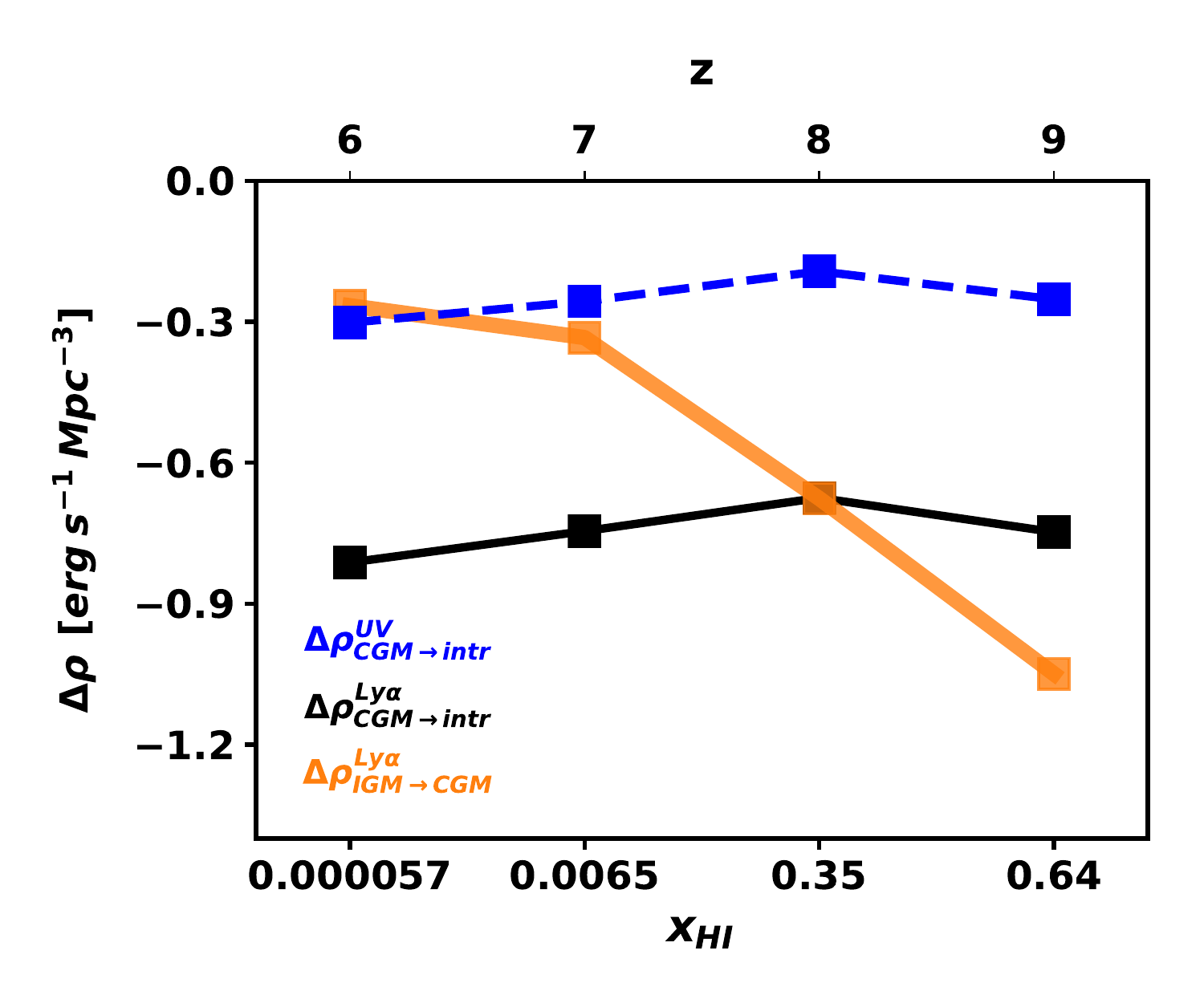}
\vskip-3ex
\caption{Evolution of the \lya and UV luminosity density decrement as a function of the IGM neutral fraction, \xhit. The top x-axis shows the redshifts of the corresponding snapshots of the \sphinxd simulation. The \lya and UV density have been integrated down to $L_{Ly\alpha}=10^{40}$ \ergs and $M_{\rm 1500}=-12$ respectively which approximatively corresponds to our completeness limit.}
\label{fig:lya-uv_density}
\end{figure}

In order to get a more quantitative assessment of the impact of the IGM, we define the decrement of the observed \lya luminosity density \citep[in erg s$^{-1}$ Mpc$^{-3}$,][]{Inoue_2018}:
\begin{flalign}
\Delta \rho^{\rm Ly\alpha}_{\mathrm{IGM} \rightarrow \mathrm{CGM}} = \mathrm{log}\rho^{\rm Ly\alpha}_{\rm IGM} - \mathrm{log}\rho^{\rm Ly\alpha}_{\rm CGM}
\end{flalign}
where log$\rho^{\rm Ly\alpha}_{\rm IGM}$ and log$\rho^{\rm Ly\alpha}_{\rm CGM}$ are the \lya luminosity densities after IGM transmission and after internal transfer, integrated down to our completeness limit ($L_{\rm Ly\alpha} \approx 10^{40}$ \ergs). The orange curve in Figure \ref{fig:lya-uv_density} shows that the decrement is nearly constant from $z=6$ to $z=7$ and decreases significantly by $\approx 0.5$ dex from $z=7$ to $z=8$ and by $\approx 1$ dex from $z=8$ to $z=9$ due to reduced IGM transmission. 

For comparison, we also plot the decrements of the dust-attenuated \lya and UV luminosity densities relative to the intrinsic ones ($\Delta \rho^{\rm Ly\alpha}_{\mathrm{CGM} \rightarrow \mathrm{intr}}$ and $\Delta \rho^{\rm UV}_{\mathrm{CGM} \rightarrow \mathrm{intr}}$; black and blue curves respectively). Both remain nearly unchanged from $z=6$ to $z=9$ which suggests that any significant detectable evolution in the \lya LF during the EoR should be fully attributed to a rapid increase of $\xhi$. Finally, we note that the offset between $\Delta \rho^{\rm Ly\alpha}_{\mathrm{CGM} \rightarrow \mathrm{intr}}$ and $\Delta \rho^{\rm UV}_{\mathrm{CGM} \rightarrow \mathrm{intr}}$ reflects the differential escape fractions from galaxies between \lya and UV photons. Stellar (non-ionising) UV continuum usually escapes galaxies more easily than \lyat, especially for more massive, dustier sources, which is a direct consequence of the enhanced probability of resonant \lya photons to be destroyed by dust grains on their way out of the galaxy \citep{verh08,hayes2011a,garel2015a}.

\subsubsection{Abundance of very faint LAEs}

Once internal RT and IGM transmission are accounted for, the dynamical range spanned by LAEs in our $10^3$ cMpc$^3$ simulation is restricted to \lya luminosities below 
$L_{\rm Ly\alpha} \approx 10^{42}$ \ergs. Thanks to its fine mass-resolution, \sphinxd is however able to resolve low-mass systems, allowing us to investigate the very faint-end of the \lya LF. As shown in Figure \ref{fig:lyalf-igm}, \lya emitters at such low levels are unfortunately still out of reach in current surveys and it is not clear to which extent the LF keeps rising at the faint-end. Still, the recent detection of extended \lya emission at $>1$ cMpc scale at $z \approx 3-5$ in the MUSE Extremely Deep Field provides clues for the existence a numerous population of ultra-faint LAEs, possibly down to $L_{\rm Ly\alpha} \approx 10^{37}$ \ergs{} and assuming a steep LF slope \citep{Bacon_2021}. %(Bacon et al., 2020).

Such sources should sit predominantly in low-mass DM haloes but, as extensively discussed in the literature, the feedback from stellar radiation can prevent the formation of galaxies in these systems due to photoheating and gas inflow suppression \citep{okamoto08}. Using a smaller \sphinx simulation run than ours (but with the same baryonic physics and BPASS library), \citet{Katz_2020} have shown that reionization has a significant impact on the gas content of dwarf galaxies at $z\gtrsim6$ but that, meanwhile, most haloes below the atomic cooling limit can remain self-shielded against ionising radiation and can thus keep forming stars even after the end of reionization.

\begin{figure}
\hspace{-0.5cm}
\includegraphics[width=0.48\textwidth]{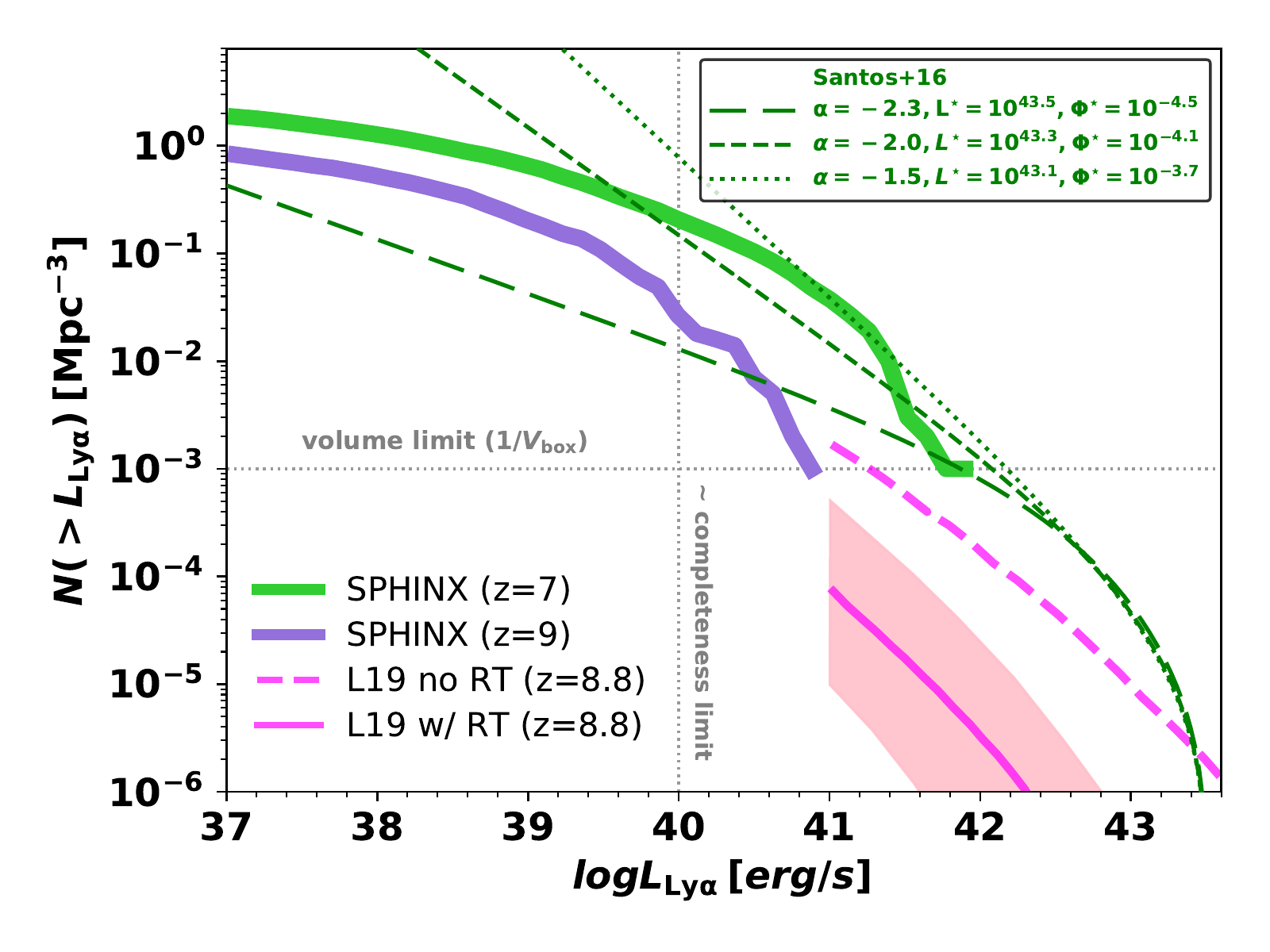}
\vskip-3ex
\caption{Cumulative \lya luminosity function at $z=7$ (green) and $z=9$ (purple) after IGM transmission in \sphinxdt. We show that the \lya LF keeps rising down to very faint \lya luminosities during the EoR. The horizontal and vertical grey lines indicate our volume and \lyat-luminosity completeness limits respectively. These imply that we cannot predict the number density of LAEs below $10^{-3}$ Mpc$^{-3}$ due to our limited simulation volume, and that we under-estimate the number density of LAEs below $L_{\rm Ly\alpha} \approx 10^{40}$ \ergs{} because we choose to only identify galaxies more massive than $M_{\star}=10^5$\msun{} (i.e. 100 star particles) in our study. At $z=9$, we compare our prediction with the LF of \citet{Laursen_2019} who model bright-end LAEs using cosmological hydrodynamic simulations post-processed with ionising and \lya transfer at $z=8.8$. Their intrinsic LF is represented by the dashed magenta curve. The solid magenta curve shows their predicted LF after \lya RT in the haloes and in the IGM and is therefore comparable to our LF at $z=9$. The pink shaded area corresponds to the 1$\sigma$ directional variation. The dotted, dashed, and solid dark green lines are the extrapolated Schechter best-fits of the observed \lya LFs from \citet{Santos_2016} at $z\approx7$ (see legend for the Schechter parameter values).}
\label{fig:cumul_lyalf}
\end{figure}

As can be seen from the cumulative \lya LFs (after IGM) at $z=7$ and 9 (Figure \ref{fig:cumul_lyalf}), very faint LAEs do exist in our simulation and we find that their cumulative number density keeps rising until $L_{\rm Ly\alpha} < 10^{37}$ \ergs{} which confirms that low-mass haloes keep forming stars efficiently. We compare our predicted \lya LF with the best-fit Schechter functions measured by \citet{Santos_2016} at $z\approx7$, assuming three different faint-end slopes, $\alpha$. Our $z=7$ LF seems to be more consistent with moderately steep values ($\alpha \approx -1.5$) but it is difficult to assess because we have restricted our sample to galaxies more massive than $10^5$ \msun. As shown in Figure \ref{fig:scal_rel_llya}, the brightest \lya luminosities in galaxies at our stellar mass threshold correspond to roughly $L_{\rm Ly\alpha} \approx 10^{40}$ \ergs{} at all redshifts considered here. This means that our LAE sample is incomplete below this value such that our LFs appear shallower than they should. Based on our simulation, the expected numbers of LAEs at $L_{\rm Ly\alpha} \gtrsim 10^{37}$ \ergs{} at $z=7$ (\xhi $=0.007$; $\approx 2$ per cMpc$^3$) and $z=9$ (\xhi $=0.64$; $\approx 0.8$ per cMpc$^3$) may therefore be seen as lower limits, suggesting that the abundance of extremely faint LAEs is high towards the end of the EoR.

We note that our predicted LFs are in slight disagreement with the results of \citet{Laursen_2019} who simulated the visibility of LAEs at $z\approx9$ using zoom-in hydrodynamics simulations applied to a large cosmological DM run. As our study, they follow the \lya radiation from their emission sites through the ISM, CGM and IGM allowing for accurate estimation of the internal RT and IGM transmission. Two relevant differences though relate to the ionising transfer, which they perform as a post-processing step, and their dynamical range which covers more massive haloes than ours on average. Our $z=9$ LF (after IGM) is only overlapping with the one of \citet{Laursen_2019} at $L_{\rm Ly\alpha} \approx 10^{41}$ \ergs{} where it roughly matches their intrinsic LF. Once they account for \lya RT, they predict an abundance of LAEs significantly smaller than in \sphinxd at this particular luminosity. The reasons for the discrepancy are unclear and could arise from incompleteness at the faint-end in the sample of \citet{Laursen_2019} or from cosmic variance effects that can be significant especially in moderate volume sizes like in \sphinxdt. 

Based on their simulation, \citet{Laursen_2019} predict that very few LAEs can be detected in the UltraVISTA survey with a 168h exposure (i.e. their probability of detecting more than one LAE is 1\%), corresponding to \lya detection limit of $\approx 10^{43}$ \ergs{} at $z=8.8$. We cannot make number count predictions at such bright \lya luminosities with \sphinxd but the significantly higher LAE number density that we predict at $L_{\rm Ly\alpha} \approx 10^{41}$ \ergs{} compared to \citet{Laursen_2019} suggests that more optimistic numbers of detections can be achieved with such deep surveys during the EoR.

\subsection{\lya equivalent widths}
\label{subsec:ew}

Defined as the ratio of \lya emission over UV continuum, the equivalent width (EW) encodes valuable information about galaxies such as the metallicity and age of the underlying stellar population \citep[e.g.][]{Hashimoto_2017}. During the EoR, the differential evolution of EWs can also be used as a proxy for IGM neutrality \citep{Mason_2018,Jung_2020}. 

\begin{figure}
\hspace{-0.5cm}  
\includegraphics[width=0.48\textwidth]{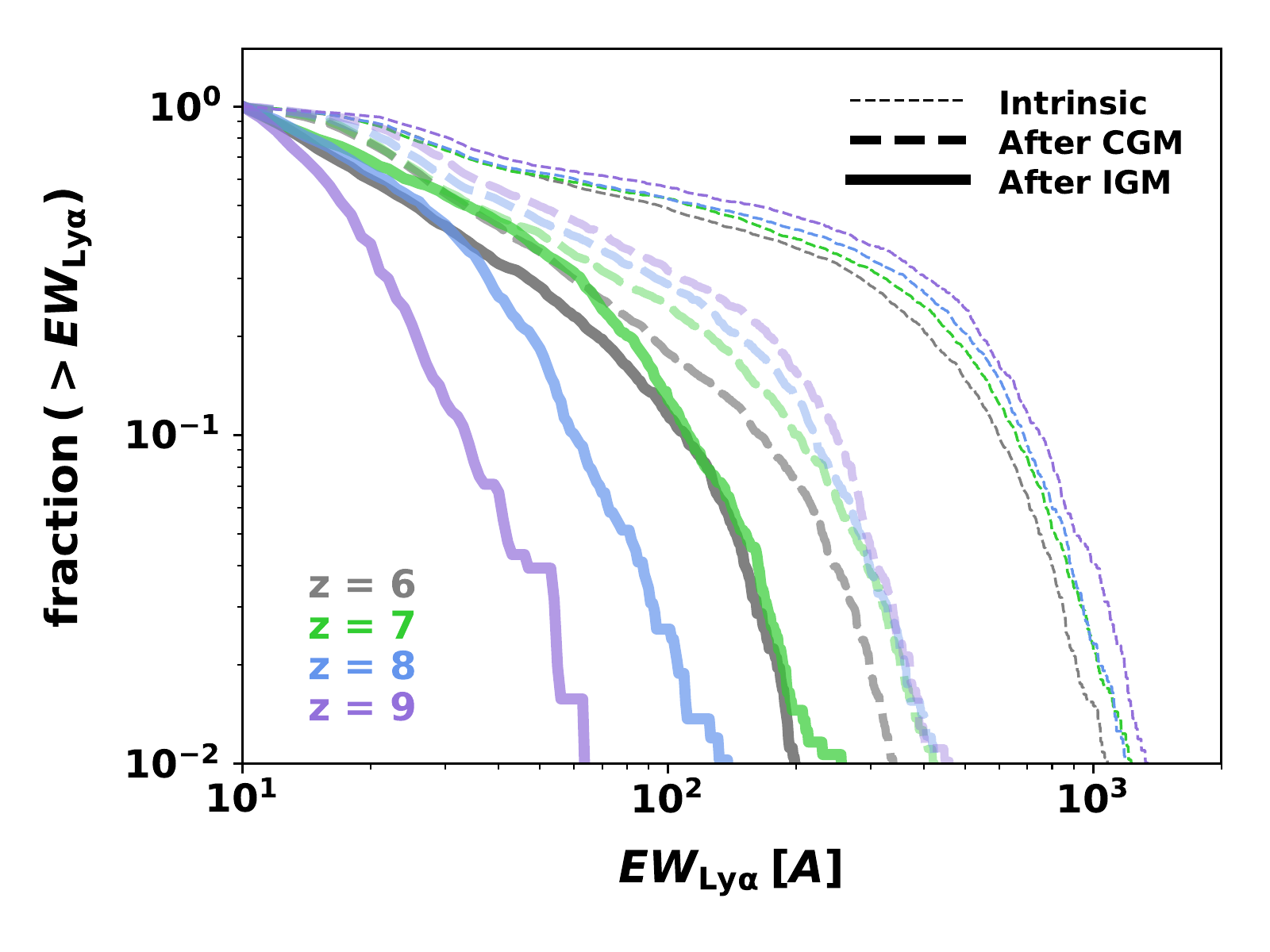}
\vskip-4ex
\caption{Redshift evolution of the cumulative \lya EW distribution showing the effects of dust attenuation and IGM transmission from $z=6$ to $z=9$. The thin dotted curves represent the fraction of galaxies with an intrinsic \lya EW larger than a given value while the thick dashed and solid curves include the effects of internal RT (after CGM) and IGM transmission (after IGM) respectively. Here, we use all galaxies from each snapshot without any UV magnitude selection.}
\label{fig:ew-evol}
\end{figure}

As is often done with observational datasets, we compute the EWs by estimating the continuum around the \lya wavelength from far-UV bands to measure the UV slope, $\beta_{\rm UV}$, and extrapolating the flux level at $1216$ \AA{} \citep{Hashimoto_2017}. In practice, we predict the intrinsic and dust-attenuated emissivities at 1500\AA{} and 2500\AA{} and we measure $\beta_{\rm UV}$ before and after internal RT in order to compute the intrinsic and dust-attenuated continuum luminosity densities at $1216$ \AA, $L_{1216}$. The \lya EWs are simply obtained as the ratio of the \lya intrinsic and dust-attenuated luminosities by these values: $EW^{\rm intr} = L_{\rm Ly\alpha}^{\rm intr} / L_{1216}^{\rm intr}$ and $EW^{\scaleto{\rm CGM}{3.5pt}} = L^{\scaleto{\rm CGM}{3.5pt}}_{\rm Ly\alpha} / L^{\scaleto{\rm CGM}{3.5pt}}_{1216}$. To estimate the IGM-transmitted EW, we multiply the dust-attenuated EW by the IGM transmission such that  $EW^{\scaleto{\rm IGM}{3.5pt}} = T_{\scaleto{\rm IGM}{4pt}} EW^{\scaleto{\rm CGM}{3.5pt}}$.

Figure \ref{fig:ew-evol} shows the resulting cumulative distributions at z $=$ 6, 7, 8, and 9 normalised to the total number of LAEs in each snapshot. A first striking aspect is the high intrinsic EW values that are reached in some galaxies. About ten percent of LAEs have $EW$ greater than $500$\AA{} and $\approx$ one percent of them produce EW above 1000 \AA. For a standard IMF and solar metallicity, the maximum EW produced through recombination in star-forming regions is about 250 \AA. However, with metallicities of $0.02Z_{\odot}$ that are plausibly more representative of low-mass galaxies at high redshift, stellar synthesis models can easily produce EWs as large as 400 \AA{} \citep[e.g.][]{Hashimoto_2017}. In our simulation, the gas-phase metallicities are comprised between $0.1Z_{\odot}$ and $0.001Z_{\odot}$ (see Figure \ref{fig:scal_rel_muv}) so values of 400 \AA{} are indeed expected. The other two factors able to boost even further the EW above 1000 \AA{} in our simulated galaxies are (i) the use of BPASS which increases the \lya emissivity for a given SF episode (see Section \ref{subsubsec:intr_lya}), and (ii) the contribution of collisional emission that can increase the global \lya photon budget (see Figure \ref{fig:lyalf}). After internal RT, the median EW is about 50 \AA{} but a small fraction of galaxies harbour very large values ($>400$\AA) at all redshifts. This seems consistent with the recent measurements of Kerutt et al. (in prep.) who report EWs up to 900 \AA{} in deep MUSE observations at $z=3-6.5$. 

The intrinsic and dust-attenuated EW distributions do not show a strong evolution with redshift but we note that the fraction of high values becomes slightly larger towards higher redshifts (see Figures \ref{fig:ew-evol} and \ref{fig:scal_rel_llya}). 
A more drastic evolution is seen when looking at the redshift evolution of the distribution of IGM-attenuated equivalent widths. While the $z=6$ and $z=7$ distributions evolve similarly after IGM transmission, the high-EW tail is cut off when the IGM neutral fraction becomes significant (i.e. at $z\gtrsim 7-8$ in our simulation). This behaviour can be interpreted as the \lya LF evolution discussed in Section \ref{fig:lyalf-zevol} where the increasing IGM neutrality starts suppressing the \lya line when \xhi becomes greater than $\approx 1\%$. Overall, our results support the idea that the evolution of the \lya EW distribution at $z > 6$ can be a used to probe the IGM neutrality during the EoR \citep{Mason_2018}.

\subsection{LAE fraction}
\label{subsec:xlae}

\begin{figure}
\hspace*{-0.4cm}  
\includegraphics[width=0.48\textwidth]{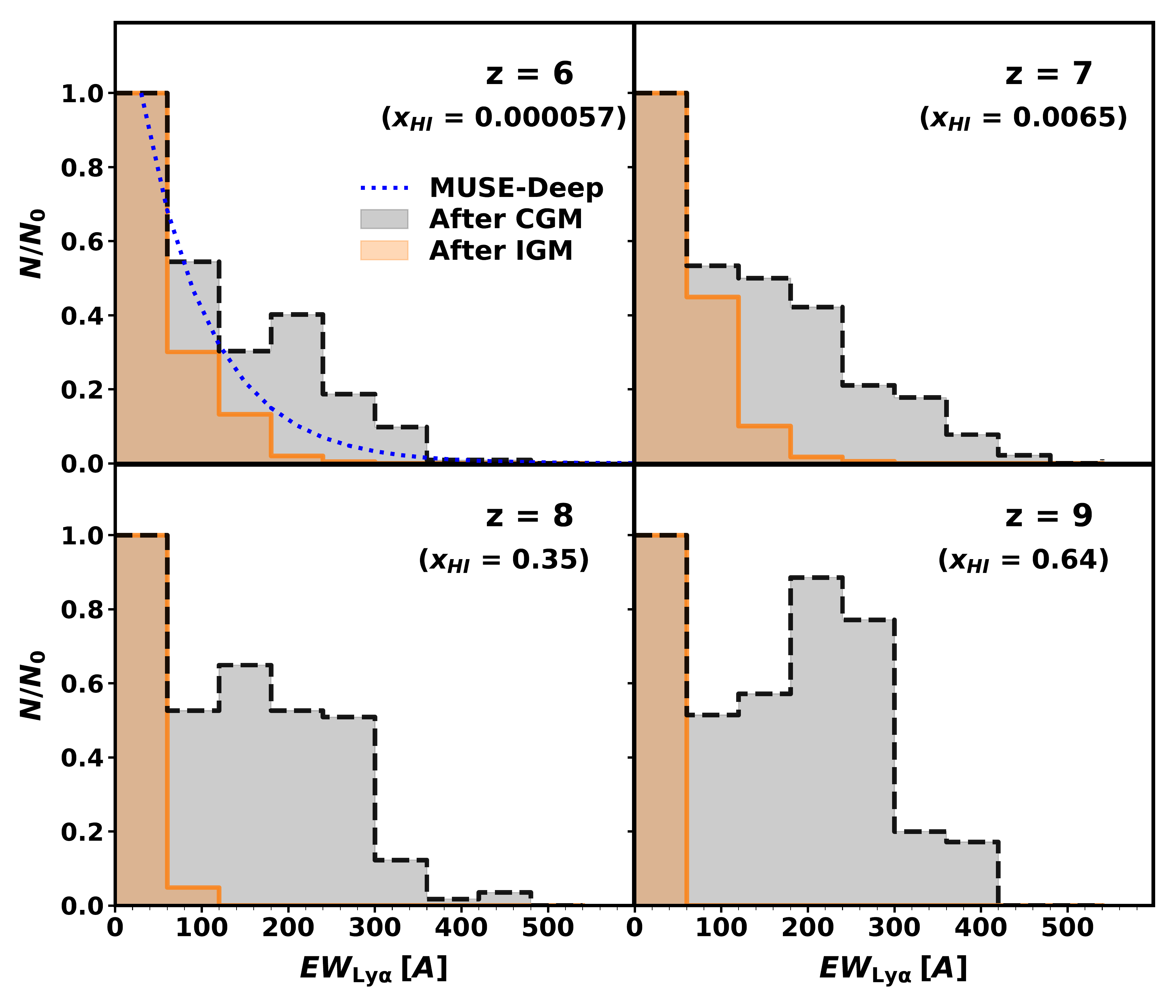}
\vskip-2ex
\caption{Equivalent width distributions at z $=$ 6, 7, 8, and 9. The grey histograms correspond to the \lya EW after internal RT while IGM-transmitted values are in orange. For comparison, we overplot the best-fit exponential distributions from the MUSE-Deep survey at $4.5 < z < 6.6$ \citep{Hashimoto_2017} ($N \propto \exp(-EW / w_0)$, where $w_0=79 \AA$). In this figure, we use a UV magnitude cut of $-14$ in order to have similar statistics as in the MUSE-Deep data (i.e. at least 100 galaxies per snapshot). The comparison is however mainly illustrative because, in spite of the $>10$h MUSE exposure-time combined with exquisite HST counterpart data, MUSE-Deep LAEs usually have $L_{\rm Ly\alpha} \gtrsim 10^{41}$ \ergs{} and $M_{\rm 1500} \lesssim -16$ whereas most of our simulated sources are fainter than these values.}
\label{fig:ew-igm}
\end{figure}

The measurement of the fraction of UV-selected galaxies that emit \lya is a commonly used diagnostic of reionisation. The LAE fraction, or \xlyat, is defined as follows:

\begin{equation}
\xlya(z) = \frac{N_{\rm LAE}(z,M_{\rm 1500},EW)}{N_{\rm 1500}(z,M_{\rm 1500})}
\end{equation}

where $N_{\rm 1500}(z,M_{\rm 1500})$ is the number of galaxies brighter than a fixed UV magnitude limit in a given redshift bin. $N_{\rm LAE}(z,M_{\rm 1500},EW)$ is a subsample of $N_{\rm 1500}(z,M_{\rm 1500})$ that corresponds to LAEs, i.e. sources with a \lya equivalent width greater than a typical threshold value ($EW >$ 25 or 50 \AA{} are two commonly used values in LAE surveys). Under the assumption that the IGM is the main cause of the apparent fading of the \lya line at $z \gtrsim 6$, \xlya should decline when \xhi increases. Such a trend has been reported by many surveys \citep[e.g.][]{schenker2012a,Pentericci_2018,Fuller_2020} based on samples of galaxies brighter than $M_{\rm 1500} \lesssim -18.5$. Given that there are only a handful of such bright sources in our simulation, we can only compute \xlya with a lower UV magnitude limit. Using a somewhat arbitrary cut of $M_{\rm 1500} = -14$ allows us to have sufficient statistics (i.e. at least 100 galaxies at each snapshot) to produce a sample size comparable to observational studies such as the MUSE-Deep survey \citep{Hashimoto_2017}. We show in Figure \ref{fig:ew-igm} the resulting EW distributions where the black and orange histograms represent the EWs after CGM and after IGM respectively. Despite the different UV selection, our predicted EW distributions reproduce reasonably well the one from the MUSE-Deep survey so we keep $-14$ as our UV detection limit for our study the LAE fraction\footnote{We note that a correlation exists (with a large scatter) between EW and $M_{\rm 1500}$ (see Figure \ref{fig:scal_rel_muv}) so varying the UV magnitude cut will affect the selected EW distribution, and hence the resulting LAE fraction. Nevertheless, we have checked that setting lower or higher $M_{\rm 1500}$ limits only impacts the overall amplitude of \xlyat, and not the shape of its redshift evolution, so this choice does not change our main conclusions (see Figure \ref{fig:xlya_vary_muv}).}.  

\begin{figure*}
\includegraphics[width=0.9\textwidth]{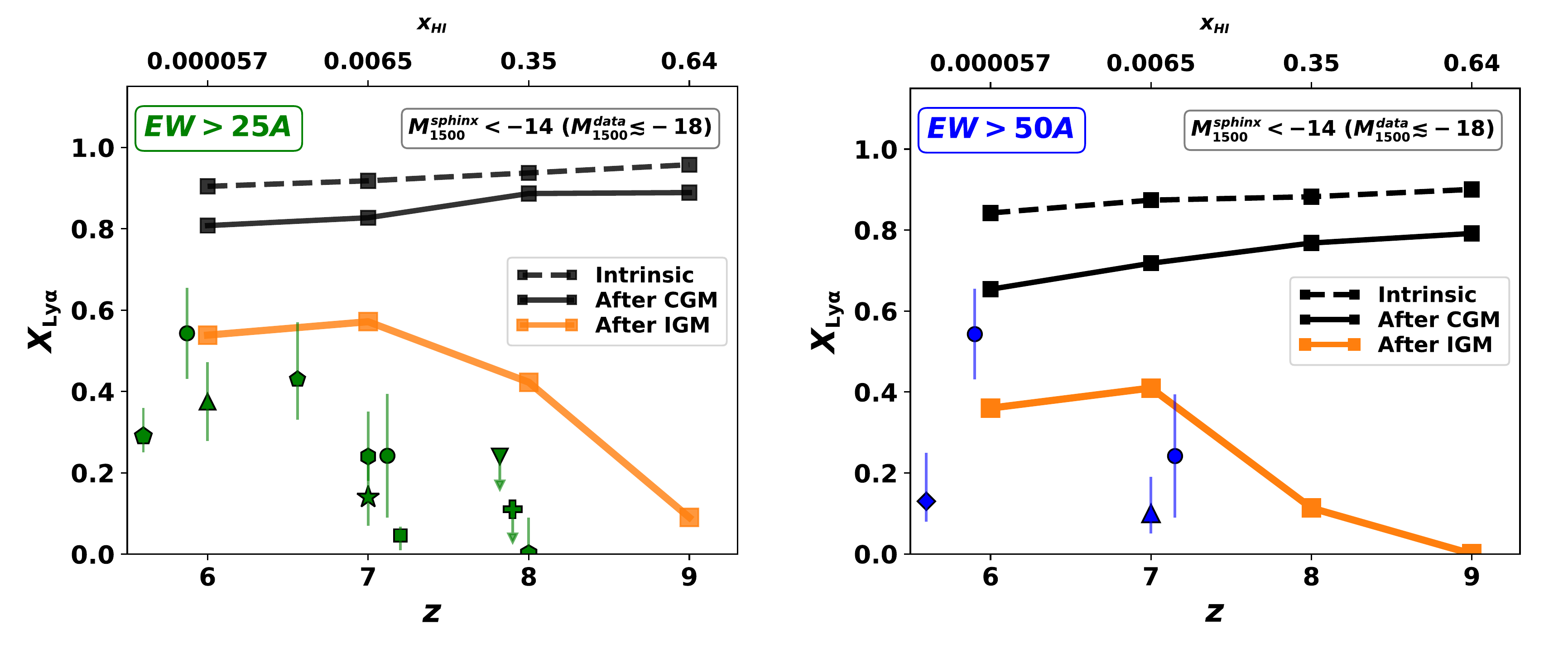}
\vskip-4ex
\caption{Fraction of LAEs in the total sample of galaxies, \xlyat, as a function of redshift. In the left and right panels, we construct the fraction of galaxies with $M_{\rm 1500} < -14$ (see text) exhibiting an EW greater than $25 \AA$ and $50 \AA$ respectively. The LAE fractions based on intrinsic (dashed black lines) and dust-attenuated (solid black lines) UV magnitudes and \lya EWs remain almost constant with $z$. The orange curves show $\xlya(z)$ with the inclusion of the effect of IGM transmission in the computation of the \lya EW. The symbols correspond to the following observational measurements: \citet[][circles]{stark10}, \citet[][upward triangle]{De_Barros_2017}, \citet[][star]{Pentericci_2018}, \citet[][hexagons]{schenker2012a}, \citet[][downward triangle]{Tilvi_2014}, \citet[][pentagons]{Fuller_2020}, \citet[][square]{Hoag_2019}, \citet[][diamond]{Kusakabe_2020}.}
\label{fig:lae_frac-zevol}
\end{figure*}

In Figure \ref{fig:lae_frac-zevol}, we compare our \xlya for weak emitters (i.e. $EW > 25$ \AA) and strong emitters (i.e. $EW > 50$ \AA) in the left and right panel respectively. Ignoring the effect of dust and IGM transmission (dashed black curve), we find for both cases that \xlya does not evolve with redshift, indicating that, on average, the intrinsic \lya strength of galaxies remains unchanged relatively to the stellar continuum. The solid black curve represents the LAE fraction by only accounting for dust (i.e. galaxies selected based on their dust-attenuated magnitudes and \lya EW). Again, \xlya does not decline but remains constant (or even slightly increases) at $z \gtrsim 6$. It is only when IGM transmission is included to compute the \lya EW that \xlya starts to drop sharply around $z \approx 7$ (orange curve), which corresponds to the transition between a fully ionised to a partially ionised Universe in our simulation. While the UV continuum is not affected by the IGM, the \lya line can be strongly altered leading to a significant reduction of the EW, and hence a clear drop in \xlyat. Nevertheless, we note that \xlya is not extremely sensitive to the evolution of \xhi. For weak emitters for instance, \xlya is reduced by a factor 5 between $z=7$ and $z=9$ while the IGM neutrality has increased by a factor of $\approx 100$ over this period. 

As explained earlier, our results are not directly comparable to observations due to the different UV magnitude selection. Nevertheless, we note that the overall shape of the LAE fraction evolution is well recovered by the simulation, especially for $EW > 25$ \AA{} where the constraints are the tightest. In that particular case, the observed \xlya declines from $\approx 50 \%$ at $z \approx 6$ to $\approx 10 \%$ at $z \approx 8$. The simulated \xlya spans a similar range as the observations but with a horizontal shift of about $\Delta z=1$, most probably due to the reionisation history in \sphinxdt. Indeed, as shown in Figure 9 of \citet{Rosdahl_2018}, the \sphinxd neutral fraction is only $\approx 0.005$ at $z = 7$ and rapidly increases to $\approx 0.35$ at $z = 8$ while the observationally estimated \xhi value is already $\approx 0.3-0.4$ at $z = 7$. This just reflects that cosmic reionisation is achieved too early in \sphinxdt. The main and remarkable point is that the predicted evolution of \xlya and the amplitude of its decline is clearly tracing the change in the global neutral fraction of the IGM.

\subsection{\lya IGM transmission}
\label{subsec:tigm}

In this section, we first investigate how the IGM transmission varies with respect to the velocity shift from the line center. Then we compare the evolution of the global, blue, and red transmissions as a function of redshift and galaxy properties.   

\subsubsection{Transmission curve as a function of wavelength}
\label{subsubsec:tlambda}

\begin{figure}
\hspace{-0.3cm}  
\includegraphics[width=0.48\textwidth]{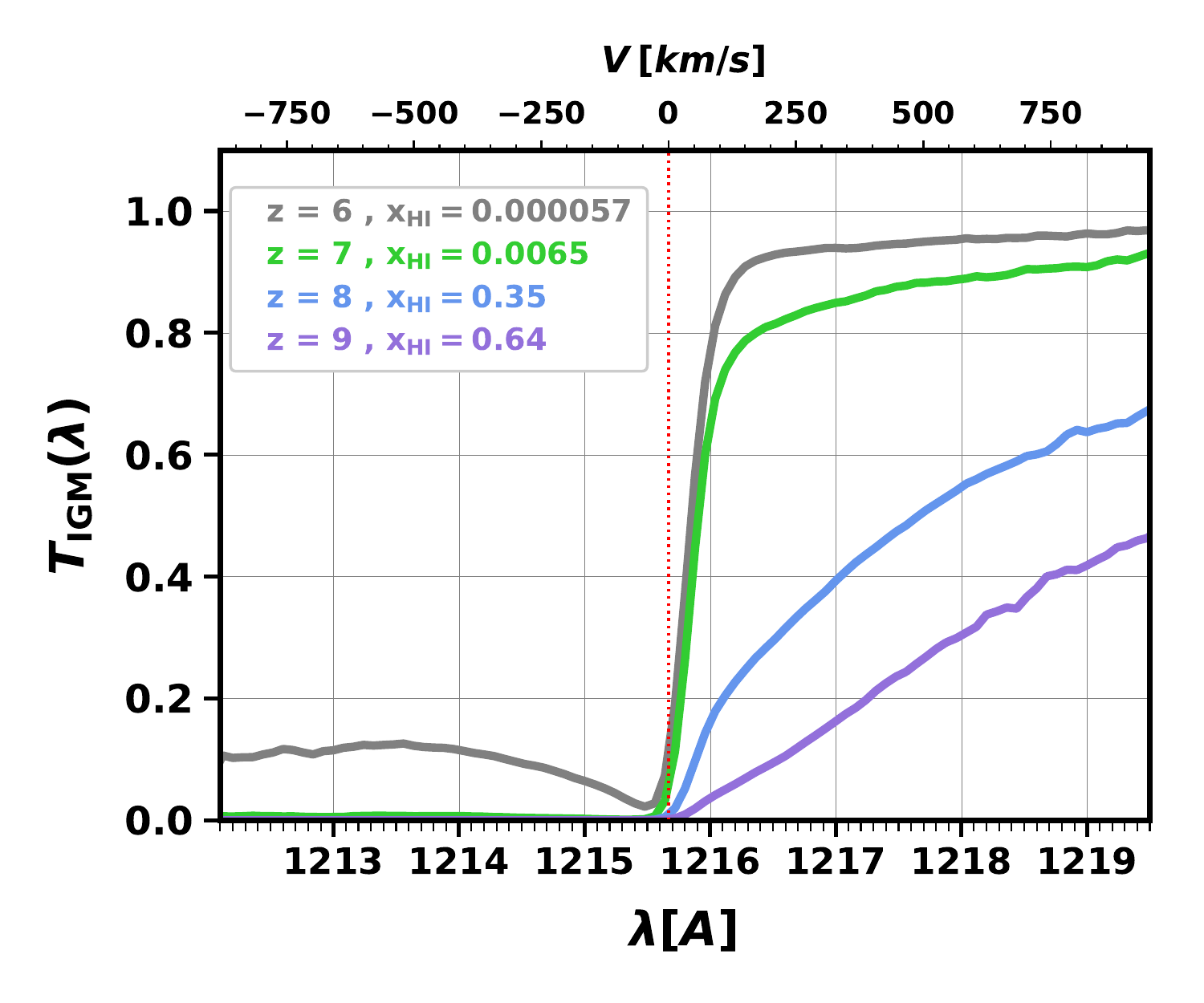}
\vskip-2ex
\caption{\lya IGM transmission as a function of rest-frame wavelength at $z=6, 7, 8$, and 9. The vertical red dotted line indicates the line centre ($V=0$). The evolution of \tigm with redshift reflects the increase of the volumetric neutral fraction \xhi towards higher z (see legend).} 
\label{fig:tigm-lambda}
\end{figure}

Here we focus on the redshift evolution of our simulated \lya IGM transmission. Figure \ref{fig:tigm-lambda} presents the wavelength dependence of \tigm for our four snapshots computed as the mean IGM transmission of \lya photons \textit{that escaped the galaxies}, \tigmt$(\lambda)$. It is worth pointing out that this definition of \tigmt$(\lambda)$ is somewhat different from what has been used previously in the literature. For instance, \citet{laursen2011a} compute the \lya IGM transmission by casting sightlines in random directions from the border of the haloes \citep[see also][]{Gronke_2020}. While this method allows to estimate accurately the average isotropic \lya transmission from a given location through the IGM, it does not account for the possibility that \lya photons escape galaxies along particular lines-of-sight, and that the direction of escape may be correlated with the local IGM distribution. By propagating only photons that can emerge from the CGM, we therefore estimate the \textit{effective} IGM transmission of \lya photons for each galaxy, as opposed to the formulation of \citet{laursen2011a}.
 
At all redshifts, we measure a strong variation of \tigm as a function of $\lambda$ with the blue side being much more suppressed than the red side. This is a well-known consequence of the \lya RT in the Hubble flow : blue photons unavoidably redshift past the resonance along their propagation through the IGM such that they will be scattered off the line-of-sight as soon as \hi is present at the corresponding distance, $d = V / H(z)$ (where $V$ is the velocity offset from line centre of a blue photon in the Hubble flow). In a fully neutral IGM, the medium is extremely optically thick and the blue part will be fully absorbed. In a partially ionised IGM, a diffuse neutral component may remain but the \lya transmission is also strongly affected by surviving dense self-shielded clouds and residual \hi within ionised bubbles \citep{Dijkstra_2017,Gronke_2020}. As can be seen from Figure \ref{fig:tigm-lambda}, a small (but non-negligible) fraction of flux is transmitted at $z=6$ blueward of \lya even though the Universe is almost fully ionised ($\xhi \approx 6 \times 10^{-5}$). We note that \tigm reaches a minimum at $V\approx -100$ \kms{} which is attributed to the increase of gas density in the vicinity of galaxies \citep[see section 5.1 in][for a detailed discussion of this effect]{laursen2011a}. When the volumetric neutral fraction becomes less than about $1\%$ (i.e. $z \approx 7$ in \sphinxdt), the blue part of the spectrum is nearly fully erased and only red photons can be transmitted. 

At $\lambda > \lambda_\alpha$, the IGM transmission increases towards higher velocity offsets as a result of the wing absorption profile of the diffuse neutral component. Closer to the line centre, the red transmission, $T_{\scaleto{\rm IGM}{4pt}}^{\scaleto{\rm red}{3.5pt}}$, can be further decreased by infalling \hi clouds which are able to resonantly scatter photons leading to an IGM absorption. We recall that here we are showing the mean transmissions but there is a strong dispersion of \tigm from one galaxy to another, especially near the line centre due to the directional variation of the occurence rate of optically thick \hi in the neighbourhood of galaxies. 

Overall, the IGM transmission is unsurprisingly dominated by the red part at all redshifts. We also clearly recover a strong evolution with the velocity offset at $\lambda > \lambda_\alpha$, with $T_{\scaleto{\rm IGM}{4pt}}^{\scaleto{\rm red}{3.5pt}}$ at ${\rm V} = 150$ \kms reaching $\approx 80\%$ ($\approx 10-20\%$) at $z \lesssim 7$ ($z=8-9$). As we will discuss in Section \ref{subsec:spectra}, this emphasises the need for realistic modelling of the internal RT since gas outflows in typical high redshift star-forming galaxies \citep{Cassata_2020} can alter and shift the \lya line at similar velocity offsets \citep{verh08}.

\subsubsection{Evolution of \lya \tigm and \fesc with galaxy properties and redshift}
\label{subsubsec:tigm_fesc}

\begin{figure}
\hspace{-0.4cm}  
\includegraphics[width=0.48\textwidth]{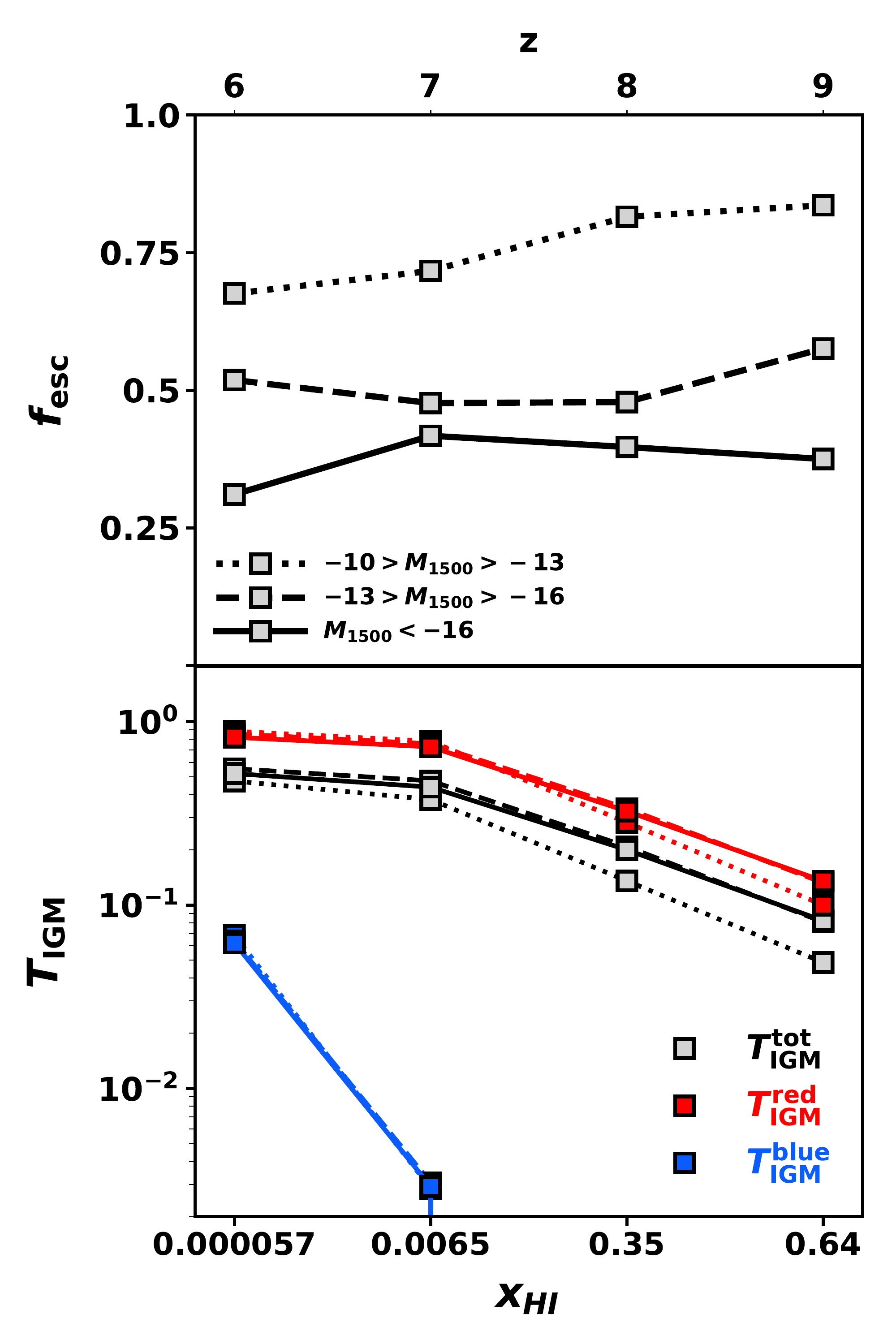}
\vskip-1ex
\caption{Evolution of the median \lya escape fraction (top) and the median IGM transmission (bottom) as a function of IGM neutral fraction and redshift. The different line styles correspond to different bins of dust-attenuated UV magnitude. In the bottom panel, we also highlight the relative evolution of IGM transmission blueward ($T_{\scaleto{\rm IGM}{4pt}}^{\scaleto{\rm blue}{3.8pt}}$) and redward of \lya ($T_{\scaleto{\rm IGM}{4pt}}^{\scaleto{\rm red}{3.5pt}}$).}
\label{fig:fesc_tigm_z}
\end{figure}

Figure \ref{fig:fesc_tigm_z} summarises the redshift evolution of the median \lya internal escape fraction \fesc and IGM transmission \tigm of galaxies split according to their UV magnitude. As mentioned earlier, the ability of \lya photons to escape through the ISM and CGM does not vary much from $z=6$ to $z=9$ on average but it strongly depends on UV magnitude, or equivalently stellar mass (see Figure \ref{fig:scal_rel_muv} for the correlation between $M_{1500}$ and $M_{\star}$). We find that \fesc increases from $\approx 25\%$ for bright UV sources ($M_{1500} \leq -16$) to $\approx 75\%$ at the very faint end ($M_{1500} \geq -13$). 

The IGM transmission on the other hand weakly varies as a function of UV magnitude. \tigm is slightly larger for the bright UV sample than for the UV faint one. There does not seem to be a very strong correlation between \tigm and galaxy properties, or as one could have expected, with environment. In patchy reionisation scenarios, as is the case in \sphinxdt, brighter galaxies form at density peaks so that they can blow larger \hii bubbles around them which can ease the transmission of \lya photons. The fact that we detect only a small environment dependency in \sphinxd is likely due to the fact our box is representative of an \textit{average} field which, by construction, contains neither big voids nor significant overdensities.  

In Figure \ref{fig:fesc_tigm_z}, we also show separately the blue and red median transmissions ($T_{\scaleto{\rm IGM}{4pt}}^{\scaleto{\rm blue}{3.8pt}}$ and $T_{\scaleto{\rm IGM}{4pt}}^{\scaleto{\rm red}{3.5pt}}$) computed respectively from $[-1000;0]$ and $[0;1000]$ \kms. As already mentioned in the previous section, \tigm is mainly determined by the transmission redward of \lyat, especially at $z \gtrsim 7$ when $T_{\scaleto{\rm IGM}{4pt}}^{\scaleto{\rm blue}{3.8pt}}$ drops to zero. At $z=6$, most of the red part of the spectrum is transmitted ($80-90\%$) and $T_{\scaleto{\rm IGM}{4pt}}^{\scaleto{\rm red}{3.5pt}}$ starts to decrease rapidly above $z\approx7$ to reach only $10-15\%$ at $z\approx9$. 

Altogether, this suggests that (i) a small but non-negligible fraction of \lya radiation can be transmitted to the observer even when \xhi is large, and that (ii) the global \lya IGM transmission is fully dominated by the contribution of the red part of the spectrum during the EoR. This reinforces the statement made in previous studies that special care must be given to the modelling of the \lya transfer at smaller scales in order to assess realistically the visibility of LAEs.

\begin{figure*}
\vskip-2ex
\includegraphics[width=0.8\textwidth]{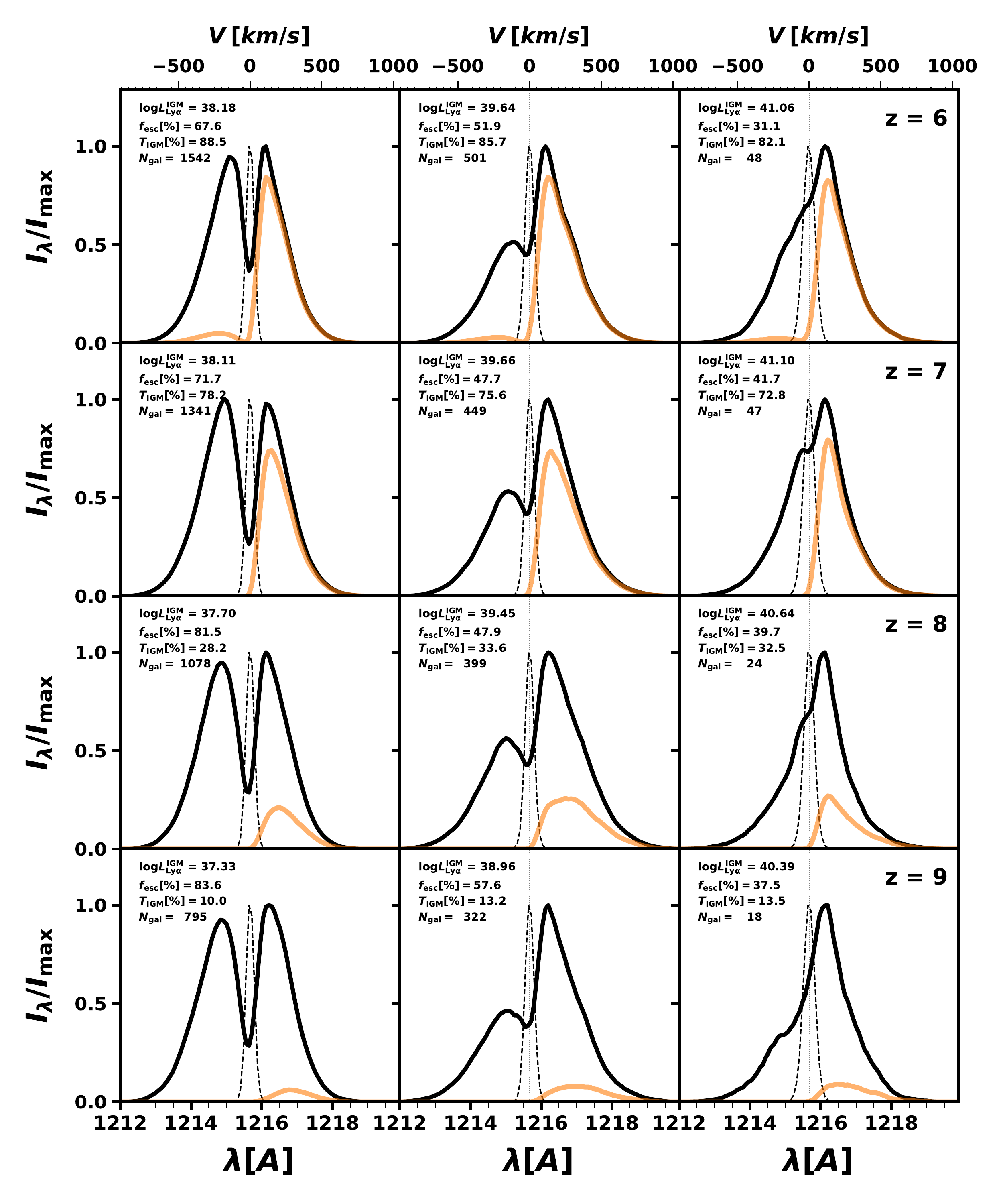}
\vskip-2ex
\caption{Median angle-averaged \lya spectra at z $=$ 6, 7, 8 and 9 (from top to bottom). The three columns correspond to various dust-attenuated UV magnitude ranges: from left to right, $-10 \geq M_{1500} \geq -13$, $-13 \geq M_{1500} \geq -16$, and $M_{1500} \leq -16$. In each panel, the thin black dashed line represents the intrinsic emission while the thick solid black and orange curves show the profiles after CGM and IGM attenuation respectively. The spectra for intrinsic and dust-obscured emission are normalised to their respective maxima whereas the IGM-attenuated spectra are normalised to the maximum of the dust-obscured line profile. The vertical dotted grey line corresponds to the \lya line center. The legend in the top left of each panel gives the median observed \lya luminosity, escape fraction, IGM transmission and the total number of galaxies used to compute the corresponding median spectrum.}
\label{fig:spectra}
\end{figure*}

\subsection{\lya spectra}
\label{subsec:spectra}

In light of the former section, we now turn our interest to the spectral shapes of the \lya line profiles. As discussed in detail in the literature \citep[e.g.][]{santos04,dijk07a,laursen2011a}, the impact of the IGM is highly dependent on the spectral morphology of the line emerging from the galaxy, especially on the velocity offset of the \lya peak with respect to the line center. From low and intermediate redshift observations, we know that typical LAEs harbour a single red asymmetric profile \footnote{These peculiar line shapes are plausibly a consequence of internal RT effects (e.g. back-scatterings in outflowing gas) which redistribute \lya photons redward of the line centre and, consequently, ease their escape from galaxies \citep{verh06,ahn2003a,dijk06}.}, sometimes associated with a smaller blue peak (a.k.a. a blue-bump). As recently shown by \citet{hayes_2020}, this general trend seems to hold up to $z \approx 5$ and the amplitude of the blue peak appears to diminish with increasing redshift because of the IGM opacity. Indeed, IGM absorption is expected to significantly impede the \lya transmission on the blue side of the resonance due to the Hubble flow. In addition, the transmission of the red part strongly varies with velocity shift over a few hundreds of \kms{} (see Figure \ref{fig:tigm-lambda}). This velocity range corresponds to the typical speeds of galactic outflows that are thought to alter the shape, amplitude and peak shift of the \lya line. In this context, assessing the spectral shapes of LAEs after internal RT is therefore essential in order to correctly predict the \lya IGM transmission during the EoR and investigate its connection with the neutral fraction \xhit.

\subsubsection{Relative impact of internal RT and IGM}

In Figure \ref{fig:spectra}, we present the median angle-averaged \lya spectra of our simulated galaxies at z $=$ 6, 7, 8 and 9 in the galaxy frame (from top to bottom) which are split into three bins of dust-attenuated UV magnitudes ($-10 \geq M_{1500} \geq -13$, $-13 \geq M_{1500} \geq -16$, and $M_{1500} \leq -16$ from left to right). In each panel, the thin dashed curves are indicative of the intrinsic Gaussian profiles centred on the \lya line centre ($V=0$). 

We first notice from Figure \ref{fig:spectra} that, in all cases, spectra after internal RT (thick solid lines) are broader than the intrinsic lines ($FWHM \approx 300-600$ \kms) and exhibit significant flux on both blue and red sides. This signature is typical of resonant scattering in optically thick and low-velocity media \citep{neufeld1990a,verh06}, suggesting that galaxies host a dense, slow, \hi gas component in the ISM or in their local environment. That said, we are showing here spectra summed over all directions which erases any directional variation and may further broaden the spectral shapes, such that nearly symmetric angle-averaged spectra do not necessarily indicate RT in static media in our case. Preliminary analysis of the spectra along individual sightlines suggest that there is a very strong variability in terms of spectral morphologies for our galaxies but that flux blueward of \lya is nearly always present in the simulation (Blaizot et al. in prep).

Interestingly, the median line shapes after internal RT presented in Figure \ref{fig:spectra} show very little variation with redshift. This indicates that the physical conditions (e.g. gas density, ionisation state, etc) at the ISM/CGM scale does not evolve much from $z=9$ to $z=6$. Regarding the variation with UV magnitude however, we find a much more significant trend. While \lya profiles in UV faint sources display nearly symmetric double-peaks centred on $V=0$ (left panels), the blue peak becomes strongly reduced towards brighter UV magnitudes. For the brightest UV bin, it is nearly completely suppressed such that the \lya line resembles a redshifted asymmetric line. It is worth pointing out that velocity offset of the red peak is about $150$ \kms, independently of the UV magnitude, which is the typical value measured in LAEs \citep[e.g.][]{Hashimoto_2015}. From the legend of Figure \ref{fig:spectra}, we see that the median \lya escape fractions become smaller towards brighter sources. Nonetheless, UV-brighter objects still correspond to higher \lya luminosities after internal RT. Altogether, our results predict that galaxies that are observable in current surveys ($M_{1500} \leq -16$ and $L_{\rm Ly\alpha}^{\rm intr} \geq 10^{41}$ \ergs) have most of their \lya flux emerging from the CGM redward of \lyat, peaking at $\approx 150$ \kms.
 
\begin{figure}
\hspace{-0.6cm}  
\includegraphics[width=0.5\textwidth]{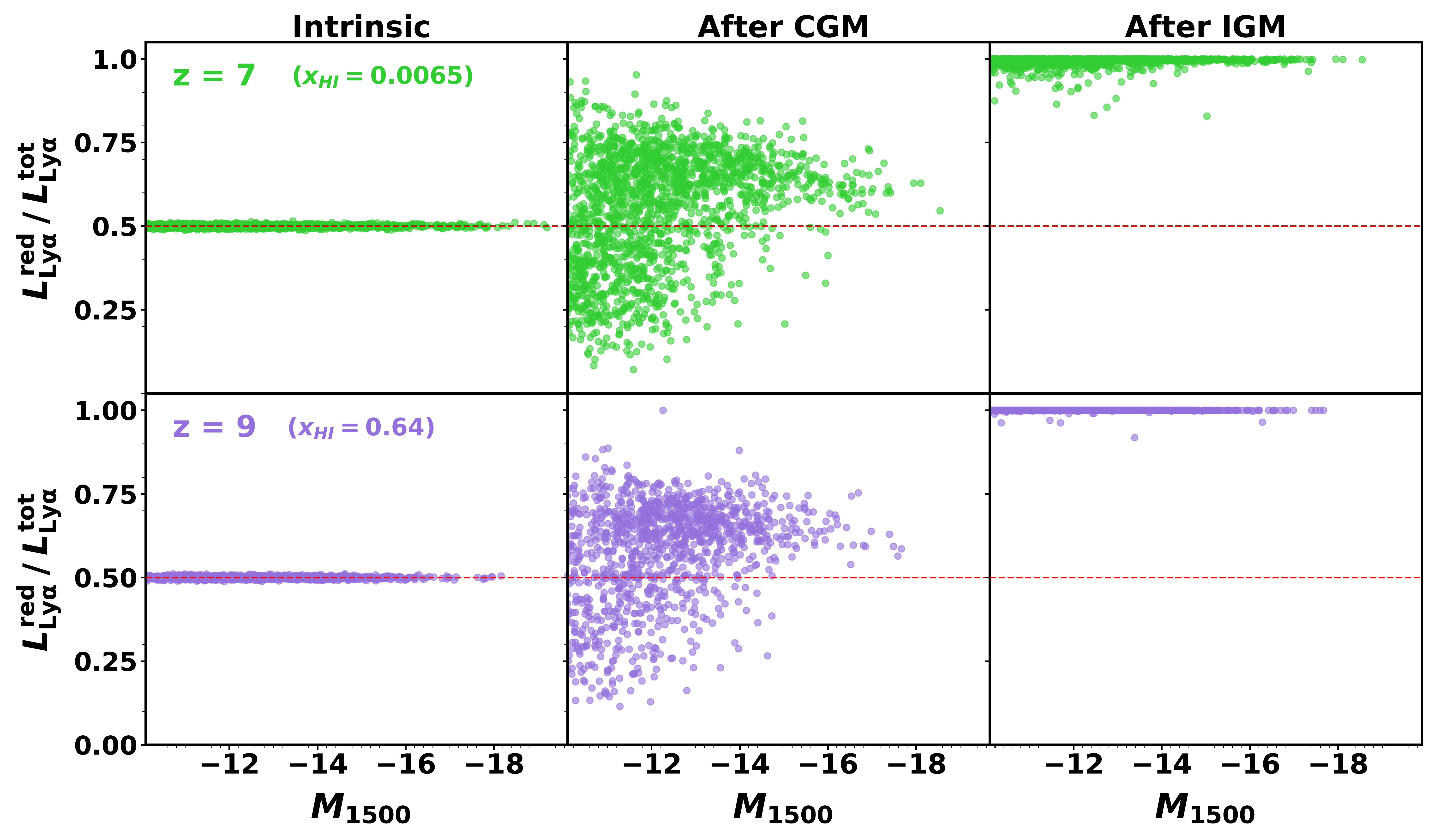}
\caption{Ratio of the red-to-total \lya flux as a function of dust-attenuated UV magnitude at z $=$ 7 (top) and z $=$ 9 (bottom). Each dot corresponds to an individual galaxy. The left, middle and right panels show this ratio for the intrinsic, dust-attenuated, and IGM transmitted emission respectively.}
\label{fig:red-ratio}
\end{figure}

The effect of the IGM on the spectra is visible in Figure \ref{fig:spectra} (orange lines): at $z=6$, the red part is almost fully transmitted whereas only a small fraction of blue photons typically remains. At this redshift, the median spectra resemble the typical observed lines, i.e. redward asymmetric or \textit{blue-bump} profiles. Unsurprisingly, the IGM-attenuated spectra strongly evolve with redshift as the IGM opacity becomes increasingly high. From $z=7$, blue photons can no longer be transmitted because the occurence of clear sightlines drops as soon as the IGM neutral fraction starts rising. A $z \geq 8$, even the red peak becomes strongly suppressed and only a weak red peak is transmitted (see Section \ref{subsec:tigm}). 

Figure \ref{fig:red-ratio} summarises the evolution of the red-to-total \lya flux ratio at $z=7$ and $z=9$ with UV magnitude for intrinsic emission (left), escaping emission (middle) and IGM-transmitted emission. We clearly see that internal RT is a major cause of the frequency redistribution of \lya photons, preferentially towards the red in UV bright galaxies. At $z=7$, the IGM significantly favors the transmission of \lya photons on the red side, although a non negligible fraction of blue photons manage to be transmitted along clear sightlines. At $z=9$ however, only photons with a sufficient red-shift avoid IGM absorption and can reach the observer.

\subsubsection{Variation of the spectral shape as a function of radius}
\label{subsubsec:spectra_r}

As detailed in the previous section, our angle-average spectra can only reproduce the observed typical shapes once we account for IGM transmission. This feature is commonly seen in \lya RT experiments in cosmological hydrodynamical simulations \citep{laursen09,Smith_2018,Mitchell_2020b}. However, single-peak red-shifted profiles are also commonly observed at low redshift (where the impact of IGM is negligible). Therefore, the failure of hydrodynamic simulations at predicting these line shapes after internal RT is most likely related to gas outflows, and especially to the lack of fast-moving neutral hydrogen predicted by state-of-art simulations of galaxy formation. As shown by \citet{barnes2011a}, the \lya line shapes (as well as the \lya spatial distribution) are very sensitive to the underlying galactic wind properties.

Single-peak red profiles with various peak shift, skewness and width seem to only be reproduced in more idealised \lya numerical experiments in which the input \lya line propagates through high-velocity ($\gtrsim 100$ \kms) and dense ($\gtrsim 10^{19}$ cm$^{-2}$) \hi outflows \citep{verh08,Hashimoto_2015,Gronke_2017}. Assessing if single red peaks after internal RT also prevail at very high redshift is still an open question. It is therefore unclear if (i) the small-scale \hi distribution/kinematics predicted by cosmological simulations is somehow unrealistic (due to e.g. poorly constrained subgrid feedback models), or if (ii) there is a redshift evolution of the ISM/CGM conditions yielding double-peak profiles which are then suppressed on the blue side by the increasingly neutral IGM. It is worth pointing out that, in the latter case, the \lya line shape could represent a very powerful probe of the IGM topology during reionisation \citep{Gronke_2020}.

\begin{figure}
\includegraphics[width=0.50\textwidth]{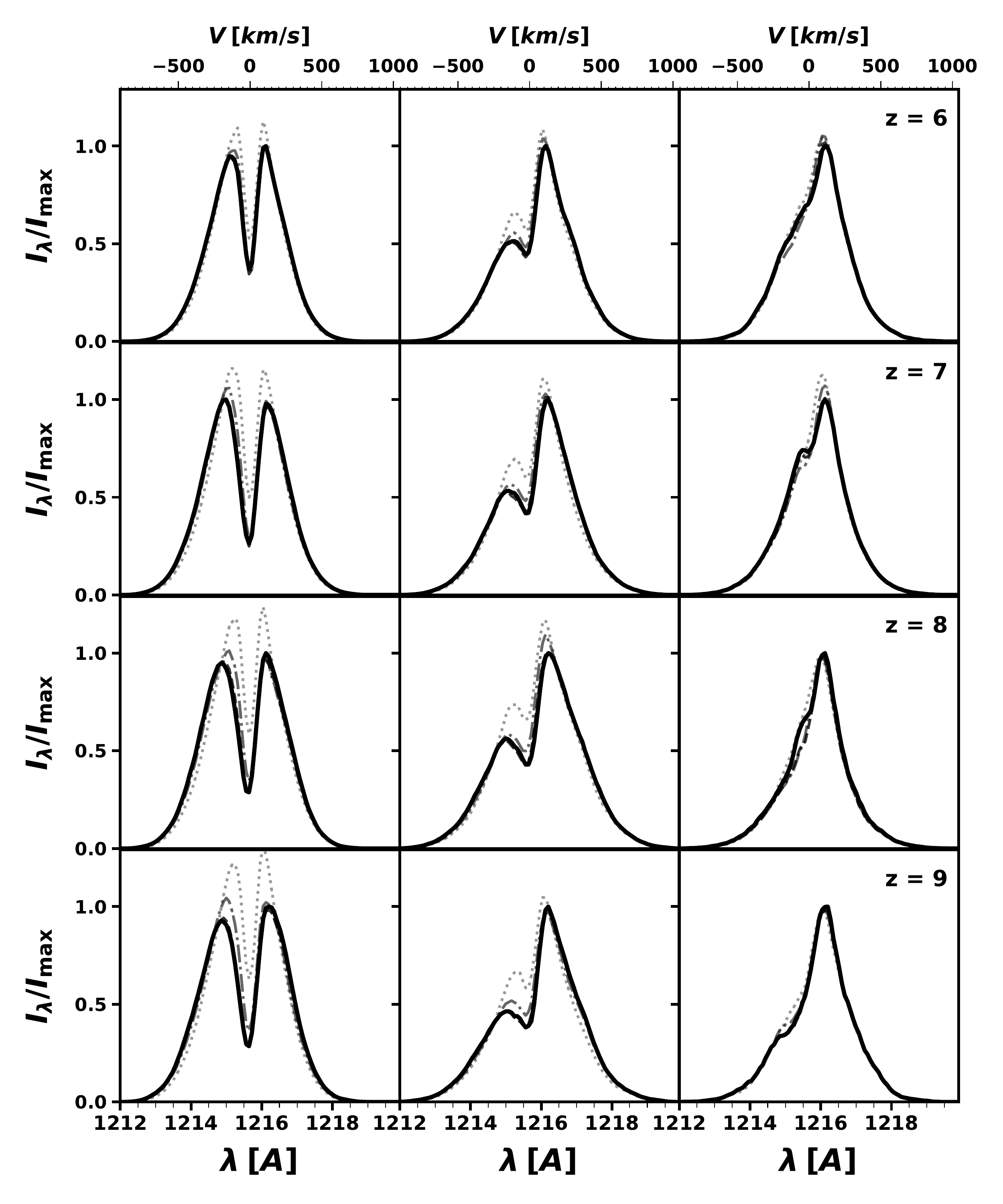}
\vskip-2ex
\caption{Variation of the median angle-averaged \lya spectra as a function of radius at z $=$ 6, 7, 8 and 9 (from top to bottom). The three columns correspond to various dust-attenuated UV magnitude ranges: from left to right, $-10 \geq M_{1500} \geq -13$, $-13 \geq M_{1500} \geq -16$, and $M_{1500} \leq -16$. In each panel, the thick solid black curves show the profiles evaluated at $R_{\scaleto{\rm CGM}{3.5pt}} = 10r_\star$, as in Figure \ref{fig:spectra}. To assess the radial evolution of the spectral shape, we also show the median angle-averaged profiles computed at $r_\star$ (dotted grey line), $2r_\star$ (dot-dashed grey line), and $5r_\star$ (dashed grey line). All spectra are normalised to the maximum of the profile measured at $R_{\scaleto{\rm CGM}{3.5pt}}$.}
\label{fig:spectraRcgm}
\end{figure}

Adding complexity to the problem, we do not fully understand either how the red single-peak \lya profile is formed in low redshift galaxies, i.e. whether it is mainly via radiation transfer effects in the ISM or in the CGM. Recently, \citet{Kimm_2019} investigated the escape of \lya photons from turbulent ISM clouds simulated with RAMSES-RT at sub-pc resolution. Quite interestingly, their findings suggest a strong variability of the \lya line shape emerging from the clouds, sometimes producing a red-dominated profile with a less prominent blue bump, which suggests that \lya spectra may be already (at least partially) in place at very small scale \citep[see also][]{Kakiichi_2019}. While the physical resolution in \sphinxd prevents us from resolving such fine structure, we can still quantify the evolution of the \lya profiles from ISM to CGM scale. To do so, we plot in Figure \ref{fig:spectraRcgm}, the median spectra after internal RT computed at $r_\star$, $2r_\star$, $5r_\star$, and $10r_\star (=R_{\scaleto{\rm CGM}{3.5pt}})$. Overall, we find very little variation in terms of shape and amplitude whatever the redshift or UV magnitude range. This demonstrates that, in our simulation, the frequency distribution of \lya photons is mostly set in the ISM. A closer look at Figure \ref{fig:spectraRcgm} reveals that the peaks are slightly broader with a larger separation at larger radii. This suggests that \lya photons do keep scattering in the CGM (as shown in Appendix \ref{appendix:rcgm}) but that the density, kinematics, and/or covering fraction of the neutral gas in the CGM are not sufficient to alter significantly the emergent spectral shapes. Besides, Figure \ref{fig:spectraRcgm} demonstrates that the shape of \lya spectra is fairly independent of the exact value of CGM scale which further validates our choice of choosing $R_{\scaleto{\rm CGM}{3.5pt}} = 10r_\star$. In light of \citet{Kimm_2019}'s study, it is worth pointing out that achieving higher resolution in the ISM \citep[or also in the CGM; see][for instance]{Tumlinson_2017,Gronke_2017b} would affect more strongly the typical line shapes. Altogether, our results and the above discussion highlight the importance of internal \lya radiative transport for interpreting LAE observations during the EoR as well as the uncertainties related to that matter. 

\subsubsection{Velocity shift of the \lya line and IGM transmission}

To illustrate the possible impact of the \lya internal RT on the visibility of LAEs during the EoR discussed in the previous section, we introduce a toy model for the \lya LF in which the spectrum emerging from the CGM is arbitrarily modified. As discussed in the previous section, our \lya spectra after CGM RT are double-peaked with a peak separation of $\approx 300 \kms$ for fainter galaxies and single-peaked with an offset of $\approx +150 \kms$ for bright ones. Here, we assess by how much the IGM transmission, and therefore the observed \lya LF, would change if different spectral shapes were assumed. 

To do so, we keep the dust-attenuated \lya luminosities the same in our toy model but we replace the \lya lines after internal RT by single-peaked Gaussian profiles with various rms widths, $\sigma_{\rm v}$, and positive velocity peak offsets, $V_{\rm peak}$. This assumption on the line shape is quite simplistic because most observed \lya lines usually appear either asymmetric or double-peaked but, to first order, the two parameters $\sigma_{\rm v}$ and $V_{\rm peak}$ are sufficient to investigate the overall effect of line broadening and red-shift on the \lya transmission by IGM. The individual IGM transmission \tigm$(\lambda)$ of each galaxy is applied to each profile to compute the IGM-transmitted luminosities. 

\begin{figure*}
\includegraphics[width=0.99\textwidth]{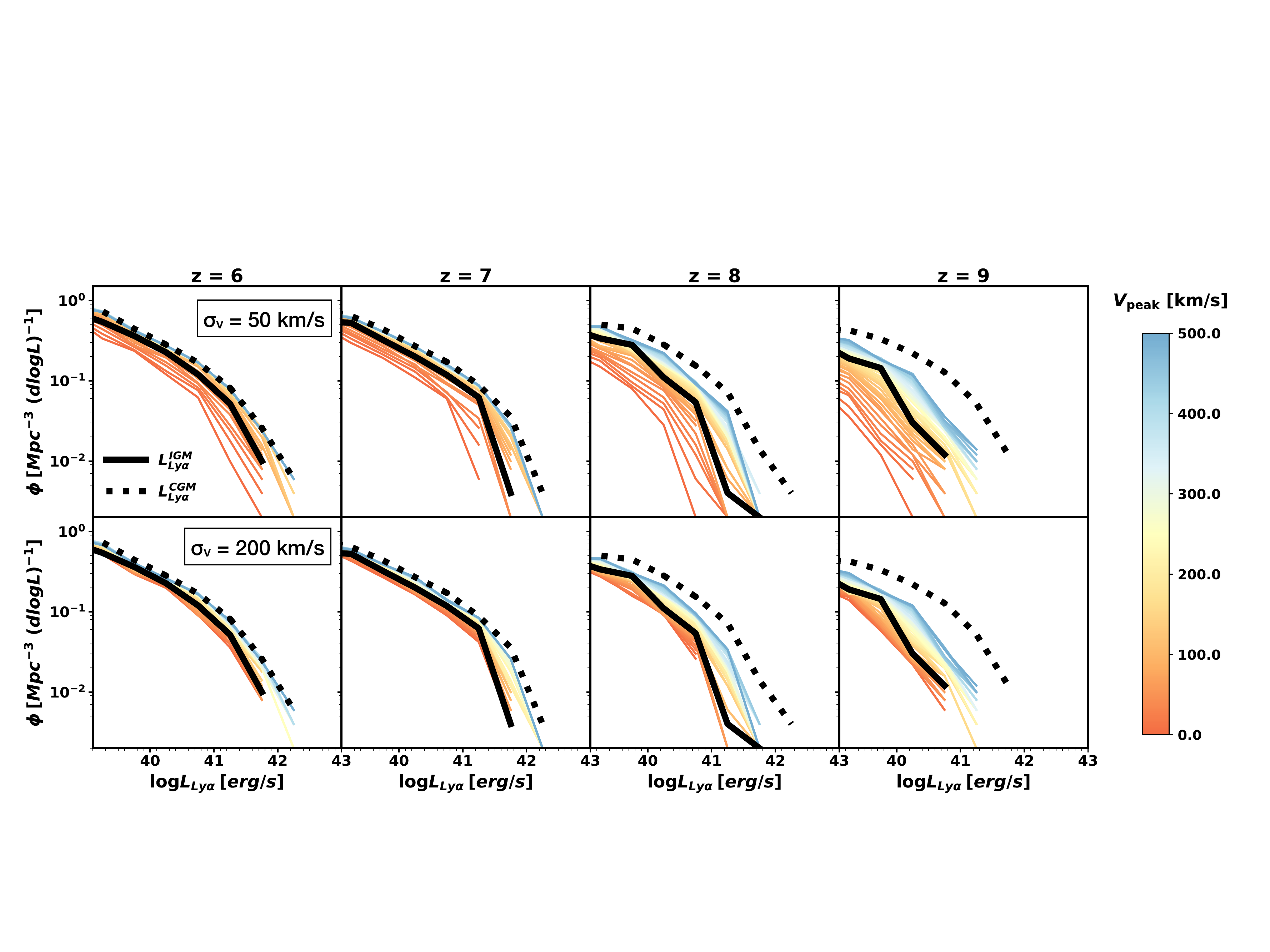}
\caption{Impact of the \lya line shape on the IGM transmission. The four panels show our fiducial \lya luminosity functions at z $=$ 6, 7, 8 and 9 for dust-attenuated (black dotted lines) and IGM-transmitted emission (solid black lines) as in Figure \ref{fig:lyalf-igm}. The coloured lines correspond to our toy model in which the line shape emerging from each galaxy (i.e. after CGM RT) is artificially replaced by a Gaussian profile with varying rms width ($\sigma_{\rm v}$) and velocity peak offset ($V_{\rm peak}$). The IGM-transmitted \lya luminosities ($L^{\scaleto{\rm IGM}{3.5pt}}_{\rm Ly\alpha}$) are then computed from the individual IGM transmission \tigmt$(\lambda)$ of each galaxy. The top and bottom panels correspond to $\sigma_{\rm v} = 50$ and 200 \kms{} respectively and the $V_{\rm peak}$ values are given by the colorbar.}
\label{fig:lyalf-toymodel}
\end{figure*}

As can be seen from Figure \ref{fig:lyalf-toymodel}, the observed \lya LF can dramatically change depending on the shape of the \lya line emerging from galaxies and the effect becomes stronger towards higher redshifts. For the parameter values assumed here, the dispersion induced on the \lya luminosities amounts to $\approx$ 0.5 dex at $z=6$ to $\approx$ 2 dex at $z=9$. Small $V_{\rm peak}$ values tend to significantly reduce the IGM transmission because, in this case, most of the \lya flux escapes galaxies near the line center where the IGM absorption is maximal (Figure \ref{fig:tigm-lambda}). Conversely, large velocity offsets (up to 500 \kms{} in our toy model) greatly favour the transmission of \lya photons to the observer. This is particularly true at $z\lesssim7$ where \tigm resembles a step-function around $V=0$ where the red part is nearly fully transmitted. At these redshifts the IGM neutral fraction is less than 0.01 and we find that the IGM becomes fully transparent to \lya photons for $V_{\rm peak} \gtrsim 300 \kms$. Note that this trend holds for the two $\sigma_{\rm v}$ assumed here, 50 and 200 \kms. These rms widths correspond to full-width-at-half-maximum of $\approx$ 120 and 470 \kms, typical of faint high-redshift LAEs (Kerutt et al., in prep). For a narrow line ($\sigma_{\rm v}=50 \kms$), the IGM transmission almost only depends on $V_{\rm peak}$ which leads to the large dispersion in the resulting LFs (top panel of \ref{fig:lyalf-toymodel}). For a broader line (bottom panel), more \lya photons can be transmitted to the red whatever the velocity peak offset, so varying $V_{\rm peak}$ has a milder effect on the LFs (as long as $\sigma_{\rm v}$ remains the same order of magnitude as $V_{\rm peak}$).

\section{Discussion}

\subsection{IGM \lya transmission and neutral fraction}

The \lya emission from galaxies has long been put forward as a possible probe of reionisation \citep{miralda-escude1998a,Haiman_2002,furlanetto2006a}. Observations show evidence that \lya emission becomes increasingly suppressed at $z\gtrsim6$ as can be inferred from the evolution of the \lya LF, LAE clustering and \xlya \citep{Zheng_2017,Itoh_2018,Ouchi_2017,schenker2012a}. 

In agreement with our present study, this is often interpreted as the imprint of the reionisation of the inter-galactic medium, but alternative explanations have been suggested. A possible scenario is of course the co-evolution of galaxy and \lya properties towards high redshift, plausibly due to variations of the gas and dust content, distribution, and kinematics \citep{dayal2012,jensen2013a,garel2015a,Hassan_2021}. The incidence of optically thick systems in the vicinity of galaxies can also have a dramatic effect on the \lya visibility \citep{bolton2012a} and reduce the number of strong emitters towards higher redshift. \citet{Sadoun_2017} pointed out the possible impact of a rapidly evolving UV background on the ionisation state of the CGM of the galaxies themselves. As the infall region becomes more self-shielded towards higher redshift, \lya scattering would produce a more extended emission component that can be partially missed by observing apertures and thus artificially induce a drop of \xlyat.
Finally, cosmic variance is undoubtedly a source of uncertainties in deep surveys, especially for measurements of \xlya that are based on spectroscopic samples. Nevertheless, this effect is unlikely to fully dominate the observed evolution of \lya properties at $z\gtrsim6$ \citep{Taylor_2013}. 

\begin{figure}
\hspace{-0.2cm}  
\includegraphics[width=0.45\textwidth]{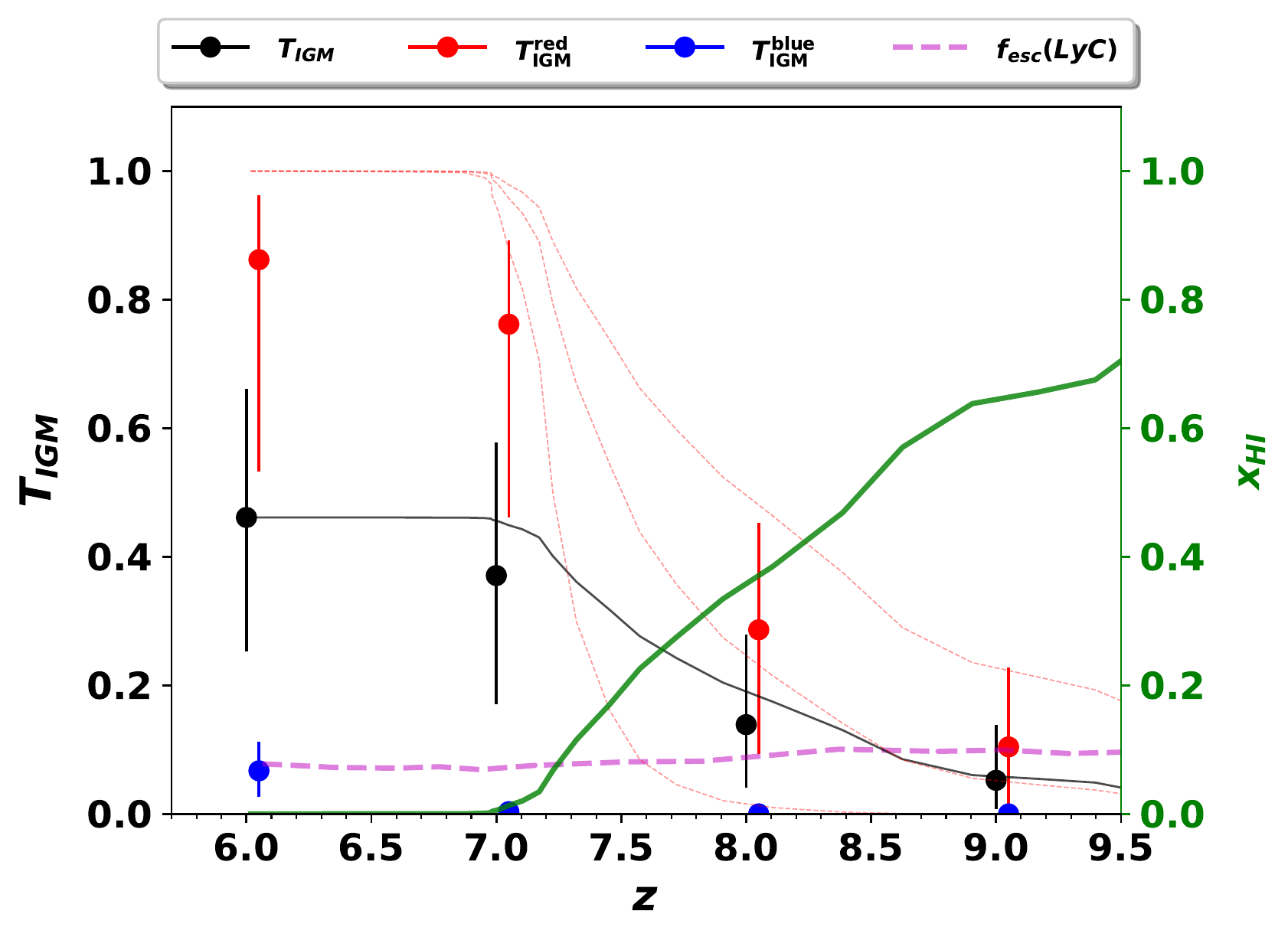}
\vskip-2ex
\caption{Comparison of the redshift evolution of \tigm with the ionising escape fraction $f_{\scaleto{\rm esc}{3.3pt}}$(LyC) and the volumetric IGM neutral fraction \xhit. The black, blue, and red circles correspond to our median \tigmt, $T_{\scaleto{\rm IGM}{4pt}}^{\scaleto{\rm blue}{3.8pt}}$, and $T_{\scaleto{\rm IGM}{4pt}}^{\scaleto{\rm red}{3.5pt}}$ respectively, and the error bars are the 10-90th percentiles. Note that the red and blue circles are shifted by $+0.05$ for clarity. The solid grey line shows the fit to \tigm as a function \xhi assuming the following functional form : \tigm $= T_{\scaleto{\rm IGM}{4pt}}^{z=6} \times (1-x_{\scaleto{\rm HI}{3.5pt}})^2$. The red dotted curves represent the Gunn-Peterson transmission for red photons assuming a velocity shift of $100, 300,$ and 600 \kms{} (from bottom to top; see text).The green curve shows the redshift evolution of \xhi in \sphinxdt. The dashed magenta curve is the average LyC escape fractions of galaxies as measured in \citet{Rosdahl_2018}.}
\label{fig:qhi_z}
\end{figure}

Our results suggest that the contributions of intrinsic galaxy properties and dust attenuation are not driving the apparent evolution of LAEs during the EoR and that only IGM transmission is at play. This does not necessarily mean however that \tigm is an obvious probe of the reionisation process. In \sphinxdt, galaxies start ionising the Universe by $z\approx15$ and \xhi drops to $50\%$ at $z\approx8.5$ and below $1\%$ at $z\approx7$ . This phase change is sustained by the inclusion of binary stars which boost the escape of LyC photons. As shown in Figure \ref{fig:qhi_z}, the mean ionising escape fraction changes only insignificantly between $z=6$ and 9 ($5-10\%$) which, interestingly, echoes the redshift evolution found for the \lya escape fraction (Figure \ref{fig:fesc_tigm_z}). Note however that the escape of \lya photons from the ISM/CGM is set by dust attenuation (which is enhanced by resonant scattering in optically thick gas) whereas the escape of ionising photons is fully driven by the \hi opacity within galaxies in \sphinx\footnote{The impact of dust on ionising RT was neglected in the \sphinx simulations \citep{Rosdahl_2018} but it would only have a very minor effect on the escape of ionising photons (i.e. probably delaying reionisation very slightly) because \hi Lyman-continuum opacities largely dominate over dust optical depths in our galaxies.}. 

Figure \ref{fig:qhi_z} also compares the evolution of \tigmt, now in linear scale, with other fundamental quantities related to cosmic reionisation. The global \lya IGM transmission computed from all galaxies at each snapshot decreases from $\approx 45\%$ at $z=6$ to $\approx 5\%$ only at $z=9$ (black circles). Over the same redshift range, the volumetric neutral fraction of the IGM \xhi varies from $\lesssim 1\%$ to $\approx 65\%$ \citep[green line; see Figure 13 in][]{Rosdahl_2018}. \tigmt(z) appears to be nicely anti-correlated to \xhit(z) and it can be well fit by a functional form (grey curve). 

As discussed in Section \ref{subsubsec:tigm_fesc}, the IGM transmission of blue photons quickly falls to zero at $z\gtrsim7$. However, the red transmission $T_{\scaleto{\rm IGM}{4pt}}^{\scaleto{\rm red}{3.5pt}}$ evolves strongly with redshift (red circles) and traces fairly well the global IGM transmission but boosted by a factor $\approx 2$. We discussed in Section \ref{subsubsec:tlambda} how \lya photons can be transmitted redward of the line centre. While photons may be scattered by infalling \hi clouds close to resonance, the damping wing of the neutral component of the IGM is mainly responsible for the overall shape of $T_{\scaleto{\rm IGM}{4pt}}^{\scaleto{\rm red}{3.5pt}}(\lambda)$. The connection between \lya transmission and \xhi therefore strongly depends on the velocity shift after internal RT and the distance to nearby neutral patches, i.e. the size of the \hii bubble in which a galaxy is embedded. 

We further illustrate this aspect by showing the expected Gunn-Peterson (GP) transmission for photons redward of \lyat, $e^{{\text -}\scaleto{\tau}{3.6pt}_{\scaleto{\rm GP}{3.7pt}}}$, where the GP opacity $\tau_{\scaleto{\rm GP}{3.7pt}}$ is inversely proportional to the velocity shift from resonance $\Delta {\rm V}$ in a partially neutral IGM \citep[$\tau_{\scaleto{\rm GP}{3.7pt}} \propto (\Delta {\rm V})^{-1}(1+z)^{\scaleto{3/2}{5.5pt}}$;][]{miralda-escude1998a,dijkstra2010a} \footnote{As detailed in \citet{Dijkstra_2017}, $\tau_{\scaleto{\rm GP}{3.7pt}} = 2.3 x_{\rm D} \Big(\frac{\Delta {\rm V}}{600 \: \kms}\Big)^{-1}\Big(\frac{1+z}{10}\Big)^{\scaleto{3/2}{5.5pt}}$ where $x_{\rm D}$ corresponds to the "patch-averaged" neutral fraction which depends on the volumetric neutral fraction \xhi in a non-trivial way. Since we only intend to illustrate the inverse scaling between GP opacity and velocity offset in Figure \ref{fig:qhi_z}, we assume here for simplicity that $x_{\rm D} = \xhi$.} . Here, $\Delta {\rm V}$ corresponds to the offset seen by the neutral IGM so it therefore includes both the contribution of outflows and Hubble flow. In Figure \ref{fig:qhi_z}, we plot the GP transmission redward of \lya for three values of $\Delta {\rm V}$ ($100, 300, $ and 600 \kms{}; red dotted lines from bottom to top) which span a similar range of velocities as our simulated spectra (see Figure \ref{fig:spectra}). We see that the $T_{\scaleto{\rm IGM}{4pt}}^{\scaleto{\rm red}{3.5pt}}$ predicted by the analytic model varies strongly depending on $\Delta {\rm V}$ but it reproduces well the general redshift evolution of the simulated transmission. While our mean simulated spectra peak at $V \approx 150$ \kms, it is the model with V $ = 300$ \kms{} that provides the best match to our results. In addition to the impact of outflows, this also plausibly highlights the connection between IGM transmission and the topology of \hii bubbles in the environment of galaxies during the EoR. This key aspect was studied recently by \citet{Gronke_2020} and we intend to investigate it in a forthcoming paper.

\subsection{Model assumptions and potential caveats}
\label{subsec:disc_caveats}
As detailed in Section \ref{subsec:sphinx}, \sphinxd provides an unprecedented trade-off between resolution (in terms of mass and physical sampling) and box size for cosmological RHD simulations which allows us to capture large-scale reionization as well as the physics and radiation transfer in resolved galaxies. The $10^3$ cMpc$^3$ simulated volume remains nevertheless relatively small which has a number of implications. First, our simulation does not contain large scale overdensities so we miss the contribution of massive galaxies (as well as active galactic nuclei) to reionisation. Regarding LAEs, bright sources are rare in \sphinxd and the highest observed \lya luminosities at $z=6$ reach $L_{\rm Ly\alpha} \approx 10^{42}$ \ergs{} which prevents us from comparing our results with most observational data. 
Due to its limited size, our simulation also underestimates the large scale fluctuations and peculiar motions of the IGM \citep[see][for a discussion on the scale needed to capture the global topology of reionisation]{Iliev_2014}, which may have an impact on our derived \lya transmissions. However, our mean \tigm at a given z seems to weakly depend on the UV magnitude (Figure \ref{fig:fesc_tigm_z}; we have also tested that the trend is similar by taking stellar mass bins instead of $M_{1500}$). Since $M_{1500}$ is, on average, tracing the environment (i.e. brighter UV galaxies are preferentially located in denser environments), our mean IGM transmissions do not appear to be very sensitive to the IGM topology fluctuations at the $10$ cMpc scale. In addition, the effect of IGM on the \lya visibility is dominated by \lya absorption in the close environment of LAEs in \sphinxdt. Indeed, we checked that varying the stopping criterion for the IGM \lya RT from 1 to 10$L_{\rm box}$ (see Section \ref{sec:igm_rt}) gives very similar results (e.g. similar Lya LFs). Altogether this suggests that the average \lya IGM transmissions in \sphinx are unlikely to be significantly affected by the box size for the population of galaxies we are looking at.

Another potential caveat in our study is that intrinsic \lya emission is only arising from the ISM ($r < r_\star$) as the contribution of CGM emission (at $r_\star < r < R_{\scaleto{\rm CGM}{3.5pt}}$) is ignored on purpose. Including photons produced in-situ in the CGM for each source up to $R_{\scaleto{\rm CGM}{3.5pt}}$ would be problematic in our setup because a non-negligible number of gas cells would be associated with more than one source. This would be particularly significant in low-mass galaxies that are strongly clustered around more massive objects. Overcounting such cells would artificially, and incorrectly, boost the intrinsic emission budget of many LAEs. We bypass this potential issue by restricting the emission region to the ISM ($r < r_\star$). Our intrinsic \lya luminosities can be seen as conservative values but we have checked that the bulk of intrinsic emission is coming from the ISM and therefore is accounted for, especially for bright galaxies for which the amount of missed \lya emission is negligible \citep[see also ][]{Mitchell_2020b}.    

Finally, a strong hypothesis in our \sphinxd run is the use of BPASS (v2.0) with 100\% binary stars. As shown in \citet{Rosdahl_2018}, this SED model has the net advantage of leading to a efficient and rapid ionisation of the Universe whereas a more standard library based on single-stars only fails to do so. We note that the Universe is almost fully ionised by $z=7$ in \sphinxd which is about $\Delta z=0.5$ earlier than suggested by observations \citep[Figure 9 in][]{Rosdahl_2018}. Interestingly, Figure \ref{fig:lae_frac-zevol} shows that our predicted LAE fraction disagrees with the observed \xlyat. Instead, our LAE fraction starts decreasing at $z\gtrsim7$ which corresponds to the epoch where the IGM neutral fraction becomes non-negligible in our simulation, suggesting that \xlya traces well the evolution of the IGM neutral phase. BPASS also significantly boosts the intrinsic budget of \lya photons produced through case B recombination (see Section \ref{subsubsec:intr_lya}). Considering 100\% binary stars is perhaps somewhat extreme and we note that models with a slightly lower fraction of binaries, and thus generating less ionising and \lya photons per starburst, may be more realistic.

While it is important to keep these aspects in mind to guide future work and to make quantitative predictions for forthcoming \lya surveys during the EoR, none of these caveats is likely to significantly affect our main results, i.e. that the LAE evolution with redshift is predominantly due to IGM absorption and not to an evolution in the ISM or CGM.

\section{Summary}

Using the non-zoom cosmological radiation-hydrodynamics \sphinxd simulation, we have investigated the redshift evolution of the \lya signatures of galaxies during the epoch of reionisation. We have mainly focused on the relative impacts of intrinsic evolution, dust-attenuation at ISM/CGM scales and IGM transmission on the visibility of LAEs at $z \geq 6$. The unique ability of \sphinxd to capture both reionisation at Mpc scale and the production and escape of ionising radiation within galaxies allows us to attempt for the first time to fill the gap between simulations of \lya RT within individual galaxy environments and \lya propagation through the IGM. 

In order to study the imprint of reionisation on \lya observables, we post-process four different snapshots with \rascas (at $z=6, 7, 8,$ and $9$) and compute \lya angle-averaged properties of galaxies. The \sphinxd volume being fully reionised by $z=6-7$ (predominantly thanks to the inclusion of binary stars in the SED modelling), our study covers a period of $\approx400$ Myr over which the IGM neutral fraction evolves from $60 \%$ to nearly $0\%$.

Even though our study is mostly restricted to faint and low-mass objects, we show that our simulation can reproduce a number of observational constraints at high redshift, in particular the stellar mass and UV/\lya luminosity functions. The detailed analysis of the \lya LFs and EW distributions in the different snapshots tells us that the redshift evolution of intrinsic and dust-attenuated properties is very mild from $z=6$ to 9, if not null. Hence, the contribution of these processes to the observed \lya suppression at $z\gtrsim6$ is predicted to be completely sub-dominant. We find however a significant reduction in terms of \lya fluxes and EW due to the increase of the IGM opacity with redshift. The inclusion of the IGM transmission provides good agreement with observational data at $z=6$. We also measure the redshift evolution of the LAE fraction \xlya and find that it is indeed a promising diagnostic to probe reionisation. While the value of \xlya varies with the UV magnitude of the selected galaxy population, its evolution with redshift is almost fully driven by the change of the ionisation state of the Universe. Looking back in time, \xlya is found to be nearly constant as long as the IGM volumetric neutral fraction \xhi is less than one percent, and to decline once the Universe becomes more and more neutral. 

In our simulation, the typical \lya escape fraction from galaxies is on average $25\%$ ($75\%$) for bright/massive sources (faint/low-mass) but remains constant from $z=6$ to 9. The global \lya IGM transmission \tigm drops from $45\%$ to $5\%$ between $z=6$ and $z=9$ and we find a positive, but barely significant, trend with UV magnitude which may suggest that the \lya observability is slightly enhanced in more massive/overdense environments. While a small fraction of the flux blueward of \lya ($5-10\%$) is transmitted at $z=6$, the blue side of the spectrum is completely erased by the IGM at $z\gtrsim7$. Individual clear sightlines might nevertheless exist at these redshifts but we decided to focus on mean properties and to leave the study of the directional variation to future work.  

The global IGM transmission is found to be mainly driven by the red part of the spectrum. The red transmission is however a strong function of the velocity shift  
of \lya photons that emerge from the galaxy. While the angle-averaged line profiles after internal RT are nearly symmetric for faint galaxies, brighter observable sources exhibit single-peak spectra red-shifted by $\approx 150$ \kms{} on average, most likely because of outflows at the ISM/CGM scale. At $z\approx6$, a significant fraction of the flux redward of \lya can be transmitted ($\approx 80\%$). At $z\approx9$, where the neutral fraction is already $\approx 60\%$, we find a mean red transmission of approximatively $10\%$ which suggests that a non-negligible number of intrinsically bright LAEs may still be visible during the EoR. Interestingly, we do not predict any break or turnover of the \lya LF at the faint-end, and we instead find that the observable number of LAEs keeps rising down to $L_{\rm Ly\alpha} = 10^{37}$ \ergs{} at least.

This first study of the full \lya modelling in a cosmological RHD simulation highlights that careful modelling of the internal \lya RT is essential to assess the impact of the IGM on the observability of LAEs during the EoR. Nonetheless, further improvements are required to draw more general conclusions. First, higher physical resolution may be needed to describe the \lya radiation transfer process in the ISM and in the CGM. Although the resolution reached in \sphinxd is already substantial for a cosmological simulation, we still have to rely on idealised or zoom simulations to assess the impact of small-scale structure on the \lya RT \citep[e.g.][]{Kimm_2019}. Second, the relatively small volume of \sphinxd is insufficient to capture the larger modes of structure formation and the large-scale topology of reionisation. Building upon the present study, we intend to make a step forward by extending our analysis to a new (eight times bigger) simulation (Rosdahl et al., in prep.).

\section*{Acknowledgements}
We thank the anonymous referee for helpful comments. TG and AV are supported by the ERC Starting Grant 757258 "TRIPLE". The results of this research have been achieved using the PRACE Research Infrastructure resource SuperMUC based in Garching, Germany (PRACE project ID 2016153539). We are grateful for the excellent technical support provided by the SuperMUC staff. The radiation transfer simulations and analysis were also performed at the Common Computing Facility (CCF) of the LABEX Lyon Institute of Origins (ANR-10-LABX-0066). TK was supported in part by the National Research Foundation of Korea (NRF-2019K2A9A1A0609137711 and NRF-2020R1C1C100707911) and in part by the Yonsei University Future-leading Research Initiative (RMS2-2019-22-0216). This work was supported by the Programme National Cosmology et Galaxies (PNCG) of CNRS/INSU with INP and IN2P3, co-funded by CEA and CNES. MGH acknowledges support from the UKRI  Science and Technology Facilities Council (grant numbers ST/N000927/1 and ST/S000623/1).

\vspace{-0.5cm}
\section*{Data availability}
The data underlying this article will be shared on reasonable request to the corresponding author. 

\vspace{-0.5cm}

\bibliographystyle{mn2e}
\interlinepenalty=10000
\bibliography{biblio_sphinx10}

%%%%%%%%%%% APPENDICES %%%%%%%%%%%%%%%%
\appendix

\section{Last scatterings in the CGM}
\label{appendix:rcgm}

As discussed in Section \ref{sec:internal_rt}, we need to define an arbitrary size for the CGM of galaxies in order to separate internal RT from IGM RT. Although we are interested in angle-averaged quantities in this study, we compute the IGM RT by assuming that any photon being scattered during its propagation in the IGM is removed from the line-of-sight, and therefore not transmitted to the observer. However, at CGM scale, photons that scatter have a probability to be re-directed back and forth on a given sightline, as demonstrated by the large projected extent of \lya emission around high-z star-forming galaxies \citep{steidel2011a,Wisotzki_2016}. Choosing a CGM scale that is too small would lead us to remove photons during the IGM RT that still have a chance to scatter back towards the observer. 

\begin{figure}
\includegraphics[width=0.41\textwidth]{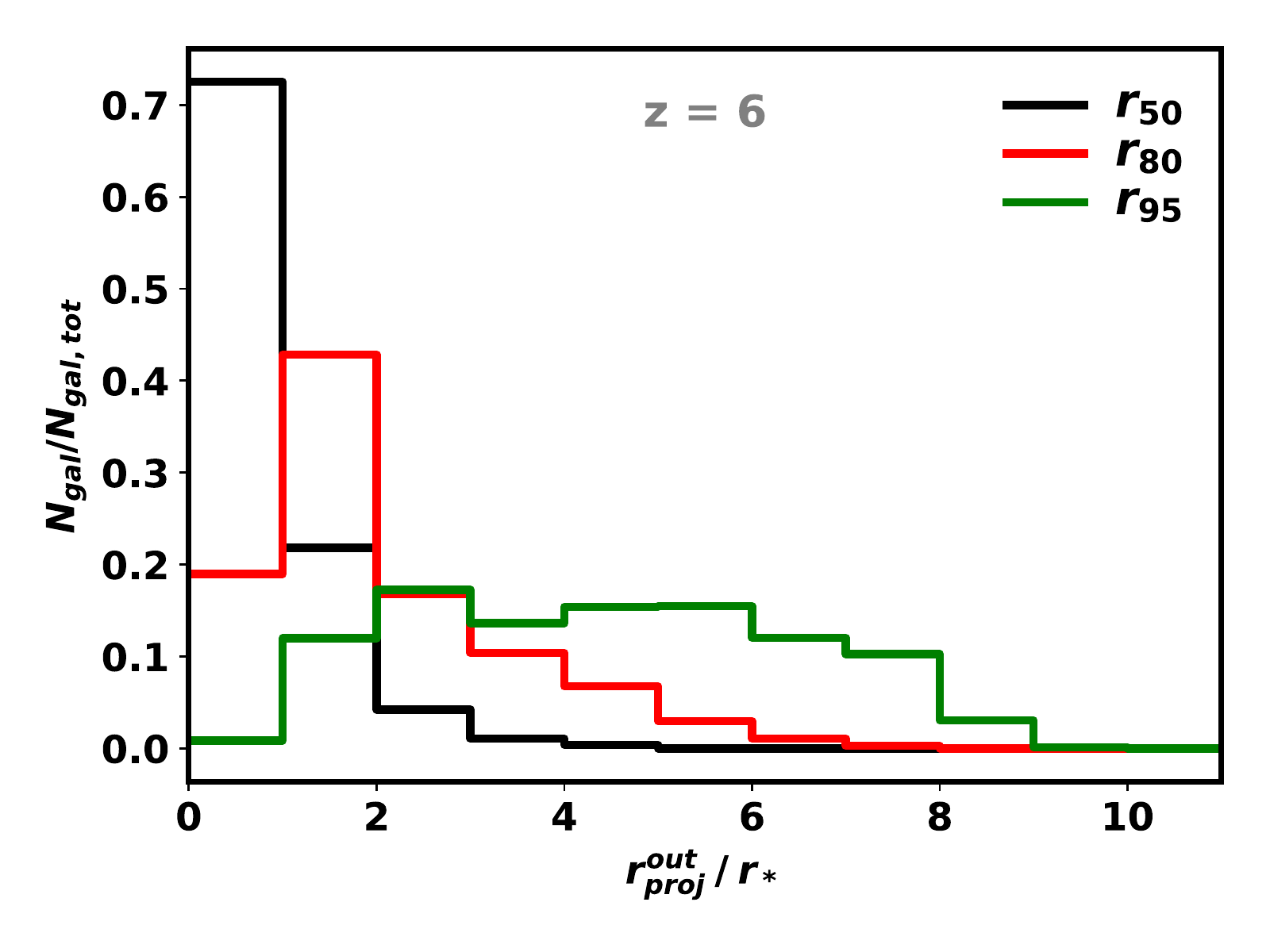}
\includegraphics[width=0.41\textwidth]{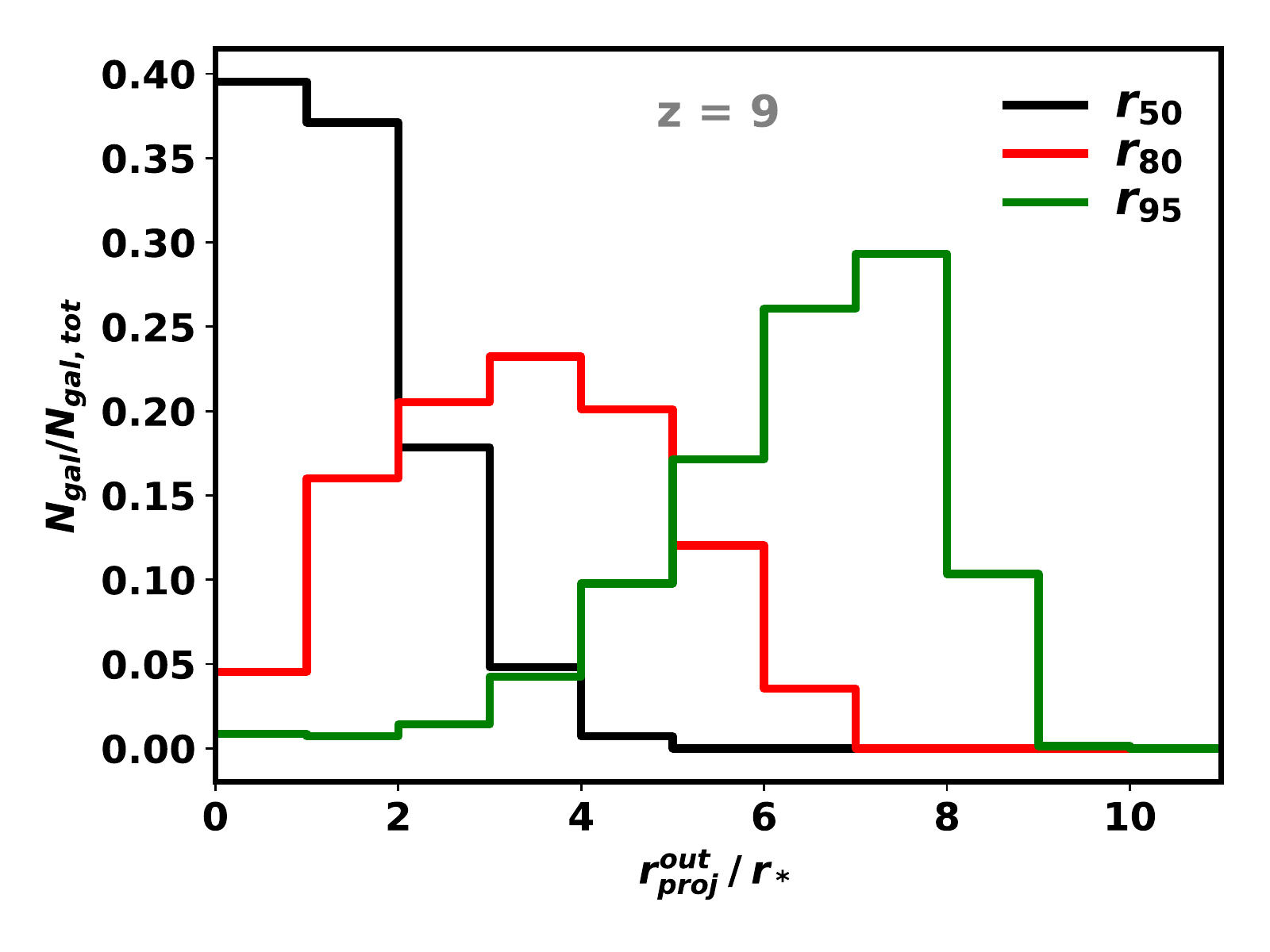}
%\vskip-2ex
\caption{Distribution of the last-scattering radii of \lya photons in the CGM. $r_{50}$, $r_{80}$, and $r_{95}$ are the scales at which 50, 80, and 95\% of the \lya photons undergo their last scattering for $z=6$ (top) and $z=9$ (bottom). For each galaxy, we compute the normed projected radius, $r^{\rm out}_{\rm proj} / r_{\star}$, at which a given fraction of \lya photons last-scatter before escaping the CGM.}
\label{fig:rcgm}
\end{figure}

By choosing $R_{\scaleto{\rm CGM}{3.5pt}} = 10r_\star$, the number of scatterings in the relatively dense CGM occurring beyond this scale should be small, as suggested by Figure \ref{fig:rcgm}. This figure shows the distribution of distances at which $n$\% of the \lya photons undergo their last scattering. In practice, for each galaxy, we compute the projected map of the last scatterings and then compute their distance to the center of the galaxy, $r^{\rm out}_{\rm proj}$. This is equivalent to imaging the CGM in \lya and stacking over all directions. We then calculate the 3D radii at which 50, 80, 90, and 95\% of the \lya photons have their last scattering and plot the distribution for all galaxies at $z=6$ (top) and $z=9$ (bottom). The \lya half-light radii (black curves) of galaxies correspond to $\approx 1-2 r_{\star}$ for most galaxies. Interestingly, photons seem to scatter further out in the CGM at $z=9$ compared to $z=6$, plausibly due the CGM being more neutral towards higher redshifts. Nevertheless, we see that the radius at which at least 95\% of \lya photons last-scatter is always below $10r_\star$ both at $z=6$ and $z=9$. It is therefore reasonable to treat \lya photon interactions with hydrogen atoms at $r > R_{\scaleto{\rm CGM}{3.5pt}}$ as IGM absorptions.

\section{Variation of the LAE fraction with UV magnitude}
\label{appendix:xlya_vary_muv}

\begin{figure}
\hspace{-0.2cm}
\includegraphics[width=0.48\textwidth]{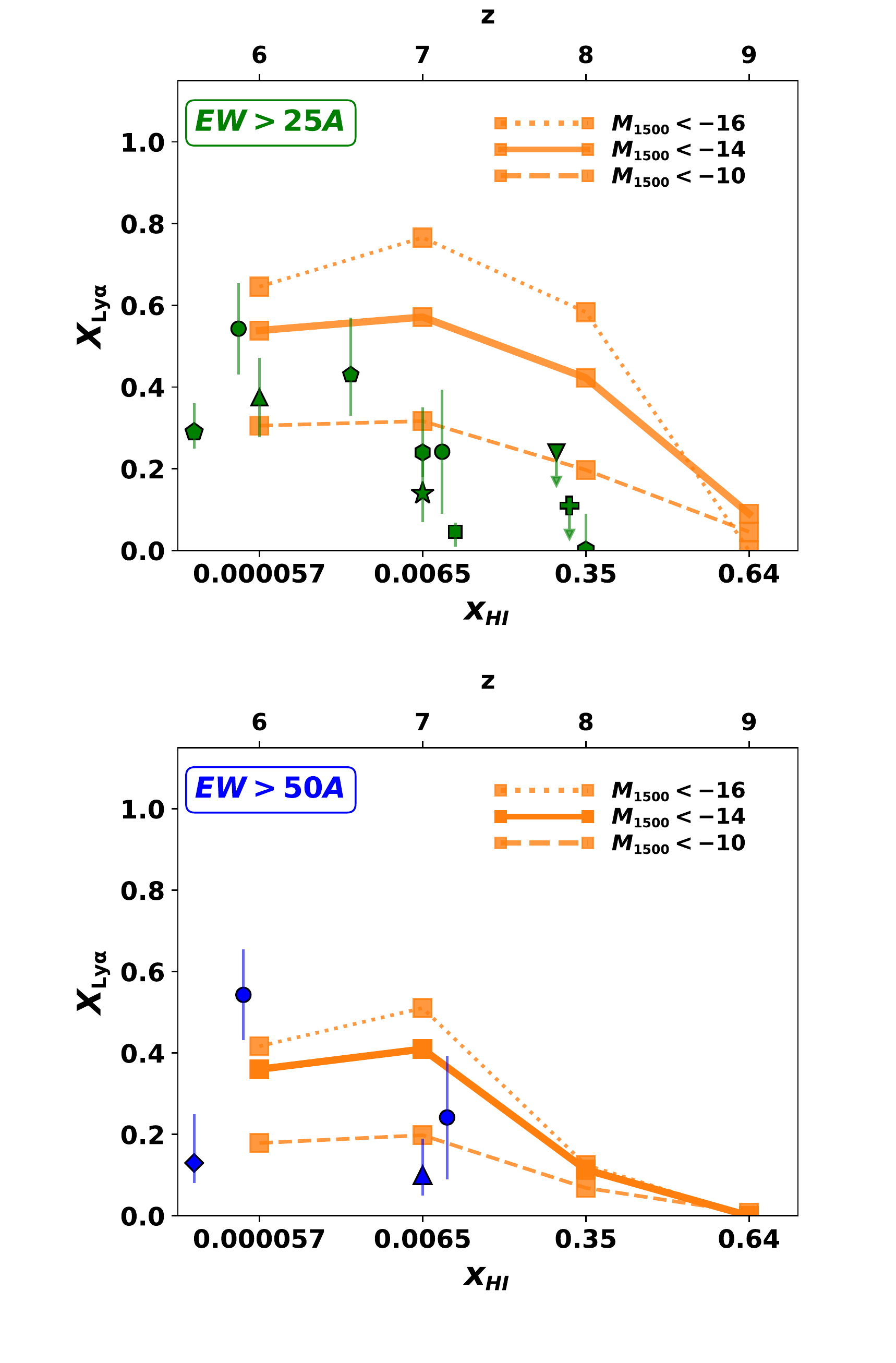}
\vskip-4ex
\caption{Variation of the LAE fraction \xlya with UV magnitude cuts. \xlya is computed from IGM-transmitted \lya luminosities and dust-attenuated UV magnitudes (i.e. the "after IGM" case). The top (bottom) panel shows \xlya with LAEs selected with a \lya EW threshold of 25 \AA{} (50 \AA). We show \xlya for three different UV magnitude cuts : $M_{1500} = -16$ (dotted line), $M_{1500} = -14$ (solid line), and $M_{1500} = -10$ (dashed line).}
\label{fig:xlya_vary_muv}
\end{figure}

In Section \ref{subsec:xlae}, we showed the LAE fraction \xlya predicted by \sphinxd using a UV magnitude cut of $M_{1500} = -14$. In Figure \ref{fig:xlya_vary_muv}, we test other values : $M_{1500} = -16$ and $-10$. We see that using a brighter (fainter) cut increases (decreases) the amplitude of \xlya at all redshifts. This is because the \lya EW are, on average, correlated with UV magnitude in our simulation (Figure \ref{fig:scal_rel_muv}). Nevertheless, it is remarkable that the general trend of \xlya as a function of redshift remains unchanged whatever the UV and EW cuts, i.e. the fraction of LAEs drops towards $z=8-9$ due to the IGM becoming more neutral. 

\section{Scaling relations}
\label{appendix:scal_rel}

\begin{figure*}
\includegraphics[width=0.9\textwidth]{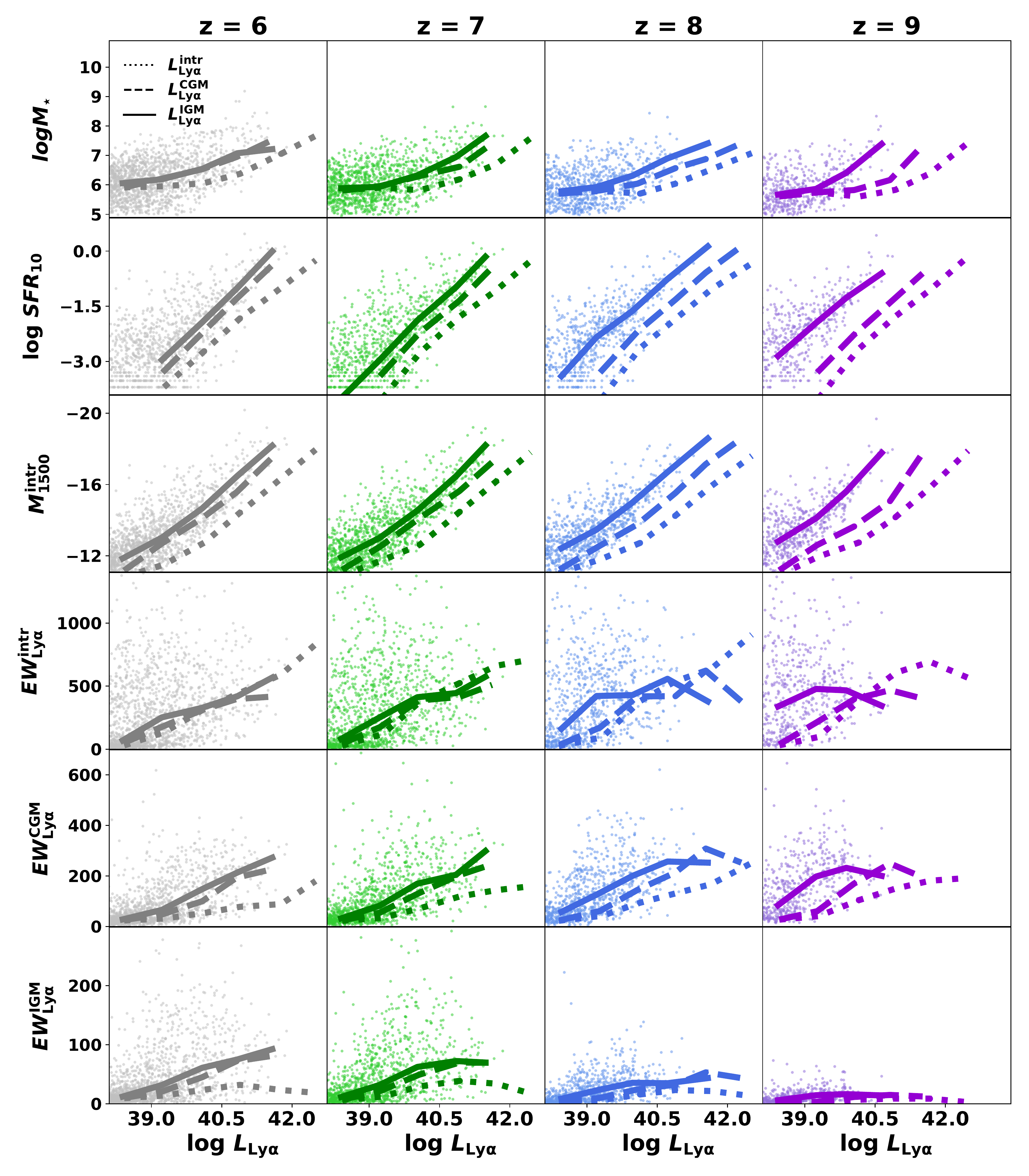}
\vskip-2ex
\caption{Scaling relations between \lya luminosities and various galaxy properties at $z=6, 7, 8,$ and 9 (columns from left to right). The dotted, dashed, and solid lines represent the median galaxy properties per bin of intrinsic, dust-attenuated, and IGM-transmitted \lya luminosities respectively. In each panel, the data points correspond to the IGM-transmitted luminosity of individual sources. $M_{\star}$ : stellar mass (in \msun), $SFR_{10}$ : star formation rate over 10 Myr (\msunyr), $M_{1500}^{\rm intr}$ : intrinsic UV magnitude, $EW_{\rm Ly\alpha}^{\rm intr/CGM/IGM}$ : intrinsic/dust-attenuated/IGM-transmitted \lya equivalent width (in \AA; rest-frame).}
\label{fig:scal_rel_llya}
\end{figure*}

Figure \ref{fig:scal_rel_llya} shows that intrinsic \lya luminosities are positively correlated with the main galaxy properties in \sphinxdt, i.e. stellar mass, SFR, and UV magnitude. Interestingly, these scaling relations hold when we account for \lya internal RT and IGM attenuation, except that \lya luminosities are shifted to lower values. As discussed in Section \ref{subsubsec:lyalf_evolution}, the $M_{\star}$-$L_{\rm Ly\alpha}$ relation gives us indications on the \lya luminosity completeness of our simulation due to our limited mass resolution. Star particles correspond to $10^3$ \msun{} and we chose, for the sake of this study, to only select galaxies with more than a hundred particles. From the top panel of Figure \ref{fig:scal_rel_llya}, we see that the highest IGM-attenuated \lya luminosities allowed in objects at our stellar mass limit ($10^5$ \msun) can reach $\approx 10^{40}$ \ergs, so we consider this value as our \lya completeness limit.

\begin{figure*}
\includegraphics[width=0.9\textwidth]{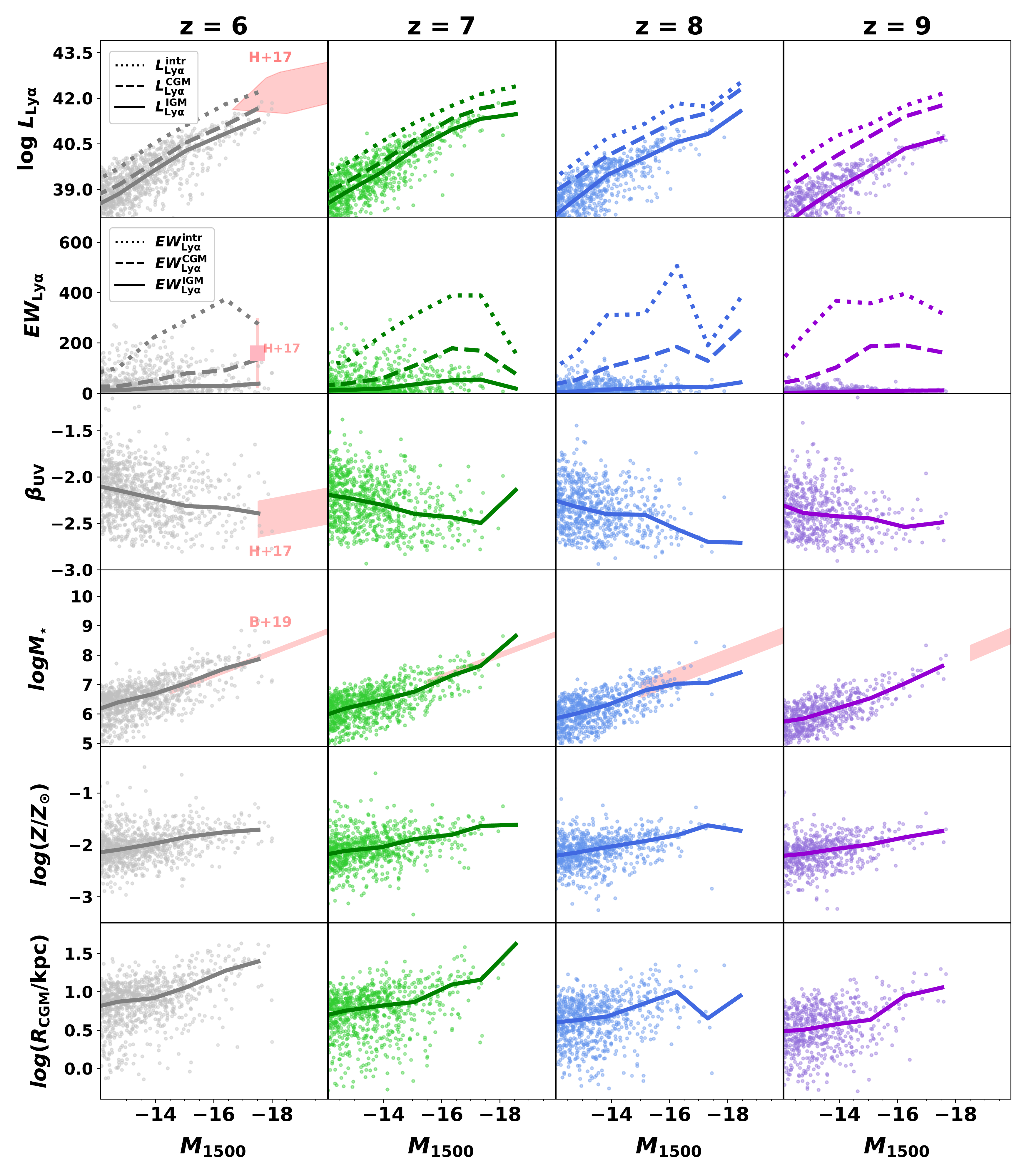}
\vskip-2ex
\caption{Scaling relations between the dust-attenuated UV magnitude, $M_{1500}$, and various galaxy properties at $z=6, 7, 8,$ and 9 (columns from left to right). In each panel, solid lines show the median relations while data points represent individual sources. In the top two rows, the dotted, dashed, and solid lines show the median relation between $M_{1500}$ and the intrinsic, dust-attenuated, and IGM-transmitted \lya luminosities (1st row) and EW (2nd row) respectively. Data points correspond to the IGM-transmitted \lya properties in these first two rows. The pink square and shaded areas represent observational contraints from \citet{Hashimoto_2017} (H+17; $5.5 < z < 6.7$) and \citet{Bhatawdekar_2019} (B+19; $z=6$, $z=7$, $z=8$, $z=9$), as labelled on the figure. $L_{\rm Ly\alpha}$ : \lya luminosity (in \ergs), $EW_{\rm Ly\alpha}$ : \lya equivalent width (in \AA; rest-frame), $\beta_{\rm UV}$ : UV slope, $M_{\star}$ : stellar mass (in \msun), $Z$ : gas-phase metallicity in units of solar metallicity $Z_{\odot}$, $R_{\scaleto{\rm CGM}{3.5pt}}$ ($ = 10r_\star$) in physical kpc.}
\label{fig:scal_rel_muv}
\end{figure*}

\lya luminosities appear to be correlated with \lya EW when looking at the median values (curves in Figure \ref{fig:scal_rel_llya}). This has the effect of predominantly reducing the number of objects at the faint-end of the \lya LF rather than the bright-end when selecting LAEs above fixed EW cuts (see Section \ref{subsubsec:igm_lya}). 
Note that there is a very strong dispersion from one object to another in the $L_{\rm Ly\alpha}$-$EW$ relation (dots). This is mainly because intrinsic EW values are sensitive to metallicity (see Figure \ref{fig:scal_rel_muv}) and to variations of the recent SF histories of galaxies, where the continuum traces young stars over the last $\lesssim 100$ Myr whereas \lya is tracing hot massive stars at shorter timescales. 

Figure \ref{fig:scal_rel_muv} presents additional scaling relations and comparisons of \sphinxd galaxy properties with observational data. It shows the relations between the dust-attenuated UV magnitudes and \lya luminosities, \lya EWs, UV slopes, stellar masses, and gas metallicities. Only our brightest sources can be compared with observations ($M_{1500} \approx -18$) but we find very good agreement with existing constraints on the link between $M_{1500}$ and $L_{\rm Ly\alpha}$, EW, $\beta_{\rm UV}$ and $M_{\star}$ \citep{Hashimoto_2017,Bhatawdekar_2019}. 

The last row of Figure \ref{fig:scal_rel_muv} shows the CGM radius of each individual galaxy. The $R_{\scaleto{\rm CGM}{3.5pt}}$ values range from a bit less than one pkpc to $\approx 30$ pkpc and, on average, are larger for brighter sources. This is of the order of the DM halo virial radii in \sphinxd \citep[i.e. with masses between $\approx 10^8$ and $10^{11}$ \msun;][]{Rosdahl_2018} which span a range between $\approx$ 1 and 20 pkpc at $z=6-9$. In addition, typical instrument apertures usually have $2^{''}$ diameter, which corresponds to $\approx 10$ pkpc at $z = 6-9$ so some objects may be more extended than these typical apertures, hence inducing potential flux losses. However, we do not expect our observed \lya luminosities to vary much as a function of the aperture size. Based on the discussion in Section \ref{subsec:disc_caveats} and \ref{subsubsec:spectra_r}, the \lya emission is only arising from the ISM (i.e. at $r < r_\star$) in our simulation, and radiative transfer in the CGM has a relatively small effect on the \lya line shapes and intensities. Moreover, the radius encompassing 80\% of the escaping \lya flux ($r_{80}$; see Figure \ref{fig:rcgm}) is 2-3 times smaller than $R_{\scaleto{\rm CGM}{3.5pt}}$ for the vast majority of galaxies. Therefore, the \lya luminosities emerging from our galaxies should only weakly depend on the exact CGM scale at which they are measured and the expected aperture flux losses are thus moderate.

\label{lastpage}

\end{document}